\RequirePackage{fix-cm}

\documentclass[twocolumn,epjc3]{svjour3}  
\smartqed  
\RequirePackage{amsmath}
\RequirePackage{amssymb}

\RequirePackage{graphicx}
\RequirePackage{mathptmx}      
\usepackage{enumitem} 

\usepackage{cite} 
\usepackage{subcaption}
\usepackage[usenames,dvipsnames,svgnames,table]{xcolor}
\usepackage[colorlinks = true,
            linkcolor = Navy,
            urlcolor  = Navy,
            citecolor = Navy,
            anchorcolor = Navy]{hyperref}

\usepackage{float}
\usepackage{subcaption} 
\usepackage{url}
\usepackage{booktabs}
\usepackage{multirow}

\usepackage{lineno}
\journalname{Eur. Phys. J. C}

\begin{document}

\title{Analysing parameter space correlations of recent 13 TeV gluino and squark searches in the pMSSM
}


\author{Alan Barr\thanksref{e1,addr1}
        \and
        Jesse Liu\thanksref{e2,addr1} 
}

\thankstext{e1}{e-mail: alan.barr@physics.ox.ac.uk}
\thankstext{e2}{e-mail: jesse.liu@physics.ox.ac.uk}


\institute{ Department of Physics, University of Oxford, 
 Oxford OX1 3RH, United Kingdom \label{addr1}
}

\date{\vskip-1.5cm 
}

\maketitle

\begin{abstract}
This paper examines unexplored correlations in the parameter spaces probed by recent ATLAS analyses for gluinos and squarks, addressing various shortcomings in the literature. Six 13~TeV ATLAS analyses based on 3.2~fb$^{-1}$ of integrated luminosity are interpreted in the 19-parameter R-parity conserving phenomenological minimal supersymmetric extension to the Standard Model (pMSSM). The distinct regions covered by each search are independent of prior, and we reveal particularly striking complementarity between the 2--6 jets and Multi-b searches. In the leptonic searches, we identify better sensitivity to models than those used for analysis optimisation, notably a squark--slepton--wino scenario for the SS/3L search. Further, we show how collider searches for coloured states probe the structure of the pMSSM dark sector more extensively than the Monojet analysis alone, with sensitivity to parameter spaces that are challenging for direct detection experiments.

\end{abstract}

\section{\label{sec:intro}Introduction}

The ATLAS and CMS collaborations pursue a rich 
programme of supersymmetry (SUSY) searches, 
but statistically significant signals remain absent~\cite{Aaboud:2016zdn,Aad:2016jxj,Aad:2016qqk,Aad:2016eki,Aad:2016tuk,Aaboud:2016tnv,Aaboud:2016dgf,Aaboud:2016lwz,Aaboud:2016uth,Aaboud:2016nwl,Khachatryan:2016xvy,Khachatryan:2016kdk,Khachatryan:2016kod,Khachatryan:2016uwr,Aad:2014wea,Aad:2013yna,Aad:2015iea,Aad:2015pfx,Aad:2015eda,Khachatryan:2015wza,Khachatryan:2015vra}. 
While conventional theoretical expectations for 
weak scale SUSY are challenged~\cite{Cassel:2009ps,Papucci:2011wy,Baer:2012up,Arvanitaki:2012ps,Arvanitaki:2013yja,Evans:2013jna,Hardy:2013ywa,Baer:2014ica,Dimopoulos:2014aua}, 
this equally motivates assessment of present experimental search strategies. 
Simplified models introduce a small number of kinematically accessible 
superpartner particles (sparticles), which are typically used to design, 
optimise and interpret collider searches~\cite{Alwall:2008ag,Alwall:2008va,Alves:2011sq,Alves:2011wf}. 

A broader framework for assessing the robustness of such strategies is the phenomenological minimal supersymmetric extension to the Standard Model (pMSSM)~\cite{Hall:1990ac,Berger:2008cq,CahillRowley:2012cb,CahillRowley:2012kx,Cahill-Rowley:2014twa,AbdusSalam:2009qd,Allanach:2011ej,Sekmen:2011cz,Strege:2014ija,deVries:2015hva}, which also facilitates dark matter (DM) interpretations~\cite{Arbey:2012na,Boehm:2013gst,Fowlie:2013oua,Cahill-Rowley:2014boa,Roszkowski:2014iqa,Bertone:2015tza,Catalan:2015cna,Aaboud:2016wna}. Notably, the ATLAS collaboration examined the sensitivity of 22 Run~1 analyses within the context of a 19-parameter pMSSM~\cite{Aad:2015baa}, while CMS undertook a similar survey using different assumptions~\cite{Khachatryan:2016nvf}. 

This paper extends the ATLAS pMSSM study~\cite{Aad:2015baa} using fast detector simulation to include combined constraints from six early 13~TeV searches, and we make this exclusion information available online~\cite{Barr:2016inz}. Instead of preliminary results with more data, we opt to use published results based on 3.2~fb$^{-1}$ of integrated luminosity, which sets the scope of our study to gluinos and light flavour squarks.
Our analysis reveals previously unexamined correlations in the pMSSM space that are already sufficiently elaborate to merit detailed evaluation of search strategies. These results are also of interest to the CMS collaboration and direct dark matter detection experiments. We organise our discussion around the following questions that address various shortcomings in the literature:

\begin{enumerate}[label=(\alph*)]
    \item \textit{How distinct are the regions of parameter space being probed by individual analyses?} Interpretations using the pMSSM often present combined constraints from multiple searches as fractions of models excluded~\cite{Cahill-Rowley:2014twa,Cahill-Rowley:2014boa,Aad:2015baa,Kowalska:2016ent,Khachatryan:2016nvf}. Overlap matrices were recently used in the literature~\cite{Aad:2015baa,Barr:2016inz} to quantify the complementarity of these searches, namely the fractional exclusion of the same subset of points by two analyses. However, this marginalisation not only obscures which analyses had greatest sensitivity to different pMSSM subspaces, but also depends on the prior distribution from the parameter scan and non-LHC constraints. 
    
    \item \textit{To what extent are analyses over-optimising to a set of simplified models, which may preclude sensitivity to a wider class of scenarios?} The pMSSM offers a greater variety of decay chains, such as those with suppressed branching fractions and placing those with more intermediate sparticles on-shell, which alter kinematics. How well these simplified model oriented searches are capturing the wider classes of signatures remains relatively unexplored.
    
    \item \textit{What neutralino DM scenarios can be probed competitively by 13~TeV collider searches for coloured sparticles?} Simplified dark matter models are stimulating the collider frontier for DM searches~\cite{Abdallah:2014hon,Abdallah:2015ter,Abercrombie:2015wmb,Boveia:2016mrp,Bauer:2016gys,Aad:2015dva,Aaboud:2016uro,Aaboud:2016qgg,Khachatryan:2014rra,Khachatryan:2016reg,Khachatryan:2016mdm}. ATLAS uses these to perform an explicit DM interpretation for the Monojet search~\cite{Aaboud:2016tnv}. However, its sensitivity is greatly altered in richer dark sectors of the pMSSM, shaped by the composition of the neutralino $\tilde{\chi}^0_1$ lightest SUSY particle (LSP)~\cite{Ellis:1983ew,Bertone:2004pz,Jungman:1995df,Feng:2000zu,Baer:2004qq} being dominantly bino, Higgsino or wino (defined in Table~\ref{tab:LSPtype} of~\ref{sec:ATLAS_pMSSM19}). Furthermore, studies of neutralino DM often combine exclusions of multiple analyses~\cite{Cahill-Rowley:2014boa,Aad:2015baa} or only focus on the electroweakino sector~\cite{Martin:2014qra,Bramante:2015una,Barducci:2015ffa,Aaboud:2016wna}, omitting potentially important coannihilation roles of other sparticles. 
      
\end{enumerate}

Section~\ref{sec:detailed} addresses question (a) by directly correlating the most sensitive of the six analyses considered with the masses of the gluino, LSP and lightest squark. We examine the individual searches that provide sensitivity to distinct regions within these mass plane projections. The regions identified are both prior-independent and reveal significantly richer information than the overlap matrices previously used in literature. 

Subsection~\ref{sec:leptonic_searches} addresses question (b) by considering the analyses that select leptonic events as a case study. We examine to what extent the simplified models used by the 1-lepton analysis~\cite{Aad:2016qqk} and the same-sign or 3-lepton (SS/3L) search~\cite{Aad:2016tuk} map onto pMSSM points, and identify scenarios beyond those considered for analysis optimisation.

Section~\ref{sec:darkmatter} addresses question (c) by ascribing DM interpretations to each 13~TeV search considered, allowing for comparisons to the Monojet search. Striking correlations are exhibited, and we discuss how distinct decay cascades are influenced by the bino, Higgsino or wino content of the LSP, while coloured sparticles may act as early universe coannihilators~\cite{Edsjo:1997bg}. Further, we identify the collider parameter space affected by recent limits from Xenon-target direct detection experiments~\cite{Tan:2016zwf,lux-si-2016}.

Section~\ref{sec:conclusion} summarises the conclusions of this work.

The pMSSM points we investigate were produced for the ATLAS study~\cite{Aad:2015baa}. The 19 parameters of the R-parity conserving MSSM, where all flavour and CP violation resides in the CKM matrix, were scanned with flat priors with sparticle mass scales capped at 4~TeV~\cite{Aad:2015baa}. However, the distinct regions of sensitivity we identify in our results are independent of sampling prior. Constraints were imposed from LEP searches~\cite{lep2:susy}, precision electroweak measurements~\cite{ALEPH:2005ab,Baak:2012kk,Aoyama:2012wk,Hagiwara:2011af,Nyffeler:2009tw,Czarnecki:2002nt,Bennett:2004pv,Bennett:2006fi,RevModPhys.80.633,Roberts:2010cj}, heavy flavour physics~\cite{Amhis:2012bh,DeBruyn:2012wk,CMS:2014xfa,Aubert:2009wt,Hara:2010dk,Adachi:2012mm,Lees:2012ju,Charles:2004jd}, DM direct detection~\cite{Akerib:2013tjd,Behnke:2012ys,Aprile:2013doa} and the Planck relic density~\cite{Ade:2015xua} upper bound\footnote{A recent independent study~\cite{Kowalska:2016ent} considered ATLAS constraints with up to 14.8~fb$^{-1}$, but on the subset of points whose neutralino relic abundance was within 10\% of Planck. We instead take the Planck measurement as an upper bound, allowing for non-SUSY contributions to DM.}. Further details of the theoretical assumptions and experimental constraints used by ATLAS~\cite{Aad:2015baa} may be found in~\ref{sec:ATLAS_pMSSM19}. We consider the 181.8k points that survive Run~1 constraints from Ref.~\cite{Aad:2015baa}, after having removed models with long-lived ($c\tau > 1$~mm as defined in Ref.~\cite{Aad:2015baa}) gluinos, squarks and sleptons as these require dedicated Monte-Carlo simulation. The methodology we employed to interpret the six 13 TeV searches using fast detector simulation is detailed in \ref{sec:13TeV_method}.

\section{\label{sec:detailed}Complementarity between early 13 TeV searches}

For the six 13~TeV (3.2~fb$^{-1}$) ATLAS searches considered in Table~\ref{tab:listSearches}, we present the regions of sensitivity between these analyses. In~\ref{sec:summary_excl_spart}, we present results using existing practices in the literature, based on fractions of models excluded (marginalised distributions) and overlap matrices. This section addresses various shortcomings of these approaches as discussed in the Introduction. Subsection~\ref{sec:most_sensitive} defines and quantifies the `most sensitive analysis' used to exclude the points in our interpretation. Following this, Subsection~\ref{sec:mass_plane_features} projects this information into 2-dimensional subspaces of the pMSSM involving gluinos, squarks and the LSP, and discusses prior features in these planes. Finally, we examine the complementary sensitivity of each analysis to distinct regions of pMSSM parameter space in Subsection~\ref{sec:compl_searches}, partitioning our discussion between analyses that veto events with leptons with those that select on them. 

Henceforth in this paper, `squark' $\tilde{q}$ refers to only the lightest superpartner of the left- or right-handed quark of the first or second generations $\tilde{q}\in\{\tilde{u}, \tilde{d}, \tilde{c}, \tilde{s}\}_{L,R}$. Similarly, `slepton' $\tilde{\ell}$ refers to the lightest superpartner of the left- or right-handed lepton of the first or second generation 
\begin{equation}
\tilde{\ell} \in \{\tilde{e}_L, \tilde{\nu}_{e L}, \tilde{\mu}_L, \tilde{\nu}_{\mu L},\tilde{e}_R, \tilde{\mu}_R\}.
\end{equation}

\subsection{\label{sec:most_sensitive}Most sensitive analyses used for exclusion}

\begin{table*}[]
\centering
\begin{tabular}{lcccc}
\toprule
Analysis    & Reference              & $N_\textrm{Lowest CLs}/N^\textrm{Excluded}_\textrm{13 TeV 3.2 fb$^{-1}$}$ & $N^\textrm{Excluded}_\textrm{13 TeV 3.2 fb$^{-1}$}/N^\textrm{Survived}_\textrm{ATLAS Run 1}$ & $N^\textrm{Excluded Uniquely}_\textrm{13 TeV 3.2 fb$^{-1}$}/N^\textrm{Survived}_\textrm{ATLAS Run 1}$ \\ 
\midrule
2--6 jets   & \cite{Aaboud:2016zdn}  & 72\%                    & 12.6\%    & 11\%   \\ 
7--10 jets  & \cite{Aad:2016jxj}     & 0.3\%                   & 0.6\%     & 0.02\% \\
1-lepton    & \cite{Aad:2016qqk}     & 1.5\%                   & 1.0\%     & 0.2\%  \\
Multi-b     & \cite{Aad:2016eki}     & 23\%                    & 4.2\%     & 3.5\%   \\
SS/3L       & \cite{Aad:2016tuk}     & 2.7\%                   & 0.5\%     & 0.4\%  \\ 
Monojet     & \cite{Aaboud:2016tnv}  & 1.1\%                   & 3.3\%     & 0.01\% \\ 
\midrule
All analyses& --                     & 100\%                   & 15.7\%    & 15.1\%   \\ 
\bottomrule
\end{tabular}
\caption{\label{tab:listSearches} List of the ATLAS 13~TeV (3.2~fb$^{-1}$) analyses used to constrain the 181.8k model points that survived Run~1. The column $N_\textrm{Lowest CLs}/N^\textrm{Excluded}_\textrm{13 TeV 3.2 fb$^{-1}$}$ denotes the fraction of the 28.5k excluded models for which the indicated analysis was the most sensitive, i.e. had the lowest CLs value $N_\textrm{Lowest CLs}$ as discussed in Section~\ref{sec:most_sensitive}; these figures may not sum to 100\% due to rounding. The column $N^\textrm{Excluded}_\textrm{13 TeV 3.2 fb$^{-1}$}/$N$^\textrm{Survived}_\textrm{ATLAS Run 1}$ is the `fractional exclusion' displaying the total percentage of points excluded by each of the analyses $N^\textrm{Excluded}_\textrm{13 TeV 3.2 fb$^{-1}$}$ out of the points that survived ATLAS Run 1 $N^\textrm{Survived}_\textrm{ATLAS Run 1}$. The right-most column quantifies the subset of points that are excluded uniquely by the corresponding analysis and not by any of the other five considered $N^\textrm{Excluded Uniquely}_\textrm{13 TeV 3.2 fb$^{-1}$}$. For all these figures, care must be taken with interpretation as they are prior dependent. Models with long-lived gluinos, squarks and sleptons are not considered.}
\end{table*}

In this study, a point is deemed excluded at 95\% confidence level if at least one analysis returned a CLs value less than 0.05, using the CLs prescription~\cite{Read:2002hq}. Of the 181.8k points that survived Run~1~\cite{Aad:2015baa}, a total of $N^\textrm{Excluded}_\textrm{13 TeV 3.2 fb$^{-1}$} = 28.5{\rm k}$ were excluded by our interpretation of the six 13~TeV analyses. We take the analysis with the smallest CLs value as the `most sensitive analysis' used to exclude the point. In Table~\ref{tab:listSearches}, we normalise the number of points satisfying this for each analysis $N_\textrm{Lowest CLs}$ to the total excluded $N^\textrm{Excluded}_\textrm{13 TeV 3.2 fb$^{-1}$}$. For example, the Multi-b search was the analysis with the lowest CLs value for 23\% of the 28.5k excluded points. Indeed, almost 95\% of the excluded points have either the 2--6 jets or Multi-b searches being the most sensitive. However, care must be taken when interpreting these fractions, as they are prior dependent and correlated with non-LHC constraints. The fractions indicate the relative number of points in forthcoming figures.

Less than 4\% of the excluded points have two or more analyses associated with the same smallest CLs value. In such cases, the `most sensitive analysis' is randomly chosen from this subset of analyses with smallest CLs value to minimise systematic selection bias. In the vast majority of these situations, this is done because the analyses share a CLs value of $0.0$.

Also displayed in Table~\ref{tab:listSearches} is the total number of excluded points by each analysis $N^\textrm{Excluded}_\textrm{13 TeV 3.2 fb$^{-1}$}$ out of those that survived Run 1 $N^\textrm{Survived}_\textrm{ATLAS Run 1}$. The relative overlap between analyses is quantified in Table~\ref{tab:overlap} in \ref{sec:summary_excl_spart}. Importantly in Table~\ref{tab:listSearches}, all analyses retain non-zero percentages in the total fraction of points uniquely excluded by each analysis and none of the other five $N^\textrm{Excluded Uniquely}_\textrm{13 TeV 3.2 fb$^{-1}$}$. This emphasises the importance of maintaining a broad programme of searches. While most of the searches optimise for gluino production, and some have dedicated signal regions for squarks, each search maintains unique sensitivity to particular classes of signatures within the MSSM, featuring distinct final states and kinematic regimes. 

\subsection{\label{sec:mass_plane_features}Features in mass plane projections}

\begin{figure*}
    \centering
        \begin{subfigure}[b]{0.70\textwidth}
         \includegraphics[width=\textwidth]{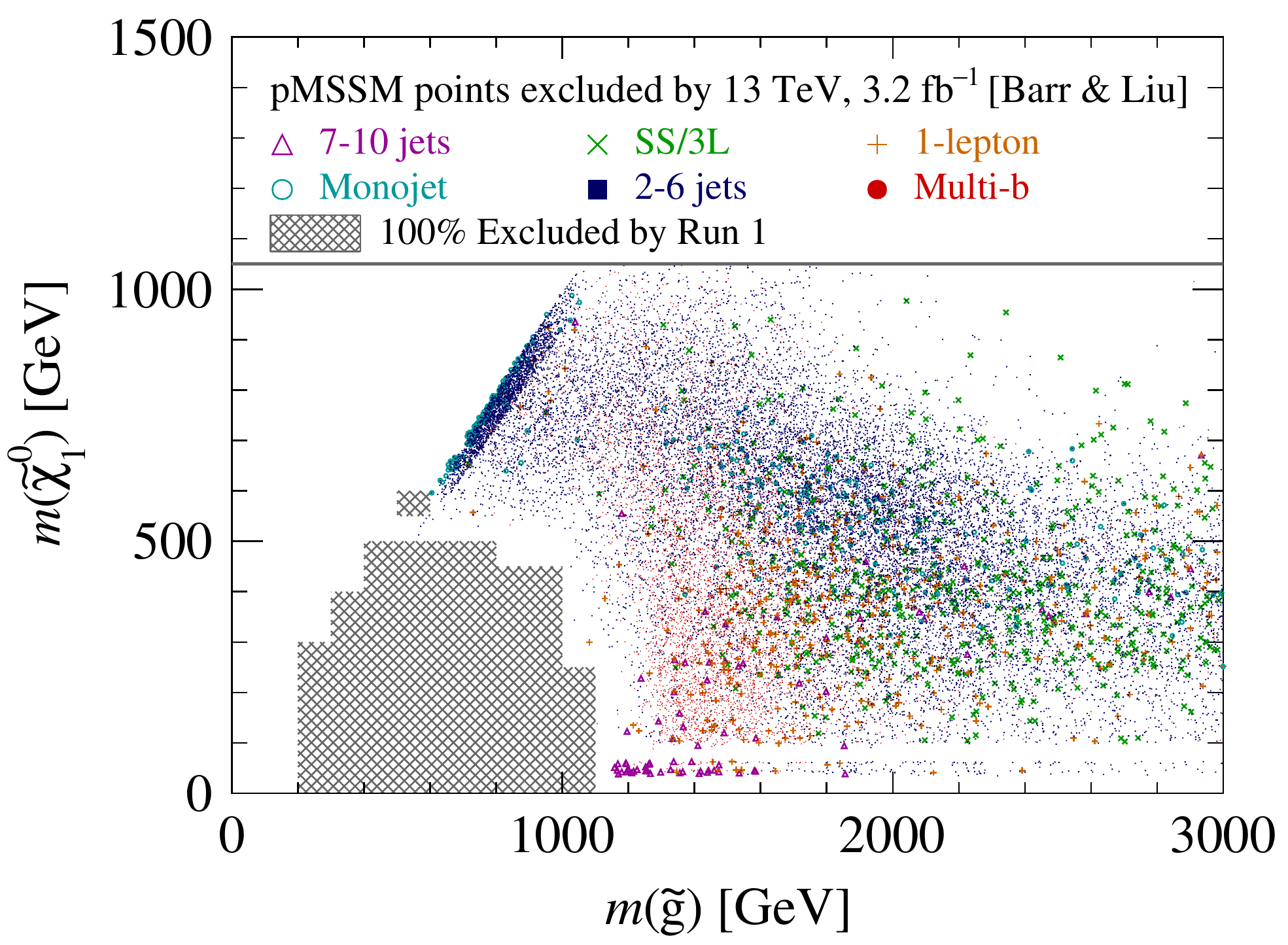}
         \caption{\label{fig:bySearch_Gl_LSP}Gluino--LSP.}
        \end{subfigure}
        \begin{subfigure}[b]{0.70\textwidth}
        \includegraphics[width=\textwidth]{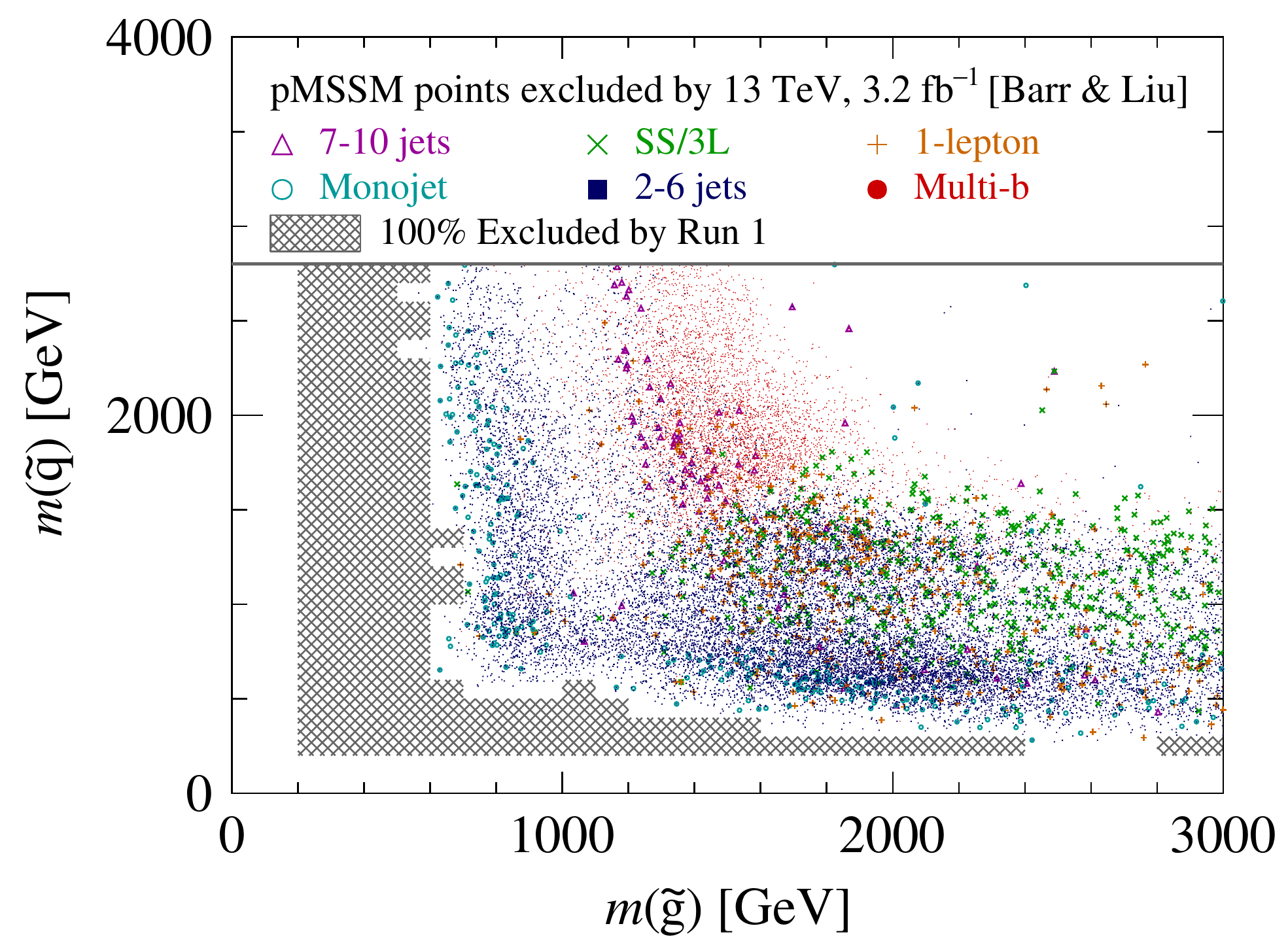}
         \caption{\label{fig:bySearch_Gl_Sqk}Gluino--Squark.}
        \end{subfigure}
    \caption{Most sensitive analysis for each of the 28.5k points excluded at 95\% confidence level by the six 13 TeV searches displayed in Table~\ref{tab:listSearches}. The exclusion information is from Ref.~\cite{Barr:2016inz}. These are projected into the mass planes described in the captions. Here $m(\tilde{q})$ denotes the lightest squark of the first or second generations. The markers are styled according to the analysis with the lowest CLs value: 7--10 jets (magenta triangle), 1-lepton (orange plus), SS/3L (green cross), Monojet (cyan ring), 2--6 jets (blue filled square), and Multi-b (red filled circle). Due to their large numbers, the markers for the 2--6 jets and Multi-b analyses are reduced in size for clarity. The hatched grey regions indicate the mass bins where all model points were completely excluded by Run~1 searches considered in Ref.~\cite{Aad:2015baa}. Models with long-lived gluinos, squarks or sleptons are removed from these figures.
    }
    \label{fig:bySearch_summary}
\end{figure*} 

\begin{figure*}
    \centering
         \includegraphics[width=0.75\textwidth]{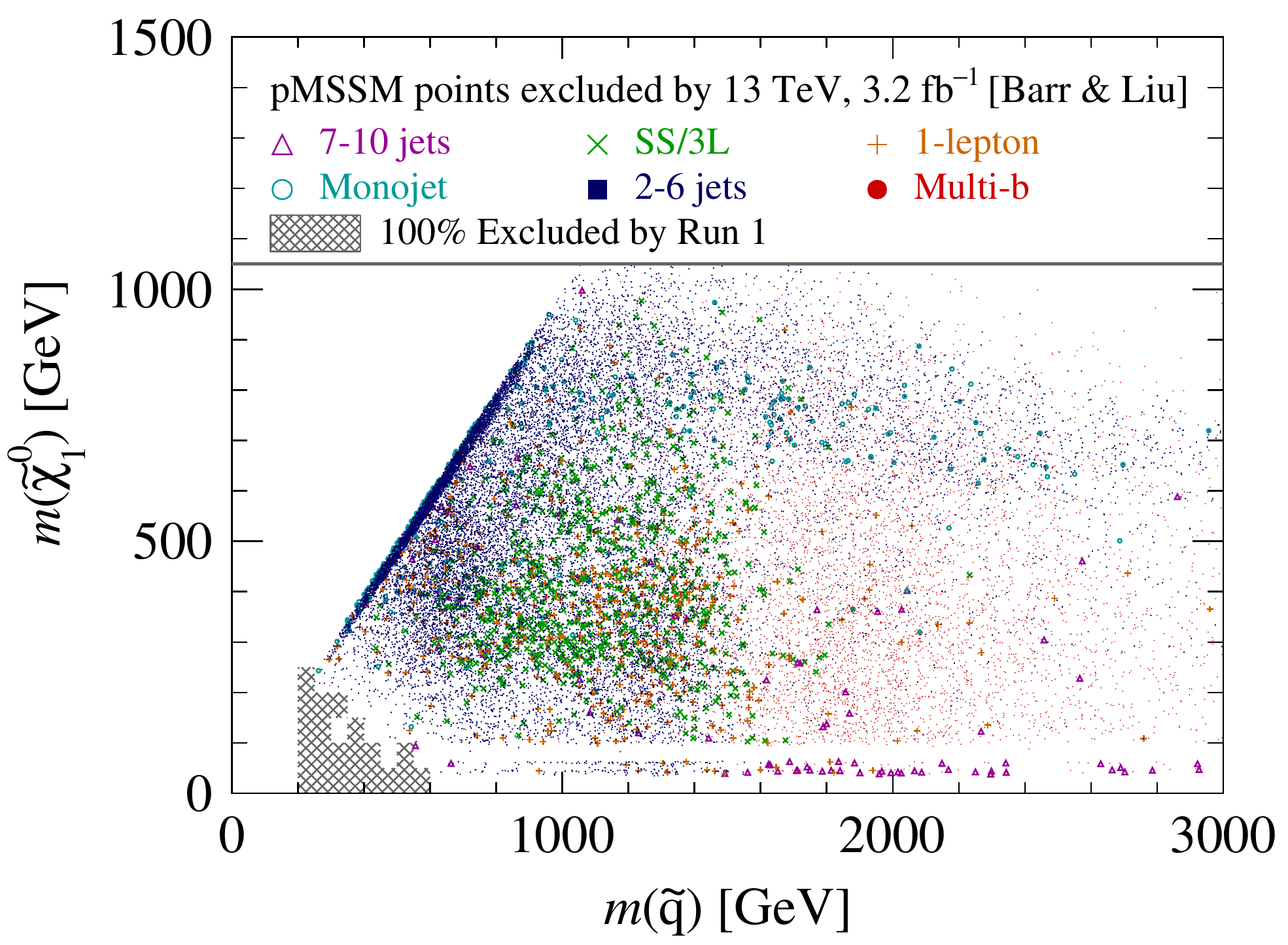}
    \caption{\label{fig:bySearch_Sqk_LSP}Same plot as Figure~\ref{fig:bySearch_Gl_LSP} but projected into the squark--LSP plane. Here $m(\tilde{q})$ denotes the lightest squark of the first or second generations. 
    }
\end{figure*} 

The distinct regions of sensitivity for each of the six searches become unambiguous when we project into various two-dimensional subspaces of the pMSSM. Figures~\ref{fig:bySearch_summary} and \ref{fig:bySearch_Sqk_LSP} display each excluded point styled and coloured according to the analysis that returned the lowest CLs value, projected into the mass planes of gluino vs LSP, gluino vs squark and squark vs LSP respectively. Due to their large numbers, the 2--6 jets and Multi-b analyses are allocated smaller markers to improve clarity of other points. 

Before discussing the sensitivity of each analysis in turn (Section~\ref{sec:compl_searches}), we note that the two-dimensional \textit{distributions} (but not the \textit{regions} to be discussed in Sections~\ref{sec:compl_searches} and \ref{sec:13TeV_on_dm}) of excluded points do depend on the flat prior and importance sampling ATLAS used to scan the parameter space~\cite{Aad:2015baa}, in addition to non-LHC constraints. This is especially apparent in the mass planes involving the LSP (Figures~\ref{fig:bySearch_Gl_LSP} and~\ref{fig:bySearch_Sqk_LSP}), and we comment on several features common to these figures to aid interpretation:
\begin{itemize}
    \item Within around 100~GeV of the grey hatched regions where 100\% of models in the mass bins were excluded by Run~1 ATLAS searches, very few points are present. The number of models close to this boundary is often less than ten\footnote{The prior distributions for these plots are shown in Figure~\ref{fig:2Dprior_distro} of the Appendix.}. Typically, such points marginally survived Run~1 constraints and many have since been excluded by the six 13~TeV searches.

    \item There is a visible break in points at $m(\tilde{\chi}_1^0) \sim 100$~GeV. Below this mass, there are few models with Higgsino- or wino-like LSP as these generally have a near-degenerate chargino, which are excluded by LEP direct searches. For the definition of the three LSP classifications, see Table~\ref{tab:LSPtype} in~\ref{sec:ATLAS_pMSSM19}.
    
    \item The points below $m(\tilde{\chi}_1^0) \sim 100$~GeV are therefore predominantly bino-like LSP models. These LSPs are close to half the mass of the $Z^0$ or Higgs boson $h^0$ so they can undergo resonant annihilation in the early universe in so-called `funnel regions' to satisfy relic abundance constraints.
    
    \item A large density of points is visible along the diagonals where the mass splitting between the LSP and the gluino or squark become small ($\lesssim 50$~GeV). The enhanced density of points in this coannihilation region has physical origin: bino-like LSPs tend to oversaturate the observed relic abundance~\cite{ArkaniHamed:2006mb} unless there is a near mass-degenerate sparticle, such as a squark or gluino, to act as an early universe coannihilator. This effect is made more apparent by using importance sampling to ensure the number of bino-like LSP models is of the same order as those of Higgsino- and wino-like LSP; see Ref.~\cite{Aad:2015baa} for details. 
\end{itemize}

\subsection{\label{sec:compl_searches}Mass plane correlations of searches for squarks and gluinos}

This subsection discusses the distinct regions of sensitivity correlated with each analysis considered. Importantly, the identified regions in the two-dimensional planes are independent of the prior distribution of points. 

\subsubsection{Searches with a lepton veto}

First, we investigate the excluded points where each of the 2--6 jets, Multi-b, 7--10 jets, and Monojet searches are most sensitive, in that order. These analyses all veto on leptons, and exhibit the most unambiguous correlations in the mass planes of the gluino, squark and LSP (Figures~\ref{fig:bySearch_summary} and~\ref{fig:bySearch_Sqk_LSP}). Generally, these analyses select events with transverse momentum imbalance of magnitude $E_\mathrm{T}^{\rm miss}$, varying jet multiplicity, together with discriminants dependent on mass scale or flavour. For full details, refer to the ATLAS references in Table~\ref{tab:listSearches}.

\paragraph{2--6 jets} 

Figure~\ref{fig:bySearch_Gl_LSP} reveals the extensive distribution in the gluino--LSP plane where the 2--6 jets is the most sensitive analysis (blue points). The 2--6 jets search uses the effective mass discriminant, with varying degrees of minimum jet multiplicity to target models where a squark or gluino directly decays to the LSP. Meanwhile, the larger jet multiplicity regions target scenarios where a chargino mediates the decay of the gluino to the LSP. 

This analysis has almost exclusive sensitivity to points with gluino $\tilde{g}$ masses below 1~TeV, where its mass splittings with the LSP in the range $25 \lesssim m(\tilde{g}) - m(\tilde{\chi}^0_1) \lesssim 500$~GeV. Larger mass splitting scenarios for sub-TeV mass gluinos are excluded by Run~1 searches (hatched grey mass regions). For regions with gluinos above 1~TeV with large gluino--LSP mass splittings, we find the 2--6 jets has reduced sensitivity compared with other analyses, especially for LSP masses $m(\tilde{\chi}^0_1) \lesssim 500$~GeV. For gluino masses $m(\tilde{g}) \gtrsim 2$~TeV, we expect reduced sensitivity to gluinos, but points in this region are correlated with excluded points involving low mass squarks. This is confirmed by comparing with the gluino--squark plane (Figure~\ref{fig:bySearch_Gl_Sqk}). Indeed, the 2--6 jets search is also predominantly the most sensitive analysis for light squarks in regions not far beyond Run~1 sensitivity. High mass gluinos in such scenarios can nevertheless contribute to production cross-sections of the squarks as a mediator via $t$-channel diagrams.

The gluino--squark plane (Figure~\ref{fig:bySearch_Gl_Sqk}) also reveals a vertical strip with a lower density of points around gluino mass between about 1~TeV and 1.2~TeV. This corresponds to a region where a lower fraction of models is excluded per mass interval (see also Figures~\ref{fig:gsqk_fracExcl} and \ref{fig:2D_prior_Gl_Sqk} in~\ref{sec:method}). This reduced sensitivity, and lack of any other dedicated analysis probing this region, is due to gluino--LSP mass being moderately small (between around 25 and 200~GeV). Such scenarios are challenging for traditional `missing energy plus jets' searches, but represent the greatest potential for high luminosity where novel techniques are being developed to target such regions~\cite{Jackson:2016mfb}.

Further interpretations for the 2--6 jets analysis apply in the squark--LSP plane (Figure~\ref{fig:bySearch_Sqk_LSP}). The search has sensitivity to a wide variety of squark mass scenarios, gradually reducing above $m(\tilde{q})\sim 1$~TeV. The strip of blue points for $m\left(\tilde{q}\right)\gtrsim 1.5$~TeV and $600 \lesssim m\left(\tilde{\chi}^0_1\right)\lesssim 900$~GeV where we would expect reduced squark sensitivity are correlated with low mass gluinos. These points largely have sub-TeV gluinos, an interpretation, again confirmed in the gluino--squark plane (Figure~\ref{fig:bySearch_Gl_Sqk}). 

\paragraph{Multi-b} 

\begin{figure}
    \centering
         \includegraphics[width=0.50\textwidth]{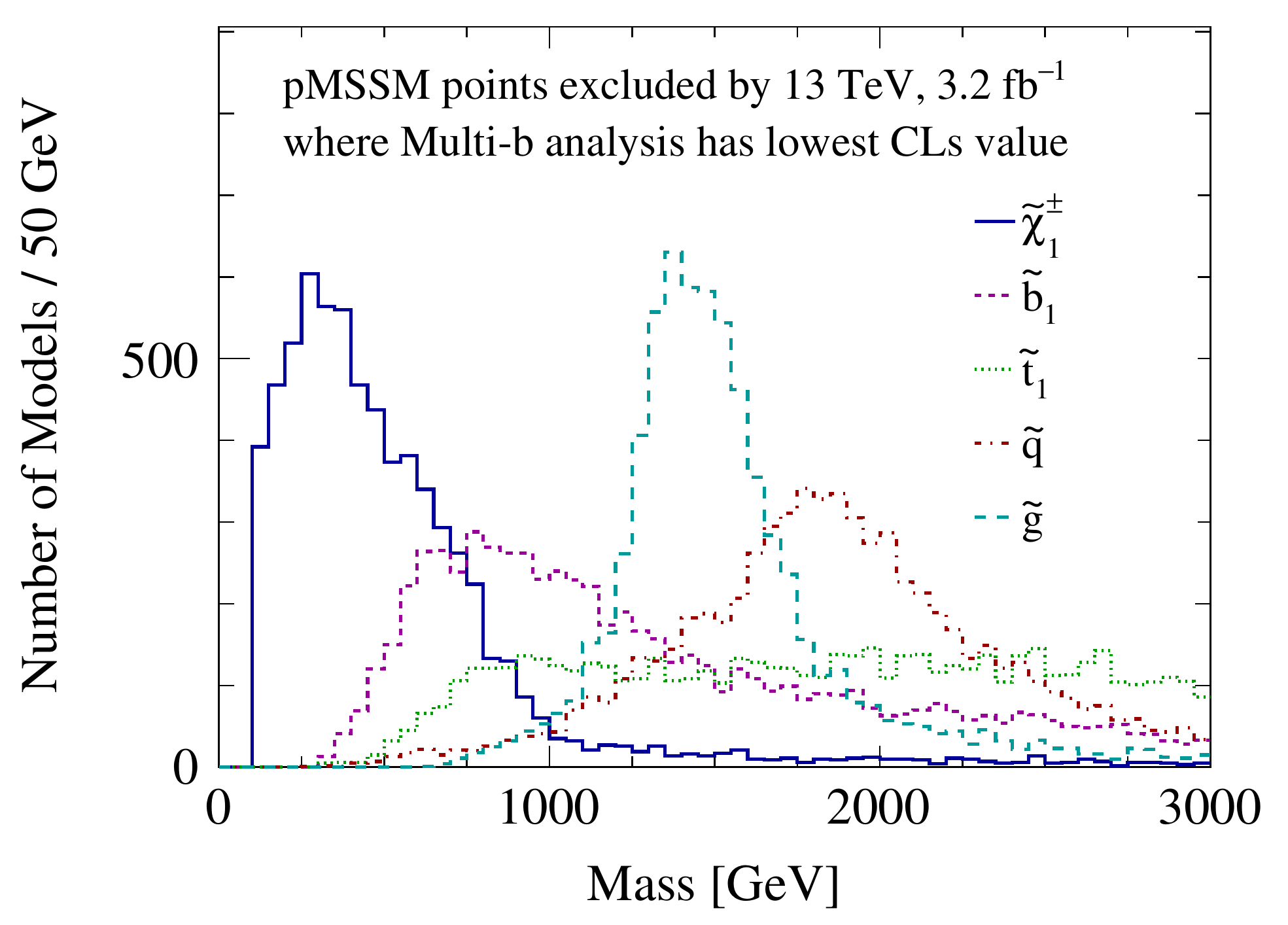}
    \caption{Distribution of sparticle masses for the excluded points where the Multi-b analysis is most sensitive. The lines correspond to the chargino $\tilde{\chi}^\pm_1$ (blue solid), lightest sbottom $\tilde{b}_1$ (magenta short-dashed), lightest stop $\tilde{t}_1$ (green dotted), lightest squark $\tilde{q}$ (red dot-dashed) and gluino $\tilde{g}$ (light-blue long-dashed).
    }
    \label{fig:multib_mass_distro}
\end{figure} 

Where the Multi-b analysis has most sensitivity (red points) is localised to regions complementary to the 2--6 jets analysis. In the gluino--LSP plane (Figure~\ref{fig:bySearch_Gl_LSP}), the region is enclosed by gluino masses of $1.2 \lesssim m(\tilde{g}) \lesssim 1.7$~TeV and LSP masses of $100 \lesssim m\left(\tilde{\chi}^0_1\right)\lesssim 500$~GeV. This sensitivity also extends to smaller gluino--LSP mass splittings, though to a lesser extent. 

The distinctiveness of this region of sensitivity remains in the gluino--squark plane (Figure~\ref{fig:bySearch_Gl_Sqk}). Here, points where the Multi-b analysis is most sensitive are highly correlated with squark masses above 1.5~TeV, largely untouched by other analyses, being strikingly separated from 2--6 jets and to some extent the 7--10 jets analysis. Figure~\ref{fig:bySearch_Sqk_LSP} also confirms such correlations of the Multi-b in the squark--LSP plane and again sensitivity to squarks above 1.5~TeV is largely from gluino rather than squark production.

These correlations arise from the Multi-b search targeting $\tilde{g}\to b\bar{b}\tilde{\chi}^0_1$ and $\tilde{g}\to t\bar{t} \tilde{\chi}^0_1$ models with heavy gluino production decaying via off-shell stops $\tilde{t}$ and sbottoms $\tilde{b}$. This analysis selects events with large $E_\mathrm{T}^{\rm miss}$ and at least three jets originating from bottom quarks. A subset of the signal regions selects loosely on boosted top quarks decaying from gluinos, including events enriched with lepton presence. 

Figure~\ref{fig:multib_mass_distro} takes the points where the Multi-b is most sensitive and illuminates the mass distributions of various pertinent sparticles. The light flavour squarks $\tilde{q}$ are centred around 2~TeV with gluinos around 1.4~TeV, where Figure~\ref{fig:bySearch_Gl_Sqk} indicates that squarks are predominantly heavier than gluinos. This fact suppresses gluino decays to light flavour quarks, which proceed via the three body $\tilde{g} \to q \bar{q} \tilde{\chi}^0_1$ process. By contrast, the mass distribution of sbottoms $\tilde{b}_1$ peaks around 800~GeV in Figure~\ref{fig:multib_mass_distro} and have a preference to be lighter than the gluinos, allowing on-shell $\tilde{g}\to \tilde{b}b$ decays to be favoured. Indeed, this demonstrates favourable sensitivity to the on-shell counterpart of the simplified models considered for optimisation by ATLAS. Meanwhile, the distribution of stop masses is relatively uniform for $m(\tilde{t}_1)\gtrsim 900$~GeV. The requirement of three or more jets originating from bottom quarks therefore favours such scenarios. We furthermore note that out of the models excluded where the Multi-b is most sensitive, 56\% models have a Higgsinos-like LSP while 22\% are wino-like, which have light charginos consistent with the corresponding mass distribution in Figure~\ref{fig:multib_mass_distro}. This preference of the Multi-b analysis for Higgsino-like LSP models can be understood by the higher Yukawa couplings to heavy flavour quarks, which also enhance decays of the gluino to bottom and/or top quarks.  

\begin{figure*}
    \centering
        \begin{subfigure}[b]{0.49\textwidth}
         \includegraphics[width=0.95\textwidth]{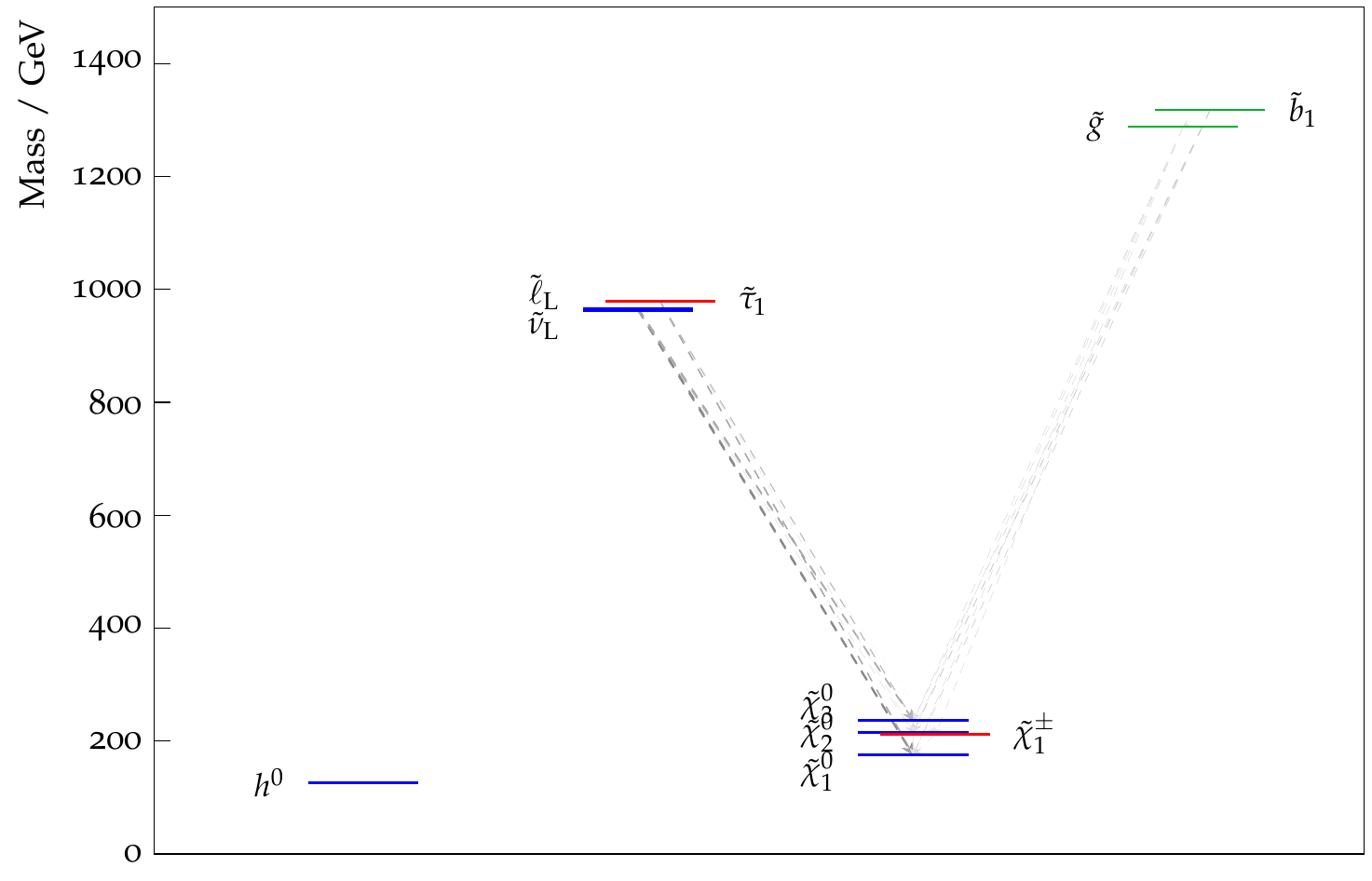}
         \caption{\label{fig:Multi_b_model}Model 148229034 (Multi-b).}
        \end{subfigure}
        \begin{subfigure}[b]{0.49\textwidth}
         \includegraphics[width=0.95\textwidth]{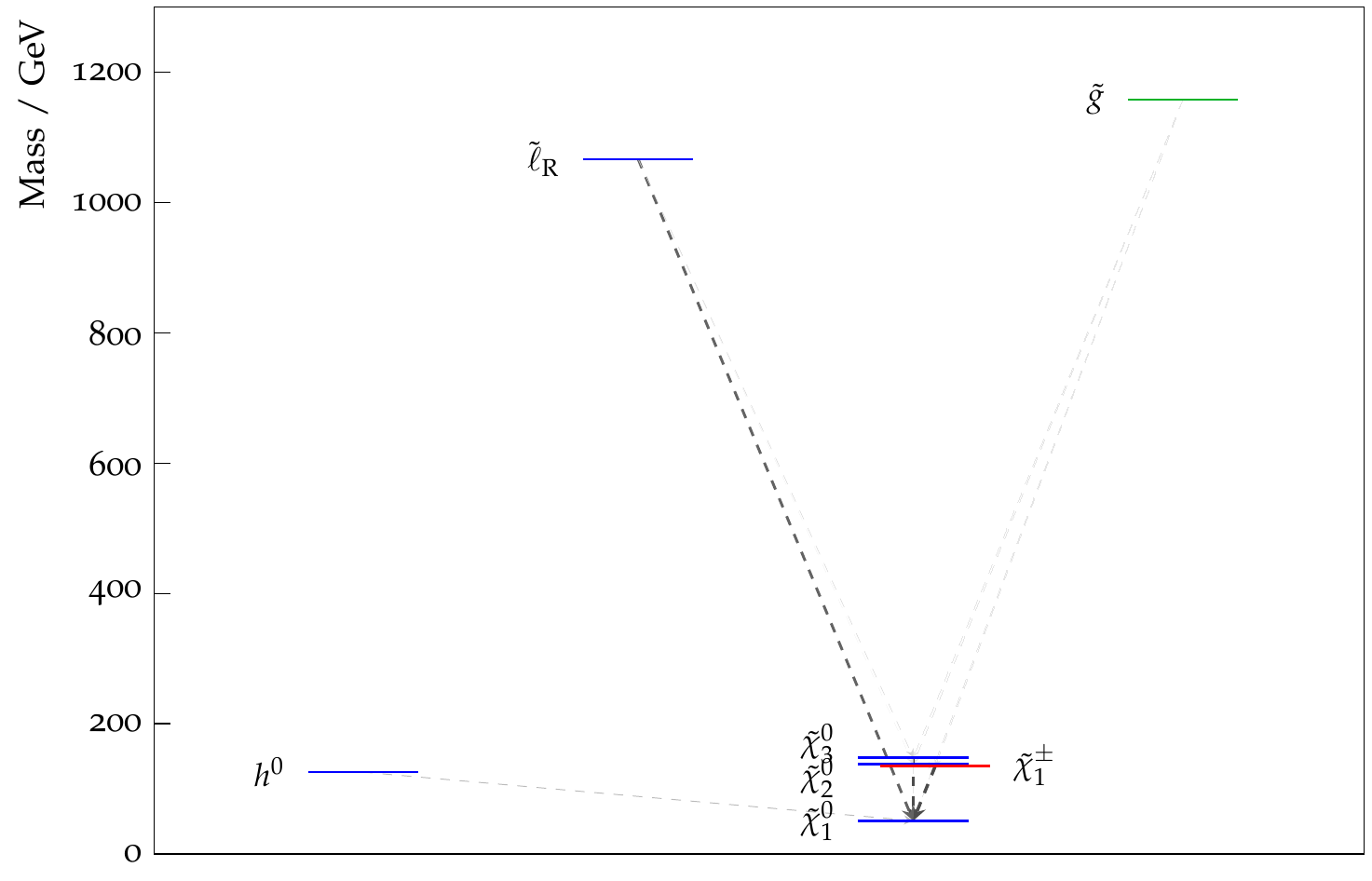}
         \caption{\label{fig:7_10_jets_model}Model 227558023 (7--10 jets).}
        \end{subfigure}
        \\[0.8cm]
        \begin{subfigure}[b]{0.49\textwidth}
         \includegraphics[width=0.95\textwidth]{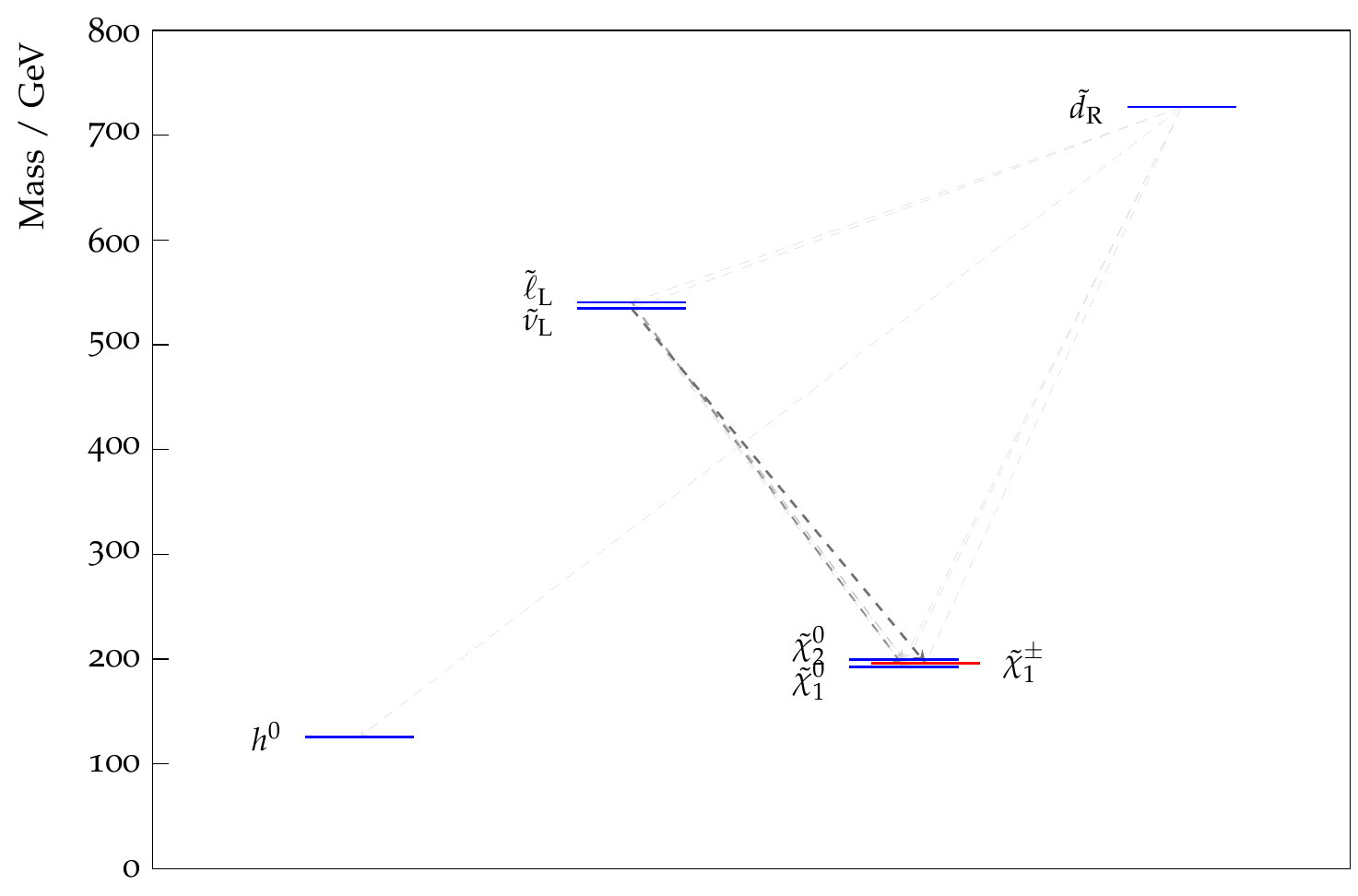}
         \caption{\label{fig:1_lepton_model}Model 13382371 (1-lepton).}
        \end{subfigure}
        \begin{subfigure}[b]{0.49\textwidth}
         \includegraphics[width=0.95\textwidth]{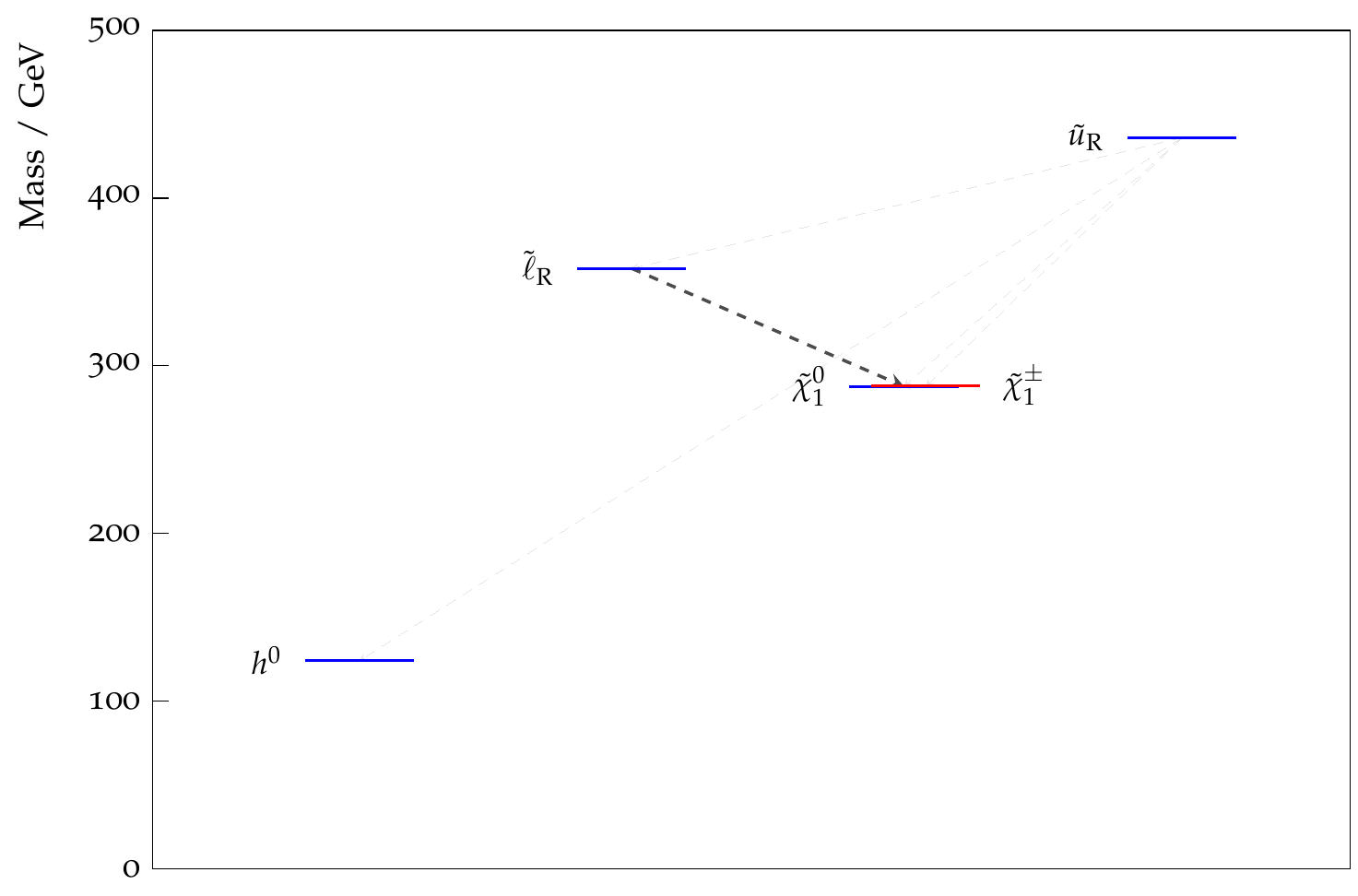}
         \caption{\label{fig:SS3L_light_Sqk}Model 11733067 (SS/3L).}
        \end{subfigure}
        
    \caption{Representative mass spectra produced using \texttt{PySLHA}~\cite{Buckley:2013jua} with model number in the captions for excluded points where the corresponding analysis in parentheses had greatest sensitivity. These show the most relevant low-mass sparticles and their decays; for the full mass spectra, see Figure~\ref{fig:full_mass_spec} in~\ref{sec:spec_plots}. The grey arrows show the branching ratios between two sparticles and proportional to its brightness. The model displayed for the 7--10 jets analysis is the one with the lightest gluino with an LSP mass below 100~GeV, while for the SS/3L search, we selected the model with the lightest squark. For the Multi-b analysis, displayed is a model selected from the region in figures~\ref{fig:bySearch_summary} where the density of red points is greatest. The model for the 1-lepton search is chosen based on one that is representative from the mass distributions in Figure~\ref{fig:1D_masses_1lepton}. 
    }
    \label{fig:mass_spec}
\end{figure*}

From the region in Figure~\ref{fig:bySearch_summary} where the density of red points is greatest, we show a representative point with model number 148229034 (Figure~\ref{fig:Multi_b_model}). This contains an LSP with a bino--Higgsino mixture at a mass of 175~GeV, a relatively low mass 1.2~TeV gluino and a 1.3~TeV sbottom enabling $\tilde{g} \to b \bar{b} \tilde{\chi}^0_1$ branching ratios to be preferred.
    
\paragraph{7--10 jets}

The concentration of magenta triangles in Figure~\ref{fig:bySearch_summary} illustrates that the 7--10 jets search is the most sensitive analysis for models having light LSPs with large mass splitting from the gluino. Most of these LSPs have masses below 100~GeV, which are bino-like as discussed in Section~\ref{sec:mass_plane_features}, and undergo early universe annihilation through a $Z^0$ or a $h^0$ boson in so-called `funnel' regions. Since this annihilation mechanism requires Higgsino or wino admixtures to proceed, such scenarios typically have light charginos $\tilde{\chi}^\pm_1$ and next-to-lightest neutralinos $\tilde{\chi}^0_2$, with relatively small mass splittings between one other. 

These types of decays are consistent with the simplified models containing long decay chains that this analysis optimises for. The observed missing transverse energy $E_\mathrm{T}^{\rm miss}$ therefore tends to be smaller than that required by the 2--6 jets or Multi-b searches. Indeed no explicit requirement on $E_\mathrm{T}^{\rm miss}$ is made by the search (instead the main discriminant is a ratio $E_\mathrm{T}^{\rm miss}/\sqrt{\sum p_\mathrm{T}^{\rm jet}}$ involving the missing energy and scalar sum of transverse jet momentum). The 7--10 jet search also has a looser requirement on jets originating from bottom quarks compared with the Multi-b analysis. Together, this allows the 7--10 jets analysis to maintain a unique coverage of models. In the gluino--squark plane (Figure~\ref{fig:bySearch_Gl_Sqk}), the 7--10 jets points occupy a similar space as Multi-b, but the longer cascade chains mean the gluino mass reach is lower. The remainder of the spectrum can be relatively decoupled. Figure~\ref{fig:7_10_jets_model} displays model number 227558023, which is representative of the models where the 7--10 jets analysis is most sensitive, where we selected the one with lowest gluino mass which had an LSP below 100~GeV.
    
\paragraph{Monojet}

The dedicated Monojet analysis selects events with an energetic jet from initial-state radiation recoiling off a system of large missing transverse momentum and up to three additional jets. In terms of SUSY models, this is optimised for scenarios where the mass of the squark is almost equal to that of the LSP, so-called `compressed scenarios'. 

In Figures~\ref{fig:bySearch_Gl_LSP} and~\ref{fig:bySearch_Sqk_LSP}, the excluded points (cyan rings) where this analysis is most sensitive involve very small mass splittings between the coloured sparticle and LSP. Though there is significant overlap in regions of sensitivity for the Monojet and 2--6 jets analyses, the former is exclusively the most sensitive analysis involve squark--LSP splittings below 30~GeV. We note that the 2--6 jets search includes a similar signal region but requiring a minimum of two jets called `2jm'. The different jet multiplicity requirements ensure the dedicated Monojet maintains a unique sensitivity to the smallest squark--LSP mass splittings, again demonstrating the complementarity of searches. We find a small number (fewer than 10) of scenarios involving small mass splittings between the lightest 3rd generation squark and the LSP in the models for which the Monojet is most sensitive.

\subsubsection{\label{sec:leptonic_searches}Searches selecting one or more leptons}

Both the 1-lepton and SS/3L searches also target production gluinos and squarks, but require one or more leptons in selected events. In Figures~\ref{fig:bySearch_summary} and \ref{fig:bySearch_Sqk_LSP} involving the gluino, squark and LSP masses, the correlations between where the 1-lepton (orange plus) or SS/3L (green cross) analyses were most sensitive are less obvious. There points tend to cluster below squark masses of $m(\tilde{q})\lesssim 1.5$~TeV, while few points are present for gluino masses $m(\tilde{g})\lesssim 1.2$~TeV where other analyses dominate. Further investigation reveals other correlations driven by the light flavour slepton and gaugino masses not apparent in these figures, which we discuss in what follows. We also identify scenarios where these searches had most sensitivity, which are beyond what ATLAS optimised for.

\paragraph{1-lepton} 

\begin{figure*}
    \centering
        \begin{subfigure}[b]{0.49\textwidth}
         \includegraphics[width=1.0\textwidth]{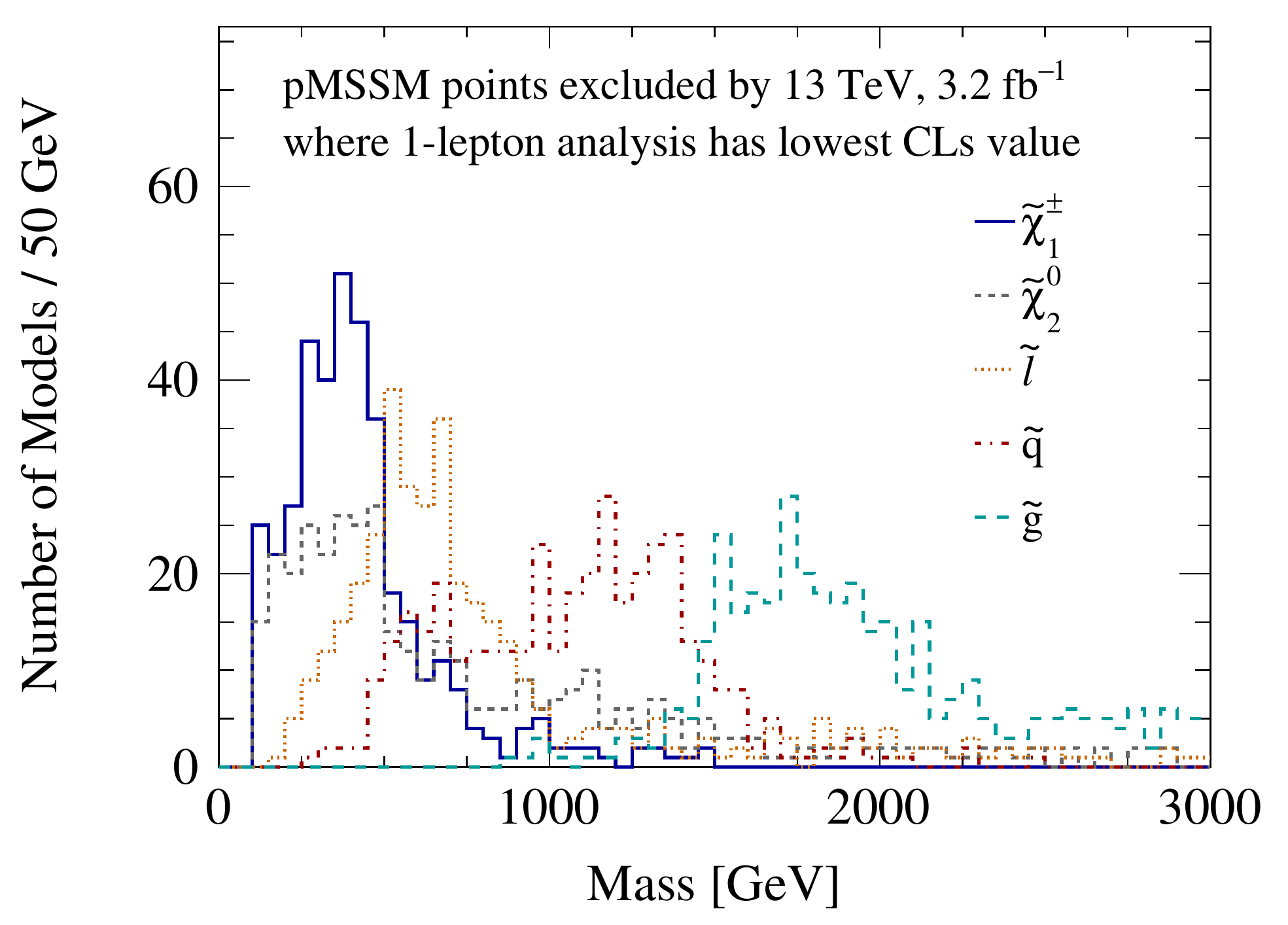}
         \caption{\label{fig:1D_masses_1lepton} 1-lepton analysis.}
        \end{subfigure}
        \begin{subfigure}[b]{0.49\textwidth}
         \includegraphics[width=1.0\textwidth]{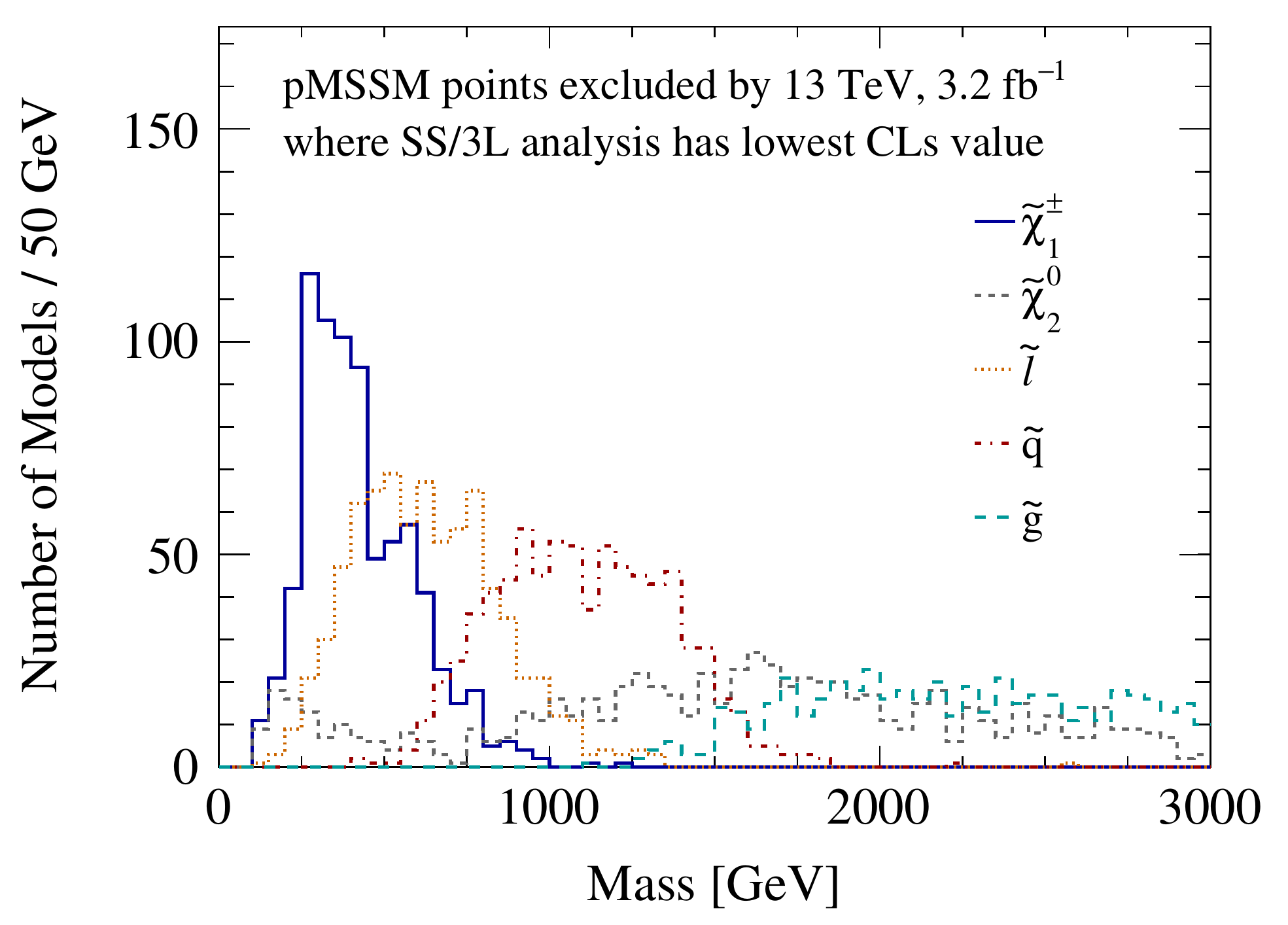}
         \caption{\label{fig:1D_masses_SS3L}SS/3L analysis.}
        \end{subfigure}
     
    \caption{Distribution of sparticle masses for the excluded points where the 1-lepton (left) and SS/3L (right) analyses are most sensitive. Presented sparticles are the chargino $\tilde{\chi}^\pm_1$ (blue solid), next-to-lighest neutralino $\tilde{\chi}^0_2$ (grey short-dashed), lightest slepton $\tilde{\ell}$ (orange dotted), lightest squark $\tilde{q}$ (red dot-dashed) and gluino $\tilde{g}$ (light-blue long-dashed).
    }
    \label{fig:1D_mass_distro}
\end{figure*}

Figure~\ref{fig:1D_masses_1lepton} shows the distribution of masses for various sparticles from the excluded points where the 1-lepton analysis had the lowest CLs value. The 1-lepton analysis requires events with exactly 1 electron or muon with various minimum jet multiplicities and large $E_\mathrm{T}^{\rm miss}$. Though the analysis was optimised for gluino production, there is also a prevalence of light squarks, whose distribution peaks around $m(\tilde{q})\sim 1$~TeV. The distributions of the chargino and to a less extent next-to-lightest neutralino $\tilde{\chi}^0_2$ are skewed towards low masses, peaking around 300~GeV, with a tail that extends above 1~TeV. The tendancy for the next-to-lightest neutralino and chargino to be light is characteristic of Higgsino content in the LSP, or a bino-like LSP with wino-like pair near in mass to the LSP. 

\begin{figure}
    \centering
    \includegraphics[width=0.50\textwidth]{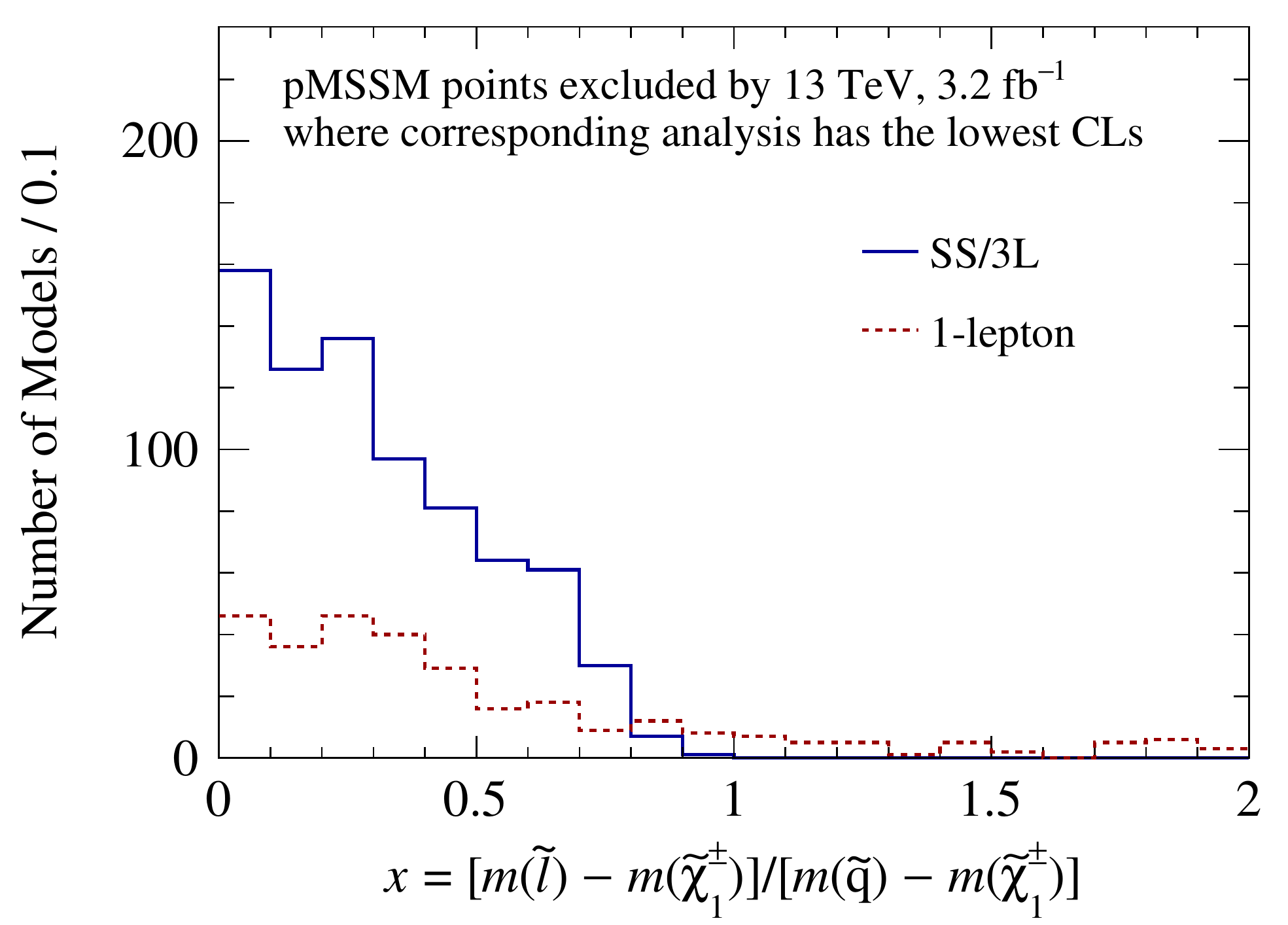}
    \caption{\label{fig:1D_mass_splitting_chrg_slep_sqk}Distribution of mass splitting for the chargino--slepton--squark $\tilde{\chi}^\pm_1$--$\tilde{\ell}$--$\tilde{q}$ system, parametrised by $x$ as defined in Eq.~\ref{eq:mass_splitting}. These are presented for the excluded points where the 1-lepton (dashed red) and SS/3L (solid blue) analyses are most sensitive.}
\end{figure}

The mass distribution also reveals a preponderance of light sleptons, peaking around $m(\tilde{\ell})\sim 600$~GeV. We find these typically reside between the coloured sparticle and the chargino--LSP states. To demonstrate this interpretation, we define the splitting parameter
\begin{align}
x \equiv \frac{m\big(\tilde{\ell}\big)-m\left(\tilde{\chi}^\pm_1\right)}{m\left(\tilde{q}\right)- m\left(\tilde{\chi}^\pm_1\right)}\,,\label{eq:mass_splitting}
\end{align}
where $m\big(\tilde{\ell}\big)$ and $m\left(\tilde{q}\right)$ are respectively the masses of the lightest slepton and squark among the first or second generations. The red dashed line in Figure~\ref{fig:1D_mass_splitting_chrg_slep_sqk} then shows the distribution in $x$ of models where 1-lepton was most sensitive in $x$, and indeed the majority of models in the histogram have $0<x<1$. 

The 1-lepton analysis interprets in a simplified model with a gluino--chargino--LSP decay chain and optimises for various mass splittings between these three sparticles. This is noteworthy given we find the 1-lepton search is most sensitive to richer mass spectra, with squarks and intermediate sleptons feature prominently. Given the mass distributions in Figure~\ref{fig:1D_masses_1lepton}, we display model number 13382371 (Figure~\ref{fig:1_lepton_model}) as a representative point to illustrate the wider sensitivity of the 1-lepton analysis. This point features a 727 GeV squark that can cascade to a slepton doublet and a 173~GeV Higgsino-like LSP. We also note that all the signal regions were involved in the exclusion of these models, indicating sensitivity to a wide variety of splittings. This addresses the question of over-optimisation raised in the Introduction (Section~\ref{sec:intro}). Despite optimising robustly for the gluino--chargino--LSP model, this analysis was sensitive to wider classes of models.

\paragraph{SS/3L}

Though there are similarities with the 1-lepton analysis, many salient differences appear for points where the SS/3L search is the most sensitive analysis. This search selects events with at least two leptons, and if there are exactly two, they are required to have the same electric charge. 

One prominent difference is that points where the SS/3L is most sensitive again have almost exclusively wino-like LSPs, and so a nearly mass-degenerate chargino. This is consistent with the mass distributions of these models in Figure~\ref{fig:1D_masses_SS3L}: compared with the 1-lepton case (Figure~\ref{fig:1D_masses_1lepton}), the SS/3L has a smaller tail of high chargino mass while the next-to-lightest neutralino $\tilde{\chi}^0_2$ distribution is no longer skewed to lower masses. We note that a future Run~2 version of the `Disappearing Track' analysis~\cite{Aad:2013yna}, not considered in this work, could be sensitive to these models, since the small chargino-LSP mass splitting in such models ensures that the chargino is typically long-lived on collider time scales. 

For these points, the gluinos are all relatively heavy and not strongly correlated with a particular mass scale, being fairly uniformly distributed for $m(\tilde{g}) \gtrsim 1.5$~TeV, in contrast to the 1-lepton discussion. This implicates that the SS/3L is not the most sensitive analysis to light gluinos, where other analyses such as the 2--6 jets are most sensitive. By contrast, squarks retain a peaked distribution centred around 1~TeV. 

Figure~\ref{fig:1D_mass_splitting_chrg_slep_sqk} demonstrates, in a similar way to the 1-lepton analysis, that the slepton mass is almost always between the chargino and the lightest squark. We find a negligible number of models (a single entry) has $x>1$, corresponding to a slepton mass being above that of the lightest squark. Thus, we find that points where the SS/3L has best sensitivity are strongly correlated with one common feature: a squark--slepton--chargino--LSP $\tilde{q}$--$\tilde{\ell}$--$\tilde{\chi}^\pm_1$--$\tilde{\chi}^0_1$ ordered mass spectrum. The squark can undergo a three-body decay to a quark, lepton and a slepton $\tilde{q} \to q  \ell \tilde{\ell}$ if the intermediate neutralino is off-shell $m(\tilde{\chi}^0_2) > m(\tilde{q})$. This hierarchical structure is displayed in model number 11733067 (Figure~\ref{fig:SS3L_light_Sqk}). Among the models where SS/3L is most sensitive, this has the lightest squark mass at 436~GeV. 

Moreover, we find one signal region `SR0b3j' was used to exclude 98\% of these models where the analysis was most sensitive. This suggests other analyses had better sensitivity to models targeted by the other three signal regions, for example the Multi-b analysis is particularly sensitive to $\tilde{g} \to t\bar{t} \tilde{\chi}^0_1$ scenarios. The `SR0b3j' signal region requires at least three leptons and was optimised to capture a $\tilde{g}$--$\tilde{\chi}^0_2$--$\tilde{\ell}$--$\tilde{\chi}^0_1$ decay chain, distinct from the $\tilde{q}$--$\tilde{\ell}$--$\tilde{\chi}^\pm_1$--$\tilde{\chi}^0_1$ scenario we just identified. 

Taken together, we draw two noteworthy conclusions from our findings for the SS/3L search. First, this analysis is the most sensitive analysis for a different scenario we identified in the pMSSM than all the simplified models those used for analysis optimisaton. Second, the exclusivity of signal region used to exclude this scenario indicates that the SS/3L lacked competitive sensitivity to points in the pMSSM corresponding to the simplified models considered.
\section{\label{sec:darkmatter}Probing the dark sector with strong SUSY searches}

In the pMSSM, the relic abundance of dark matter is predominantly shaped by the gaugino and Higgsino content of the LSP. The ATLAS Monojet search makes an explicit DM interpretation in a simplified model, but the complexity of the pMSSM dark sector necessitates other collider searches to provide decisive sensitivity. Complementing electroweak SUSY searches~\cite{Aaboud:2016wna}, this section sets out to demonstrate that 13~TeV ATLAS searches for coloured sparticles can indirectly probe different dark matter scenarios in the pMSSM. Subsection~\ref{sec:13TeV_on_dm} ascribes DM interpretations to each of the 13~TeV searches and manifests their sensitivity to distinct regions of DM parameter space. We also consider the impact of recent direct detection results from the PandaX-II/LUX experiments~\cite{Tan:2016zwf,lux-si-2016} on gluino and squark masses in Subsection~\ref{sec:dd_xs_on_Gl_Sqk}. Present constraints from indirect detection experiments searching for LSP annihilation signals in galactic halos are expected to have little sensitivity on the pMSSM space considered\footnote{Such indirect detection constraints were examined in Ref.~\cite{Cahill-Rowley:2014boa}, which finds that current constraints from Fermi-LAT are not expected to constrain the pMSSM space. Nevertheless, it was found that the future Cerenkov Telescope Array (CTA)~\cite{Doro:2012xx} is projected to have sensitivity to high mass LSPs scenarios not accessible to direct detection and collider searches.}.

\begin{figure*}
    \centering
        \begin{subfigure}[b]{0.70\textwidth}
         \includegraphics[width=\textwidth]{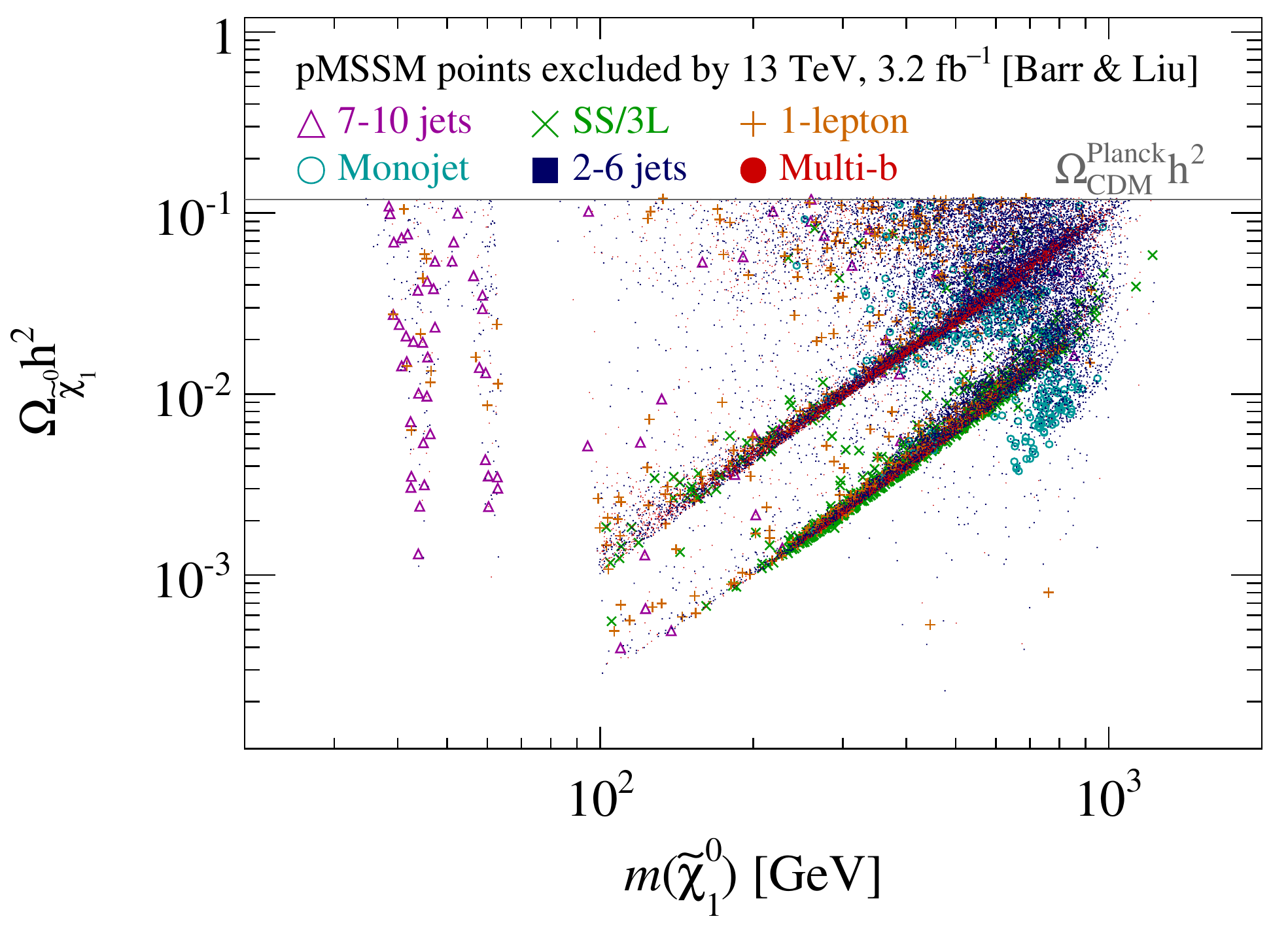}
         \caption{\label{fig:rel_dens_byAnalyses}Relic density.}
        \end{subfigure}
        \begin{subfigure}[b]{0.70\textwidth}
         \includegraphics[width=\textwidth]{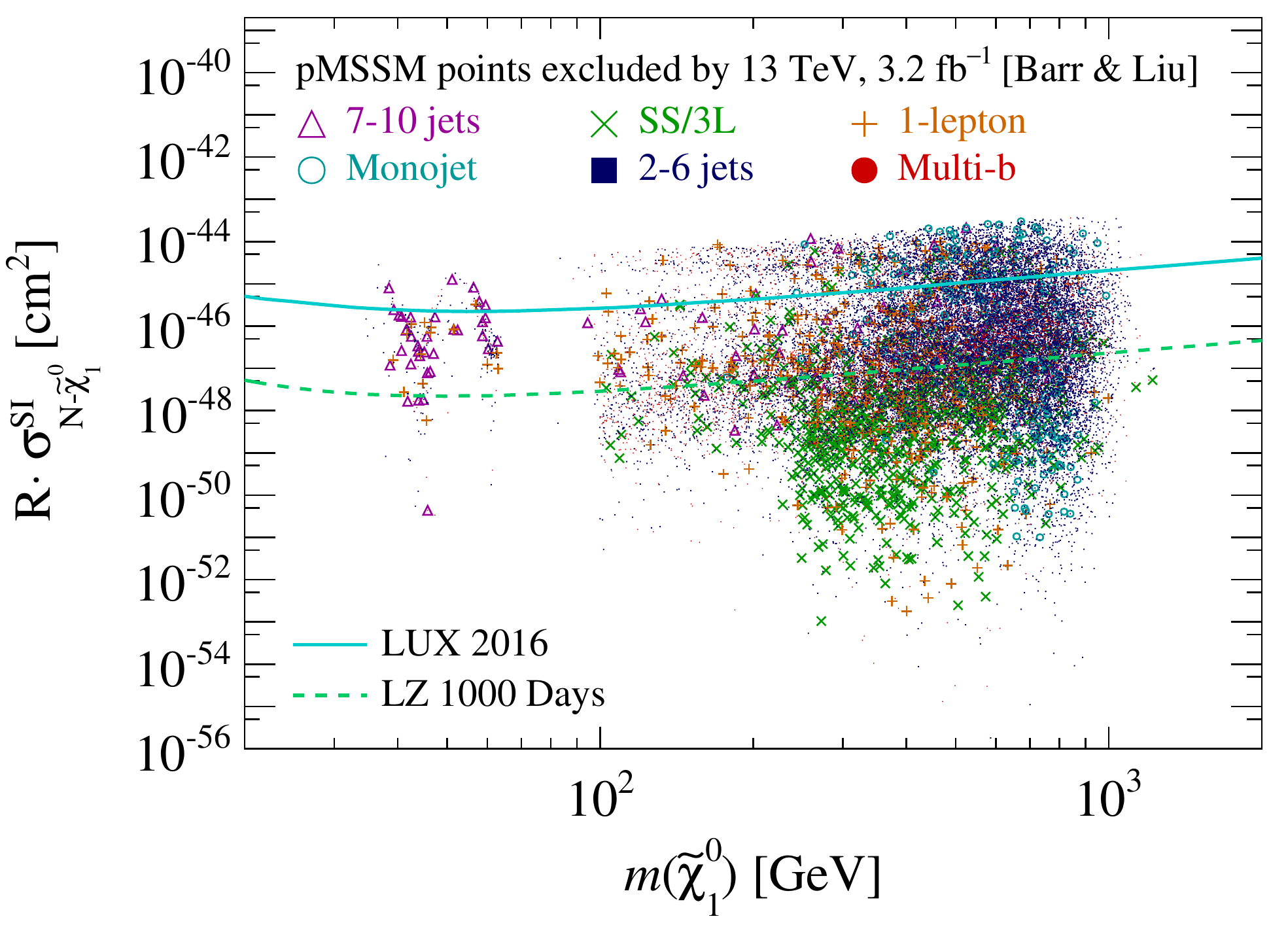}
         \caption{\label{fig:si_xs_byAnalyses}Spin independent cross-section.}
        \end{subfigure}
    \caption{Most sensitive analysis used to constrain each of the 28.5k points excluded at 95\% confidence level by the six 13 TeV searches displayed in Table~\ref{tab:listSearches}. A marker is styled according to the analysis with the lowest CLs value, as described in Figure~\ref{fig:bySearch_summary}. This is projected into the planes of LSP mass $m(\tilde{\chi}^0_1)$ against relic density (Figure~\ref{fig:rel_dens_byAnalyses}) and spin-independent LSP--nucleon  cross-section $\sigma^{\rm SI}_{N\textrm{-}\tilde{\chi}^0_1}$ (Figure~\ref{fig:si_xs_byAnalyses}). The grey line overlayed on Figure~\ref{fig:rel_dens_byAnalyses} indicates the observed relic density $\Omega_{\rm CDM}^{\rm Planck}h^2$ measured by Planck~\cite{Ade:2015xua}. In Figure~\ref{fig:si_xs_byAnalyses}, the cross-section is scaled by the ratio $R= \Omega_{\chi_1^0}h^2/\Omega_{\rm CDM}^{\rm Planck}h^2$ of the neutralino relic density $\Omega_{\chi_1^0}h^2$ to that observed $\Omega_{\rm CDM}^{\rm Planck}h^2$. Overlayed on Figure~\ref{fig:si_xs_byAnalyses} are upper limits at 90\% confidence level on the cross-section observed by the LUX 2016~\cite{lux-si-2016} (solid light blue) results, together with the projected limit from 1000 days of data taking by the LZ experiment~\cite{Akerib:2015cja} (dashed green). The central value of the limit from LUX 2016 is twice as strong as that recently reported by PandaX-II~\cite{Tan:2016zwf}. 
    }
    \label{fig:dd_si_frac_excl_byAnalyses}
\end{figure*}

\begin{figure}
    \centering
        \begin{subfigure}[b]{0.49\textwidth}
         \includegraphics[width=\textwidth]{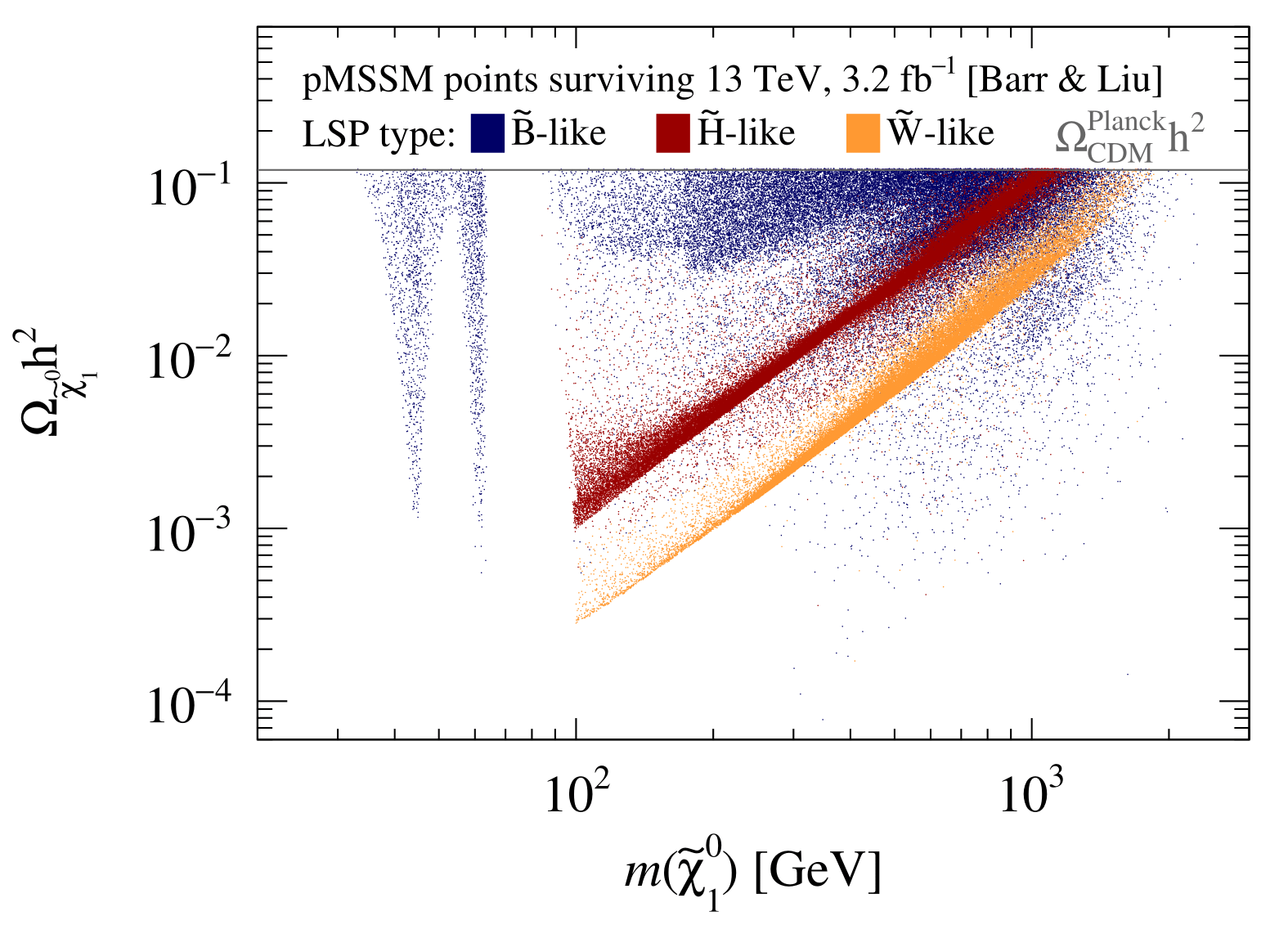}
          \caption{\label{fig:reldens_post_13TeV}Relic density.
          }
        \end{subfigure}
        \begin{subfigure}[b]{0.49\textwidth}
         \includegraphics[width=\textwidth]{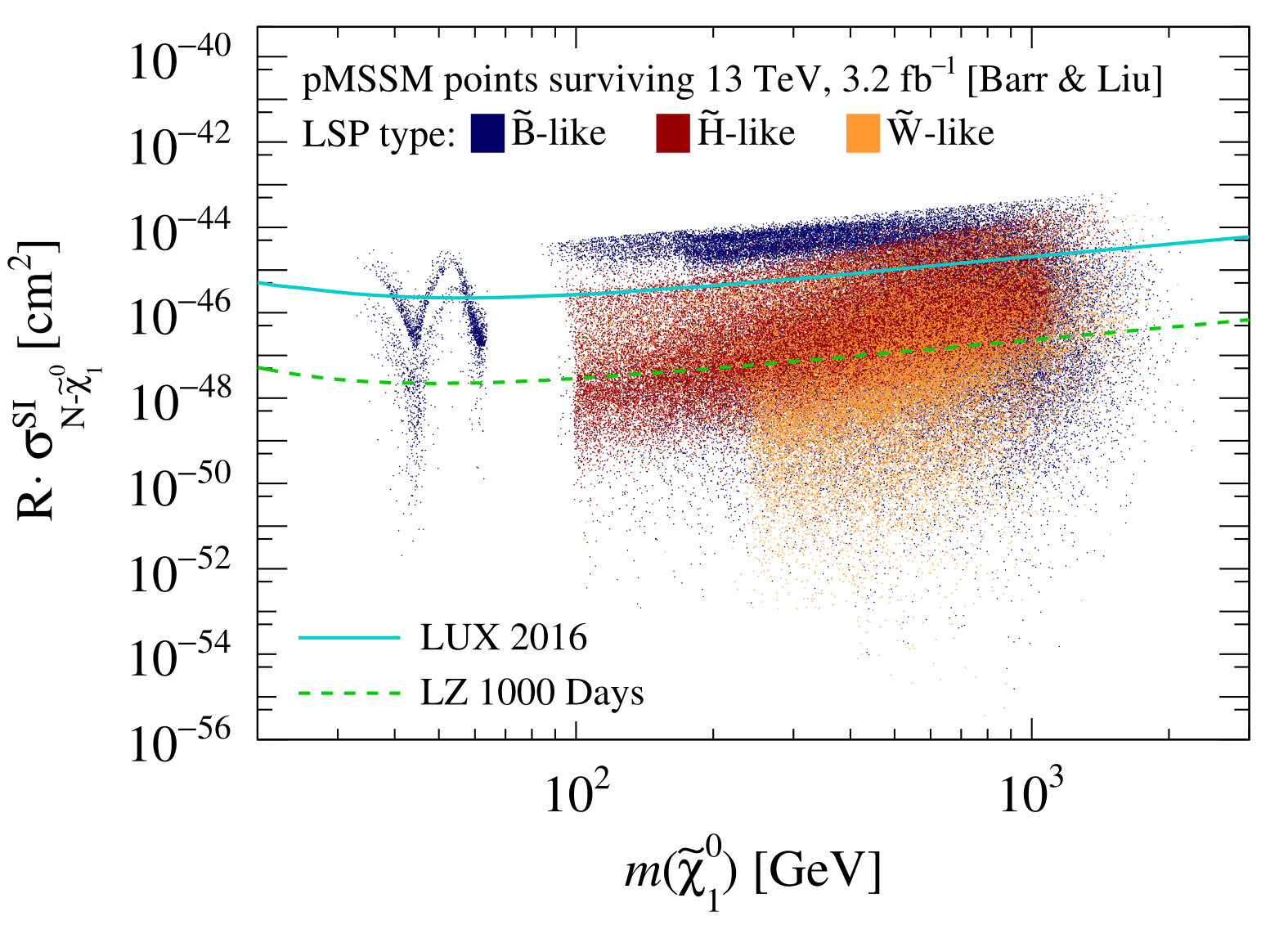}
          \caption{\label{fig:si_xs_post_13TeV}Spin independent cross-section.}
        \end{subfigure}
    \caption{Model points surviving the constraints from the six 13~TeV ATLAS searches considered in Ref.~\cite{Barr:2016inz}, projected into relic density (Figure~\ref{fig:reldens_post_13TeV}) and spin independent LSP-nucleon cross-section (Figure~\ref{fig:si_xs_post_13TeV}) vs LSP mass $m(\tilde{\chi}^0_1)$. The points are coloured by the composition of the neutralino LSP being dominantly bino (blue), Higgsino (red) or wino (orange) as defined in Table~\ref{tab:LSPtype}. 
    }
    \label{fig:LSPtype}
\end{figure}

\subsection{\label{sec:13TeV_on_dm}Impact of 13~TeV constraints on dark matter observables}

\subsubsection{\label{sec:relic_dens}Relic density}

Figure~\ref{fig:rel_dens_byAnalyses} shows each of the the 28.5k points excluded by the six 13~TeV SUSY searches considered, styled and coloured according to the analysis that had the lowest CLs value, projected into the plane of LSP mass vs relic density. Again, the strong correlations in this plane with the six searches are unambiguous. To facilitate interpretation, we discuss the most salient underlying processes that shape several features in this projection:
\begin{itemize}
    \item The LSP relic abundance $\Omega_{\chi^0_1}h^2$ is set by early universe thermal freeze-out, the hallmark of the weakly interacting massive particle (WIMP) paradigm. We do not require the neutralino to be the sole constituent of dark matter, as other well-motivated candidates such as axions can contribute~\cite{Bertone:2004pz}, making the points considered more general than Ref.~\cite{Kowalska:2016ent}. Thus the Planck measurement of the cold dark matter (CDM) abundance~\cite{Aad:2015baa} only serves as an upper bound.
    
    \item The composition of the LSP strongly influences the early universe annihilation mechanism of the LSP and the resulting cosmological relic density. Figure~\ref{fig:reldens_post_13TeV} illustrates the relic density against the mass of the LSP that survived the constraints from six 13~TeV analyses, coloured by the dominant composition to the LSP, as defined in Table~\ref{tab:LSPtype}. Notably, the Higgsino- and wino-like LSP models are concentrated along a straight diagonal line in the plot. This is because the thermally averaged cross-section $\langle \sigma v\rangle \propto m^{-2}_{\tilde{\chi}_1^0}$~\cite{Jungman:1995df}, and the relic density $\Omega_{\tilde{\chi}^0_1}h^2$ is therefore nearly proportional to the LSP mass squared. Mass splittings between the LSP and coannihilating chargino are typically a few GeV ($\sim 100$s~MeV) for Higgsino- (wino-) like LSPs. There are also no Higgsino- or wino-like LSPs below around 100~GeV. This is because such models have near-degenerate charginos to the LSP, which are excluded by LEP lower bounds.
    
    \item By contrast, bino-like LSP models have suppressed early-universe thermally-averaged annihilation cross-sections\footnote{Pure binos do not couple to any gauge or Higgs bosons. This is seen from the couplings $g_{Z\tilde{\chi}_1^0\tilde{\chi}_1^0}$ ($g_{h\tilde{\chi}_1\tilde{\chi}_1}$) of the neutralino to the $Z^0$ and (Higgs) bosons are given at tree-level by~\cite{Djouadi:2001kba}
    \begin{align}
    g_{Z\tilde{\chi}_1^0\tilde{\chi}_1^0} &= \frac{g_2}{2\cos_W}\left(\left|N_{13}\right|^2 - \left|N_{14}\right|^2\right),\\
    g_{h\tilde{\chi}_1^0\tilde{\chi}_1^0} &= g_2\cos\alpha\left(N_{11} - N_{12}\tan\theta_W \right)\left(N_{13}\tan\alpha + N_{14}\right).\label{eq:higgsLSPcoupling}
    \end{align}
    Here we have the neutralino mixing matrix elements $N_{ij}$ defined in Table~\ref{tab:LSPtype}, the SU(2) gauge coupling $g_2$, Weinberg angle $\theta_W$ and the Higgs mixing parameter $\alpha$. Pure binos have $N_{12} = N_{13} = N_{14} = 0$.
    }, leading to larger relic abundance, as generally seen in Figure~\ref{fig:reldens_post_13TeV}. Therefore to satisfy the Planck bound, bino-like LSPs must either have non-negligible mixing with winos and/or Higgsinos, or there must be a near-degenerate next-to-LSP to act as a coannihilator. In Ref.~\cite{Aad:2015baa}, the natures of the coannihilators were displayed in the planes involving relic density. No distinction was made  between light flavour squarks and gluinos, yet they have different phenomenological roles as coannihilators. We therefore elucidate this in Figure~\ref{fig:coloured_coannih}, differentiating between coannihilators being gluinos (light orange), light flavour squarks (dark blue) and 3rd generation squarks (green). We observe that light gluino coannihilators are more stringently excluded, with few points $m(\tilde{g}) \lesssim 800$~GeV than for squarks due to powerful Run~1 constraints. 
    
    \item Focusing now on electroweak particles (`Other', light grey), for LSP masses $m\left(\tilde{\chi}^0_1\right) \lesssim 250$~GeV, the coannihilation mechanism is predominantly via uncoloured sparticles due to stringent LHC constraints on squarks and gluinos. The two peaks centred around $m\left(\tilde{\chi}^0_1\right) \approx 45$ and 63~GeV involve resonant annihilation through a $Z^0$ or Higgs boson. Meanwhile, for $90 \lesssim m\left(\tilde{\chi}^0_1\right) \lesssim 250$~GeV, the coannihilators are predominantly slepton or gauginos, which are bounded from below by the LEP limit. 
\end{itemize}

Returning to the discussion of points excluded by individual analyses in the relic density plane (Figure~\ref{fig:rel_dens_byAnalyses}), the strong SUSY searches considered are sensitive to models with a wide variety of $\Omega_{\tilde{\chi}^0_1} h^2$. As previously discussed in Section~\ref{sec:compl_searches}, the 7--10 jets search is particularly sensitive to models with  $m\left(\tilde{\chi}^0_1\right) \lesssim 100$~GeV, where bino-like LSPs are associated with the $Z^0$ and $h^0$ funnel region. Meanwhile, the SS/3L analysis is most sensitive to wino-like LSPs models while the Multi-b analysis had preferential sensitivity to Higgsino-like LSP scenarios, as indicated by the clustering of green crosses and red dots along the wino and Higgsino respective diagonal bands (compare with Figure~\ref{fig:reldens_post_13TeV}). 

The 1-lepton analysis has sensitivity away from the Higgsino and wino diagonal bands, for LSP masses below about 300~GeV, where slepton and gaugino coannihilators are prevalent. The extensive presence of blue dots and cyan rings away from these bands shows respectively that the 2--6 jets and Monojet searches are particularly sensitive to gluinos and squarks that have small mass splitting with the bino-like LSP. Crucially, these impact scenarios where such coloured sparticles are the coannihilators (compare with Figure~\ref{fig:coloured_coannih}) and therefore indirectly probe LSP masses higher than those currently accessible by direct electroweakino searches at the LHC. Nevertheless, the 2--6 jets analysis also has sensitivity covering Higgsino and wino-like LSP points, with clusters along the Higgsino and wino bands. 

\subsubsection{\label{sec:direct_detection}Direct detection}

\begin{figure}
    \centering
        \begin{subfigure}[b]{0.49\textwidth}
         \includegraphics[width=\textwidth]{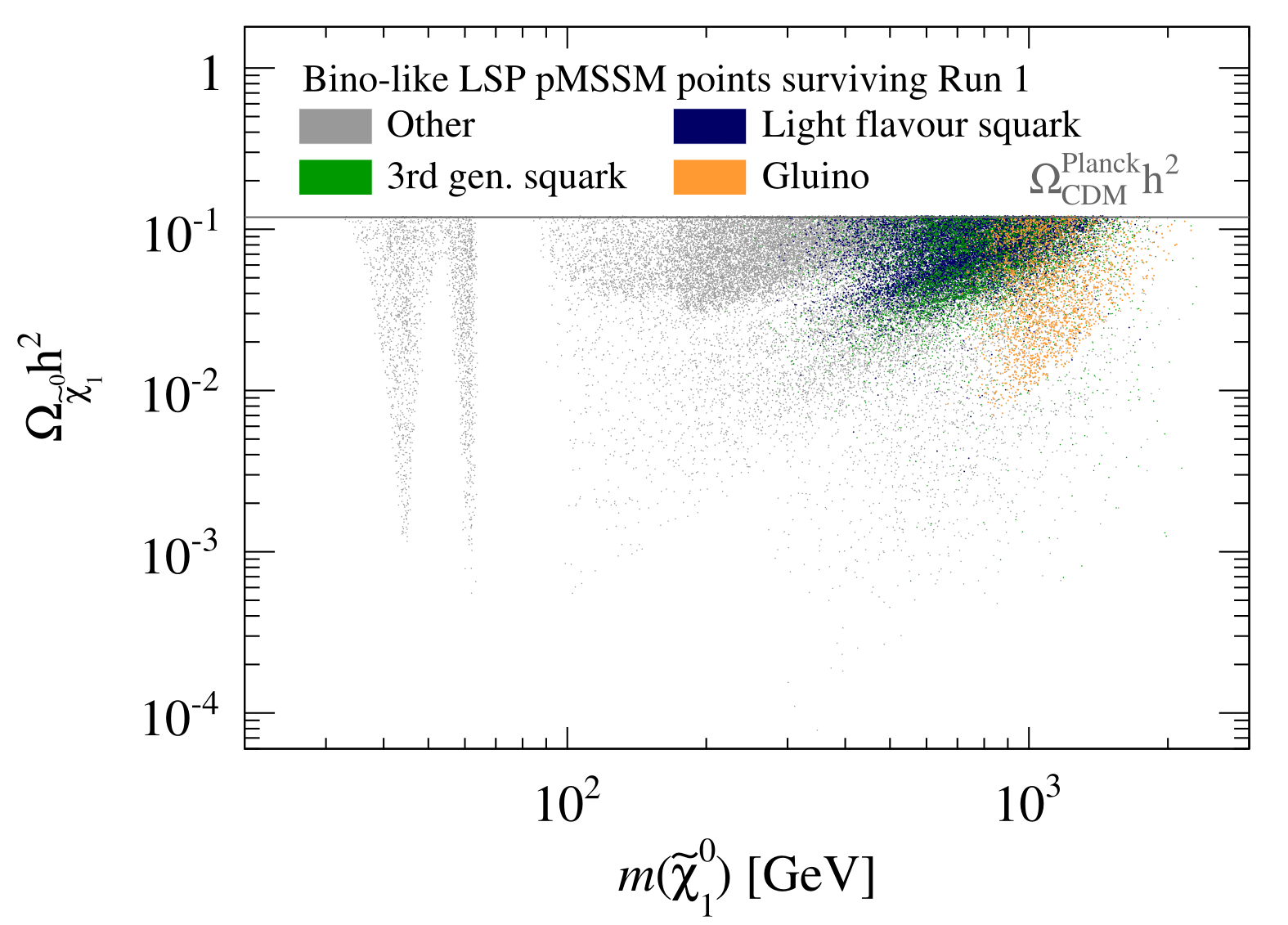}
          \caption{\label{fig:coann_reldens}Relic density.
          }
        \end{subfigure}
        \begin{subfigure}[b]{0.49\textwidth}
         \includegraphics[width=\textwidth]{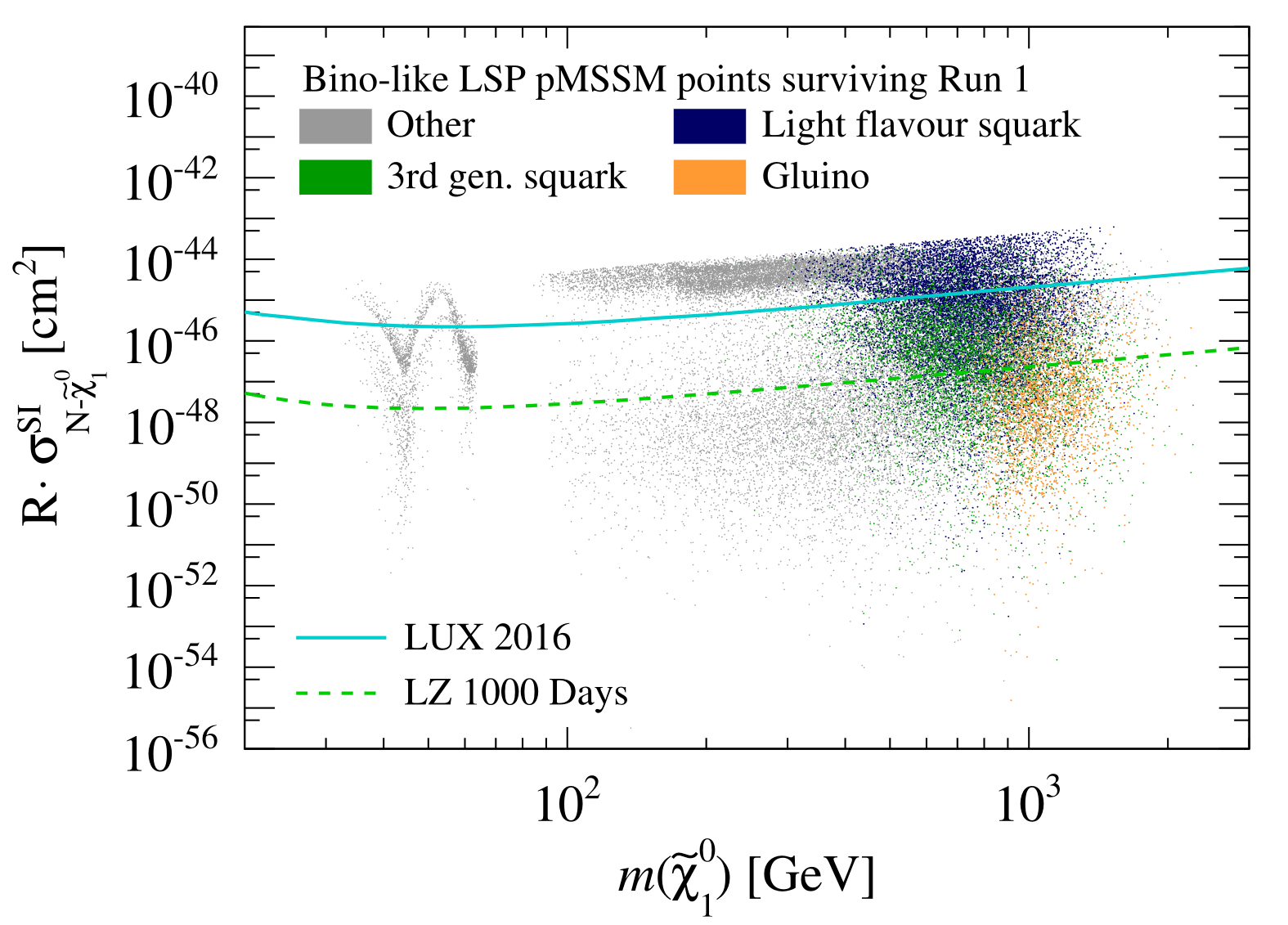}
          \caption{\label{fig:coann_si_xs}Spin independent cross-section.}
        \end{subfigure}
    \caption{The strongly interacting coannihilators of points with a bino-like LSP surviving Run~1 constraints, projected into relic density (left) and spin independent LSP-nucleon cross-section (right) vs LSP mass $m(\tilde{\chi}^0_1)$. The points are coloured according to dominant coannihilator for the LSP: gluino (light orange), light flavour squarks (dark blue), 3rd generation squarks (green) and `Other' refers to (co)annihilation mechanisms involving the electroweak sector. 
    }
    \label{fig:coloured_coannih}
\end{figure} 

The neutralino LSP contributes to the local dark matter density and can induce nuclei recoils in direct detection experiments. Figure~\ref{fig:si_xs_byAnalyses} illustrates the points excluded by the most sensitive of the six ATLAS searches, allowing us to compare with direct detection sensitivity for each analysis. This projects into the plane defined by the spin-independent\footnote{It is found in Ref.~\cite{Aad:2015baa} that recent upper limits on spin-dependent cross-sections lack sensitivity to this set of pMSSM points, although future sensitivity can provide complementary probes to spin-independent limits~\cite{Cahill-Rowley:2014boa}.} cross-section $\sigma^{\rm SI}_{N\textrm{-}\tilde{\chi}^0_1}$ of the LSP interacting with a Xenon nucleus, normalised per nucleon, and LSP mass. Again we discuss several features, with reference to other projections where appropriate, to aid interpretation:
\begin{itemize}
    \item Direct detection experiments typically interpret results assuming the LSP fully saturates the cold dark matter (CDM) relic abundance measured by Planck. As the LSP in the pMSSM need not be the sole constituent of dark matter, we rescale the direct detection interaction cross-sections $\sigma^{\rm SI}_{N\textrm{-}\tilde{\chi}^0_1}$ by a factor 
    \begin{align}
    R \equiv \Omega_{\tilde{\chi}^0_1}h^2/\Omega_{\rm CDM}^{\rm Planck}h^2.\label{eq:Rfactor}
    \end{align}
    This accounts for the reduction in direct detection sensitivity due to a lower local density of neutralino LSPs. 
    
    \item Recently, the PandaX-II~\cite{Tan:2016zwf} and LUX~\cite{lux-si-2016} collaborations presented results that extend sensitivity for WIMP masses \mbox{$m\left(\tilde{\chi}^0_1\right) \gtrsim 20$~GeV} by a factor of 2 and 4 respectively beyond the LUX 2015 result~\cite{Akerib:2015rjg}. As a guide to the sensitivity of direct detection, the observed limits from LUX 2016~\cite{lux-si-2016} and the projected sensitivity of LZ~\cite{Akerib:2015cja} are overlayed in Figure~\ref{fig:si_xs_byAnalyses}. 
    
    \item In defining the permitted pMSSM points, ATLAS conservatively increased by a factor of four the upper limit on the spin-independent cross-section $R\cdot\sigma^{\rm SI}_{N\textrm{-}\tilde{\chi}^0_1}$ from LUX 2013~\cite{Akerib:2013tjd} to account for uncertainties in nuclear form factors~\cite{Aad:2015baa}, before rejecting points in the pre-selection. This pre-selection constraint carves out the points at the highest $R\cdot \sigma^{\rm SI}_{N\textrm{-}\tilde{\chi}^0_1}$ cross-sections around $10^{-44}$~cm$^2$. 
    
    \item Figure~\ref{fig:si_xs_post_13TeV} reveals points that survived constraints from the six 13~TeV searches considered in this study, in the plane of spin-independent cross-section against LSP mass. The points are coloured according to the whether the dominant contribution to the LSP is the bino, Higgsino, or wino. The composition of the LSP has a significant effect on $R\cdot \sigma^{\rm SI}_{N\textrm{-}\tilde{\chi}^0_1}$. This is due both to the couplings of the LSP to nucleons and the relic density suppression (Eq.~\ref{eq:Rfactor} and Figure~\ref{fig:reldens_post_13TeV}). Notably, most wino-like LSPs have suppressed direct detection cross-sections given the small coupling to the Higgs boson.

\end{itemize}

Returning to the discussion of the regions of sensitivity from the ATLAS searches (Figure~\ref{fig:si_xs_byAnalyses}), points above the blue solid line are independently excluded by both LUX 2016 and the 13~TeV ATLAS searches considered. In this context, collider searches are free from astrophysical uncertainties, and therefore provide powerful cross-checks should either approach report tentative signals. On the other hand, the cosmological lifetime of the $\tilde{\chi}^0_1$ dark matter candidate can only be verified by non-collider means. For LSP masses $m\left(\tilde{\chi}^0_1\right)\gtrsim 1$~TeV, the number of models excluded by the ATLAS searches decreases rapidly. There remain many points surviving in this regime (Figure~\ref{fig:si_xs_post_13TeV}), and those with the high $R\cdot \sigma^{\rm SI}_{N\textrm{-}\tilde{\chi}^0_1}$ are within direct detection sensitivity.

The 2--6 jets analysis (blue points) is sensitive to a large class of models, particularly those with gluino or squark coannihilators. As this difference has important phenomenological consequences, we display the strong sector coannihilators in Figure~\ref{fig:coann_si_xs} for bino-like LSPs of points that survived Run~1 constraints. In this projection, it is evident that coannihilation points involving light flavour squarks have enhanced cross-section $R\cdot \sigma^{\rm SI}_{N\textrm{-}\tilde{\chi}^0_1}$ compared with gluinos. This is due to the $s$-channel diagram involving the quarks and LSP scattering via an intermediate squark, a point we will elaborate further in Section~\ref{sec:dd_xs_on_Gl_Sqk}. Thus, the 2--6 jets and Monojet analyses share sensitivity to many squark coannihilator scenarios with LUX 2016. 

As discussed in previous sections, the 7--10 jets is most sensitive to light mass LSPs with significant bino content. Many of the points excluded by this search are below the current LUX 2016 sensitivity. Meanwhile, the Multi-b analysis tends to favour scenarios Higgsino-like LSP scenarios where squark masses are above 2~TeV. There is particular sensitivity to a region centred around cross-section from 10$^{-46}$~cm$^{2}$ to 10$^{-49}$~cm$^{2}$ and LSP mass of 300 to 700~GeV (Figure~\ref{fig:si_xs_byAnalyses}). Many Higgsino-like LSP models inhabit this region and are beginning to be probed by LUX 2016, but the majority of points where the Multi-b analysis is most sensitive are below the LUX limit. 

Notably, ATLAS strong SUSY searches are sensitive to scenarios with direct detection cross-section $R\cdot \sigma^{\rm SI}_{N\textrm{-}\tilde{\chi}^0_1}$ well below even the projected sensitivity of LZ based on 1000 days of data taking~\cite{Akerib:2015cja}. The SS/3L reach into this regime is especially prominent, being most sensitive to wino-like LSP models with highly suppressed cross-sections $R\cdot \sigma^{\rm SI}_{N\textrm{-}\tilde{\chi}^0_1}$. Many gluino coannihilator scenarios occupy this region (Figure~\ref{fig:coann_si_xs}) and are dominantly probed by the 2--6 jets analysis. The projected LZ sensitivity is within an order of magnitude of the irreducible neutrino background `floor', which is a challenging regime for Xenon-target direct detection experiments.   

Concluding this subsection, we demonstrated the important complementarity of strong SUSY searches for probing models beyond both Monojet-like collider interpretations and the reach of direct detection experiments. This enables colliders to indirectly constrain bino-like LSPs with coloured coannihilators, in addition to Higgsino LSP scenarios for example before electroweak SUSY searches gain direct sensitivity. This motivates construction of simplified DM models based on such interpretations, but is beyond the scope of this work. 

\subsection{\label{sec:dd_xs_on_Gl_Sqk}Impact of direct detection constraints on squarks and gluinos}

\begin{figure*}
    \centering
        \begin{subfigure}[b]{0.49\textwidth}
         \includegraphics[width=\textwidth]{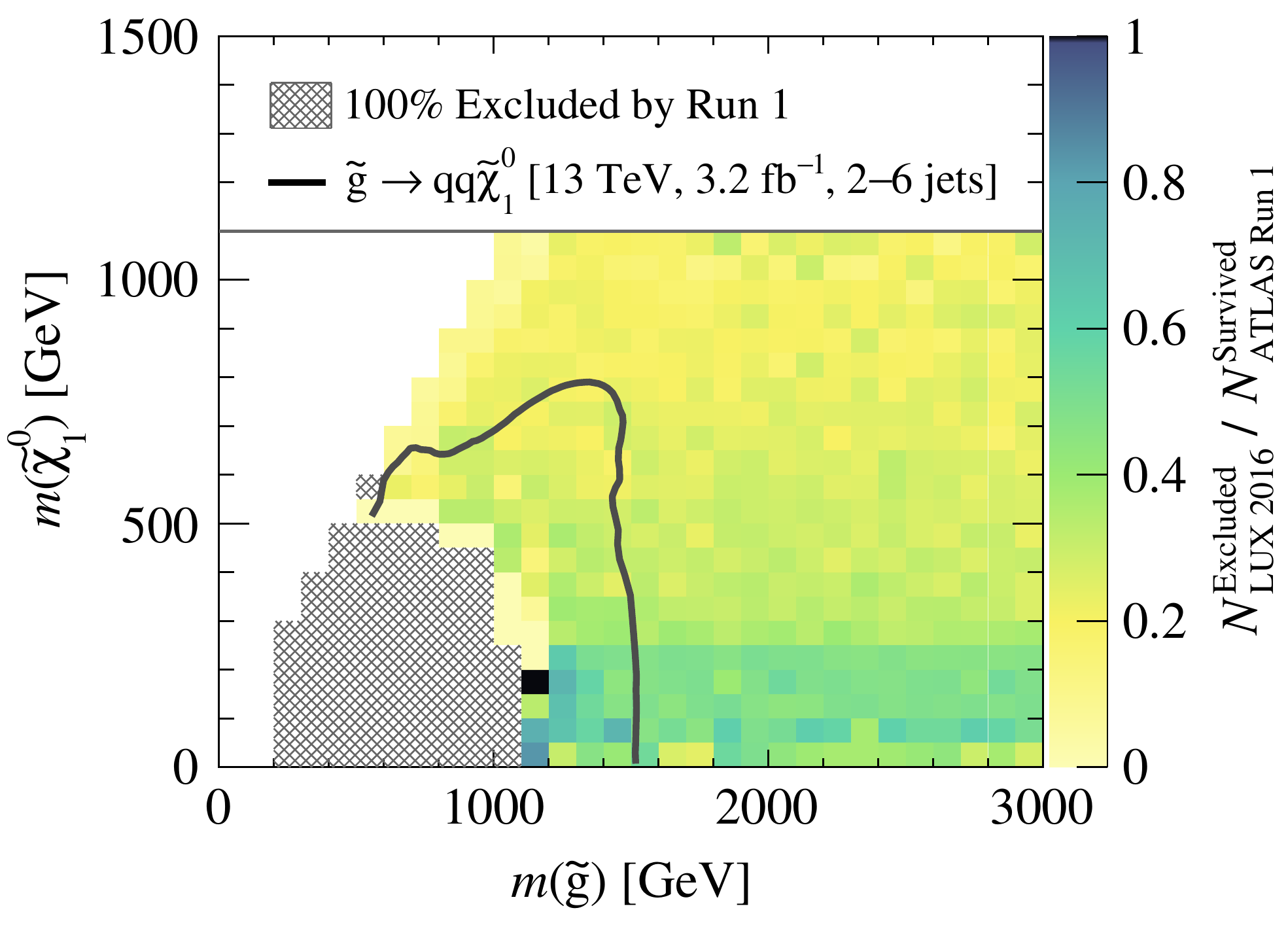}
         \caption{\label{fig:LUX_on_Gl_LSP}Gluino vs LSP.}
        \end{subfigure}
        \begin{subfigure}[b]{0.49\textwidth}
         \includegraphics[width=\textwidth]{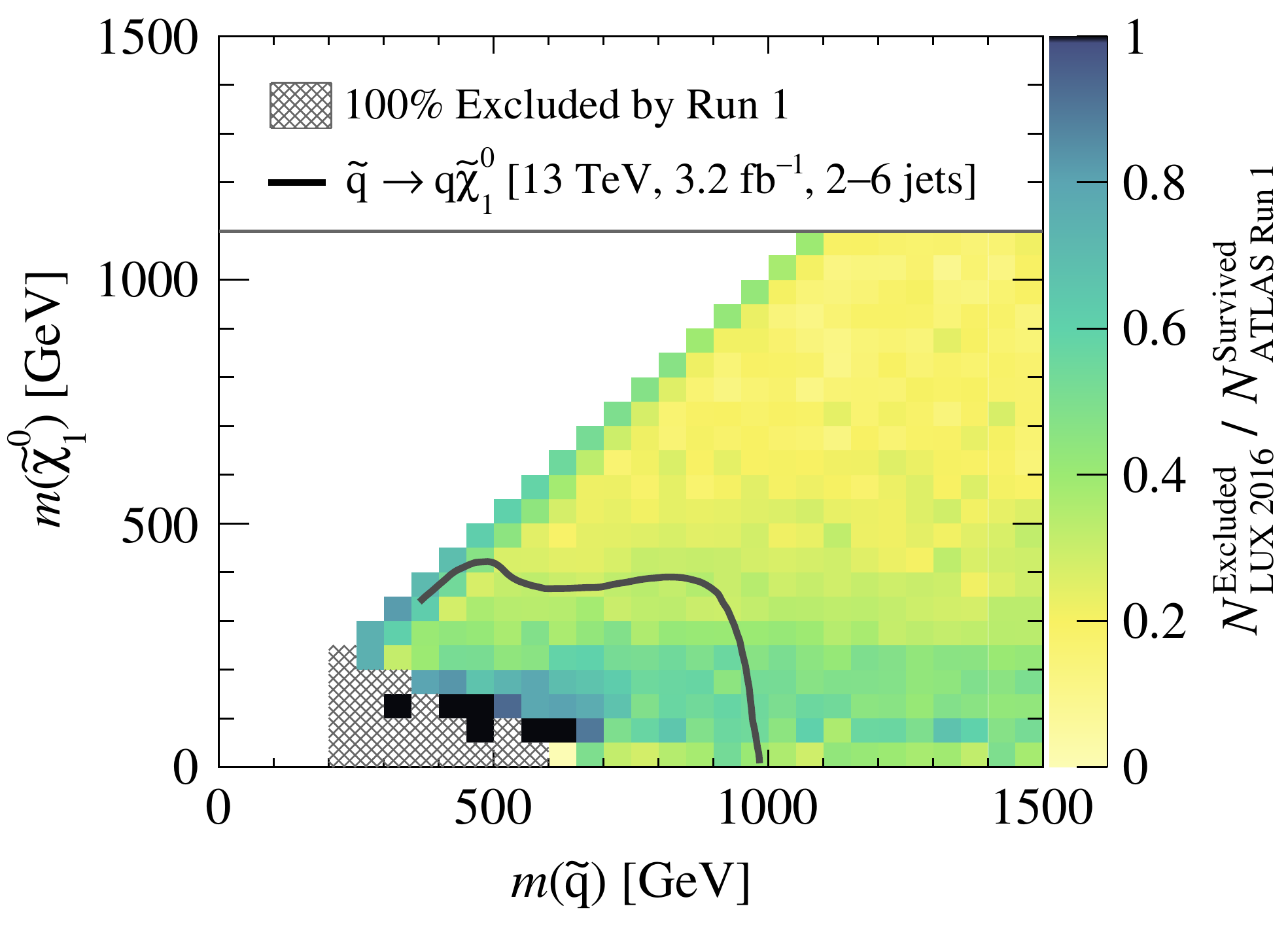}
         \caption{\label{fig:LUX_on_Sqk_LSP}Lightest squark vs LSP.}
        \end{subfigure}
    \caption{Fraction of models excluded by the observed 90\% confidence limit on the spin-independent LSP-nucleon cross-section from the LUX 2016 result~\cite{Akerib:2015rjg}, projected into the mass plane of the gluino (left) and lightest squark (right) vs LSP. Here, $m(\tilde{q})$ is the lightest squark of the first or second generations. The colour scale for each mass bin indicates the number of models excluded by LUX 2016~\cite{lux-si-2016} $N_\textrm{LUX 2016}^{\rm Excluded}$ normalised to the number of models that survived ATLAS Run~1 constraints $N_\textrm{ATLAS Run 1}^{\rm Survived}$ such that no 13~TeV analyses are considered. Black bins denote 100\% exclusion. Hatched grey bins indicate all the points are excluded by Run~1 ATLAS searches. Points with long-lived gluinos, squarks and sleptons are not considered in these figures. The overlayed grey solid lines correspond to the direct decay simplified models for gluino (Figure~\ref{fig:LUX_on_Gl_LSP}) and squarks (\ref{fig:LUX_on_Sqk_LSP}) from the 2--6 jets search~\cite{Aaboud:2016zdn}. 
    }
    \label{fig:dd_si_frac_excl_Sqk_LSP}
\end{figure*} 

Finally, turning the question around, we explore the impact of LUX 2016 constraint on the parameter space of squarks and gluinos relevant to LHC searches. Considering the set of points that survived ATLAS Run~1 constraints~\cite{Aad:2015baa}, we deem any point with scaled cross-section $R \cdot \sigma^{\rm SI}_{N\textrm{-}\tilde{\chi}^0_1}$ above the 90\% confidence level upper limit observed by LUX 2016 to be excluded. We take the observed central value of the upper limit as reported by LUX 2016~\cite{lux-si-2016}, without rescaling to account for the nuclear form factor uncertainties\footnote{Had we weakened this constraint by a factor of four~\cite{Aad:2015baa}, the key qualitative features in the discussion are unaffected; only the numerical fraction of models excluded is reduced to 18.0\%.} (as was done in Ref.~\cite{Aad:2015baa}). The upper limit derived by LUX 2016 excludes 30.3\% of such pMSSM points constraints (with long-lived gluinos, squarks and sleptons removed). Figure~\ref{fig:si_xs_post_13TeV} shows that LUX is particularly sensitive to pMSSM points with bino-like LSP, due in part to the relic density suppression of Higgsino and wino-like models. 

Figure~\ref{fig:LUX_on_Gl_LSP} projects the fractional exclusion\footnote{Although $N_\textrm{LUX 2016}^\textrm{Excluded} / N_\textrm{ATLAS Run 1}^\textrm{Survived}$ is prior-dependent, we will momentarily discuss this, while the forthcoming comparison of gluino and squark coannihilation regions has minimal prior dependence.} by LUX 2016 into the gluino vs LSP plane. There is a band of modest exclusion for LSP masses $m(\tilde{\chi}^0_1)\lesssim 250$~GeV, which is relatively uncorrelated with gluino masses. This apparent enhancement of sensitivity is partly an artefact due to the systematic oversampling of coannihilators for bino-like LSPs in Ref.~\cite{Aad:2015baa}. This region is rich in gaugino and slepton coannihilators due to their weak LHC constraints compared with strongly interacting sparticles. Due to smaller annihilation cross-sections of this electroweak process in the early universe, the relic abundance $\Omega_{\tilde{\chi}^0_1}h^2$ for a given bino-like LSP mass is larger. These models are thus not scaled down as far by the $R=\Omega_{\tilde{\chi}^0_1}h^2/\Omega_{\rm CDM}^{\rm Planck}h^2$ factor.

A prominent feature of the points excluded by LUX appears when projecting into the squark vs LSP plane (Figure~\ref{fig:LUX_on_Sqk_LSP}). The diagonal region where the mass splitting between the LSP and squark is small\ $m(\tilde{q}) - m(\tilde{\chi}^0_1) \lesssim 50$~GeV shows a distinctly higher exclusion fraction; this was absent in the gluino vs LSP plane. Furthermore, mass bins with higher exclusion fraction are correlated with lower squark masses. This is particular salient along the diagonal, where points are predominantly squark coannihilators. The band of modest exclusion fraction for LSPs with $m(\tilde{\chi}^0_1)\lesssim 250$~GeV remains, as with the gluino vs LSP plane.

Such contrasting features between gluinos and squarks highlight the importance of the $s$-channel squark exchange diagram in direct detection
\begin{align}
    \tilde{\chi}^0_1 + q \to \tilde{q} \to \tilde{\chi}^0_1 + q.
\end{align}
Here, a squark $\tilde{q}$ mediates the LSP $\tilde{\chi}^0_1$ scattering off light flavour quarks $q$ inside the nucleons. Resonant scattering occurs when the LSP and squarks are nearly mass degenerate, enhancing the sensitivity of LUX along the squark--LSP diagonal of Figure~\ref{fig:LUX_on_Sqk_LSP}. Thus, direct detection experiments are particularly sensitive to bino-like LSP scenarios with a squark coannihilator. The cross-section $\sigma^{\rm SI}_{N\textrm{-}\tilde{\chi}^0_1}$ decreases with heavier squarks due to the \mbox{$\sim 1/\left(m_{\tilde{q}}^2 - m_{\tilde{\chi}^0_1}^2\right)$} suppression in the propagator. 

These observations are consistent with the LSP--nucleon cross-section vs LSP mass plane with coannihilation mechanism identified (Figure~\ref{fig:coann_si_xs}). Squarks of the first or second generation (dark blue) have particularly enhanced direct detection cross-sections. The points with gluino coannihilators (light orange) feature $R\cdot \sigma^{\rm SI}_{N\textrm{-}\tilde{\chi}^0_1}$ primarily below $10^{-46}$~cm$^2$ and will only begin to be probed at the direct detection frontier by future experiments such as LZ. We also note that coannihilation points involving third generation squarks (light blue) also tend to have suppressed direct detection cross-sections compared with squarks, due to the negligible 3rd generation content in nucleons. 

Taken together, this highlights the important implications of LHC searches for squarks in the context of direct detection experiments, especially when the squark--LSP mass splitting is small. Yet these squark coannhilation scenarios are challenging for colliders, where direct detection experiments can provide a complementary probe.
\section{\label{sec:conclusion}Conclusion}

In this work, we interpreted six published 13~TeV (3.2~fb$^{-1}$) ATLAS SUSY searches for gluinos and light flavour squarks in a 19-parameter pMSSM. The purpose of this study was to analyse previously unexamined correlations between the most sensitive analyses with distinct regions of pMSSM parameter space. Our study addressed various shortcomings in the literature, presented under three questions in the Introduction (Section~\ref{sec:intro}), which we now summarise.

Firstly, we examined these correlations in collider parameter spaces, providing substantially richer information than overlap matrices used in the literature. For the two-dimensional projections into gluino, LSP and squark masses, the separation in regions probed by the 2--6 jets and Multi-b analyses were particularly distinct. The Multi-b was the most sensitive analysis for models with larger gluino--LSP mass splittings, where the 2--6 jets search began to lose sensitivity. The regions identified are independent of the priors in the pMSSM points. Further, while the Monojet and 2--6 jets share substantial overlapping sensitivity, the tighter jet requirements of the former is needed for the scenarios where the coloured sparticle and LSP are near mass-degenerate.  

Secondly, we identified classes of models beyond those ATLAS used for optimisation. Arguably the most striking realisation of this was the SS/3L search. Despite optimising to four distinct simplified models, we found one signal region to be most sensitive to a different scenario in the pMSSM not considered by the ATLAS search. It involved a light flavour squark cascading to a slepton and wino-like LSP with a nearly mass-degenerate chargino, which could be used by the experimental collaborations to refine future searches. Meanwhile, though the 1-lepton search optimised for a single simplified model, we showed it was sensitive to scenarios that included squark production, as well as intermediate sleptons and gauginos.   

Finally, while ATLAS performed an explicit DM interpretation for their Monojet search, our study manifested the prominent role other searches for coloured sparticles have when the dark sector is beyond non-minimal regimes as in the pMSSM. Bino-like LSPs may rely on coloured coannihilators to be consistent with the observed relic abundance, which are probed by the 2--6 jets and Monojet analyses being sensitive to small squark--LSP and gluino--LSP mass splittings. Light flavour squarks enhance LSP--nucleon scattering cross-sections, and squark lower mass bounds can still be below 500~GeV in the pMSSM. In addition, the SS/3L analysis had preferential sensitivity to wino-like LSP scenarios, which are particularly challenging for direct detection experiments. 

Using our findings to design novel search strategies or interpretations was beyond the scope of this work and is deferred to future studies. It would also be of interest to perform similar assessments for the third generation squark and electroweak sectors including long-lived sparticles, interpret searches based on simplified models of DM, as well as develop surveys for non-minimal SUSY scenarios. Previously unprobed regions of the pMSSM will be explored further as LHC luminosity continues to rise. 
\begin{acknowledgements}

We are grateful to Moritz Backes, Fady Bishara, Claire Gwenlan, Will Fawcett, Will Kalderon, Olivier Lennon, John March-Russell and Mike Nelson for helpful discussions, together with the computing support of Dennis Liu and Kashif Mohammad. This research is supported by the Science and Technology Facilities Council (STFC). The authors would like to acknowledge the use of the University of Oxford Advanced Research Computing (ARC) facility in carrying out this work~\cite{richards_2015_22558}. \textit{Disclaimer:} this is not the work of, nor endorsed by, the ATLAS or CMS collaborations.
\end{acknowledgements}

\bibliographystyle{atlas-style}
\bibliography{./bib/atlas,./bib/cms,./bib/dm,./bib/pmssm,./bib/software,./bib/susy,./bib/tools,./bib/nonlhc}

\appendix 
\section{\label{sec:method}Theoretical framework and experimental constraints}

This Appendix reviews the underlying assumptions and constraints applied to the set of pMSSM points used in this study. The points were produced by ATLAS in collaboration with external expertise~\cite{Berger:2008cq,CahillRowley:2012cb,CahillRowley:2012kx,Cahill-Rowley:2014twa}, which we review in~\ref{sec:ATLAS_pMSSM19}; for full details, see Ref.~\cite{Aad:2015baa}. In~\ref{sec:13TeV_method}, we give an overview of the method used
to obtain exclusion information from interpreting six early 13~TeV ATLAS searches and summarise the results in~\ref{sec:summary_excl_spart}. Finally,~\ref{sec:prior_distro} displays the prior distributions after Run~1 constraints, before any 13~TeV results are imposed.

\subsection{\label{sec:ATLAS_pMSSM19}Review of the ATLAS pMSSM19}

Starting with the MSSM, minimal flavour violation was imposed and CP violation was restricted to the CKM phase in the quark sector. The neutralino was required to be the lightest supersymmetric particle (LSP) and R-parity is exactly conserved. Table~\ref{tab:LSPtype} displays the three categories of LSP type by their dominant composition, as defined in Ref.~\cite{Aad:2015baa}. Due to their different resulting phenomenology, ATLAS employed importance sampling~\cite{Aad:2015baa} to ensure approximately equal proportions of each LSP type are selected. 

The 19 resulting parameters were scanned with flat priors, taking an upper limit on mass scales to be 4~TeV. The resulting model points were subjected to the following non-LHC constraints. LEP lower mass bounds~\cite{lep2:susy} were imposed together with precision measurements from the electroweak isospin splitting parameter $\Delta \rho$~\cite{Baak:2012kk}, $g-2$ of the muon~\cite{Aoyama:2012wk,Hagiwara:2011af,Nyffeler:2009tw,Czarnecki:2002nt,Bennett:2004pv,Bennett:2006fi,RevModPhys.80.633,Roberts:2010cj}, $Z^0$ invisible width~\cite{ALEPH:2005ab}, branching fractions of heavy flavour states~\cite{Amhis:2012bh,DeBruyn:2012wk,CMS:2014xfa,Aubert:2009wt,Hara:2010dk,Adachi:2012mm,Lees:2012ju,Charles:2004jd} and Higgs boson mass~\cite{Aad:2012tfa} at the time of pMSSM points generation were applied~\cite{Aad:2015baa}. Dark matter constraints were subsequently considered. To account for uncertainties in nuclear form factors, ATLAS quadrupled the upper limit on the spin-independent cross-section for LSP--nucleon interaction from LUX 2013~\cite{Akerib:2013tjd} before applying this to the points. Similar constraints for the spin-dependent cross-section of LSP--proton interactions were applied from COUPP~\cite{Behnke:2012ys}, and XENON100~\cite{Aprile:2013doa} for the LSP--neutron cross-section. The LSP was not assumed to be the sole constituent of dark matter so only an upper limit was set for the LSP relic abundance $\Omega_{\tilde{\chi}^0_1}h^2$, taken as $\Omega_{\rm CDM}h^2 = 0.1208$, which is the central value plus twice the reported uncertainty from the Planck measurement~\cite{Ade:2015xua}. 

The 310.3k pMSSM points surviving all non-LHC constraints underwent evaluation against 22 relevant Run~1 ATLAS searches for supersymmetry using 7 and/or 8~TeV data using full ATLAS software and reconstruction; for full details, see Ref.~\cite{Aad:2015baa}. A total of 40.9\% of points were excluded at 95\% confidence level. The top row of Table~\ref{tab:generatedfrac} quantifies the number of models that survive Run~1 ATLAS constraints by LSP type.

\begin{table}[]
\centering
    \begin{tabular}{lccc}
    \toprule 
     LSP type & Definition  \\ 
     \midrule
    `Bino-like' $\tilde{B}$     & $N_{11}^2 > \textrm{max}\left(N_{12}^2, N_{13}^2 + N_{14}^2\right)$ \\
    `Wino-like' $\tilde{W}$     & $N_{12}^2 > \textrm{max}\left(N_{11}^2, N_{13}^2 + N_{14}^2\right)$  \\
    `Higgsino-like' $\tilde{H}$ & $\left(N_{13}^2 + N_{14}^2\right) > \textrm{max}\left(N_{11}^2, N_{12}^2 \right)$\\
    \bottomrule
    \end{tabular}
\caption{\label{tab:LSPtype} Definition of neutralino $\tilde{\chi}^0_1$ LSP categories from Ref.~\cite{Aad:2015baa}. In the neutralino mixing parameter $N_{ij}$, the first index denotes the neutralino mass eigenstate $\tilde{\chi}^0_i$ while the second indicates its dominant composition in the order $\left(\tilde{B}, \tilde{W}, \tilde{H}_1, \tilde{H}_2\right)$.}
\end{table}

\begin{table}[]
\centering
    \begin{tabular}{lccc}
    \toprule 
     Models & Bino & Wino & Higgsino \\ 
     \midrule
    Viable after ATLAS Run~1 & 61.6k & 43.8k & 78.4k\\
    Without long-lived & 59.9k & 43.6k & 78.3k \\
    Without LL, with $\sigma_{\rm tot} \geq 5$~fb                        & 
    48.7k      & 29.7k & 52.8k  \\ \bottomrule
    \end{tabular}
\caption{\label{tab:generatedfrac} Viable model points before Run~2 constraints. These are classified by the dominant contribution to the LSP being bino, wino or Higgsino. Long-lived (LL) gluinos, squarks and sleptons with $c\tau > 1$~mm require dedicated Monte-Carlo simulation and are omitted from this study. Event simulation was performed on non-LL models with total strong sparticle production cross-section $\sigma_{\rm tot} \geq 5$~fb.  }
\end{table}

\subsection{\label{sec:13TeV_method}Interpretation of early 13~TeV searches}

We now summarise the methodology of Ref.~\cite{Barr:2016inz} to apply the combined constraints of six published 13~TeV ATLAS searches considered to the 183.8k pMSSM points that survived Run~1. Long-lived squarks, gluinos and sleptons ($c\tau > 1$~mm as defined in Ref.~\cite{Aad:2015baa}) make up 1.9k of these points. These require dedicated Monte-Carlo simulation and are removed from consideration of 13~TeV sensitivity. Of the remaining 181.8k points, 71.4\% have total inclusive production cross-section of any two coloured sparticles $\sigma_{\rm tot}$ greater than 5~fb at leading order. 
This is enumerated in the bottom row of Table~\ref{tab:generatedfrac} and underwent particle event generation to be described. The remaining 28.6\% was deemed not to have sensitivity with 3.2~fb$^{-1}$ of integrated luminosity and therefore not excluded.

We used \textsc{MadGraph5}~2.3.3~\cite{Alwall:2011uj,Alwall:2014hca} for particle event generation involving any two coloured sparticles from two protons, interfaced with \textsc{Pythia} 6.428~\cite{Sjostrand:2006za} for hadronisation and showering, using the \textsc{CTEQ6L}~parton distribution functions~\cite{Pumplin:2002vw}. Up to one additional parton in the matrix element and the MLM prescription~\cite{Mangano:2006rw} was used to match jets, setting the \textsc{MadGraph} minimum parton $k_T$ parameter to 100~GeV and \textsc{Pythia} jet measure cutoff at 120~GeV, in accord with Ref.~\cite{Aad:2015baa}. The \textsc{Delphes}~3.3.2~\cite{deFavereau:2013fsa} fast detector simulator was employed with \textsc{Fastjet} 3.1.3~\cite{Cacciari:2008gp, Cacciari:2011ma}, using the anti-$k_T$ clustering algorithm with cone parameter $R = 0.4$, to parametrise the performance of the ATLAS detector. 

Table~\ref{tab:listSearches} lists the six Run~2 analyses, whose event selection were implemented in the \textsc{MadAnalysis5}~1.3~\cite{Conte:2012fm,Conte:2014zja} recasting framework. We adapted codes from the Public Analysis Database~\cite{Dumont:2014tja,MArun1ATLAS:2-6jets,MArun1ATLAS:7+jets} where available, and write our own otherwise. The \textsc{RecastingTools} package was used for limit setting using the CLs prescription~\cite{Read:2002hq}, where a point is deemed excluded if the CLs value is below 0.05. To validate our code, we ensured at least one benchmark point had cutflows agreeing to better than 30\% and the observed simplified model limits were reproduced to within the uncertainties published by ATLAS (see~\ref{sec:validation} for further details).

\subsection{\label{sec:summary_excl_spart}Summary of sparticle masses excluded by early 13~TeV searches}

\begin{figure*}
    \centering
    \begin{subfigure}[b]{0.49\textwidth}
          \includegraphics[width=\textwidth]{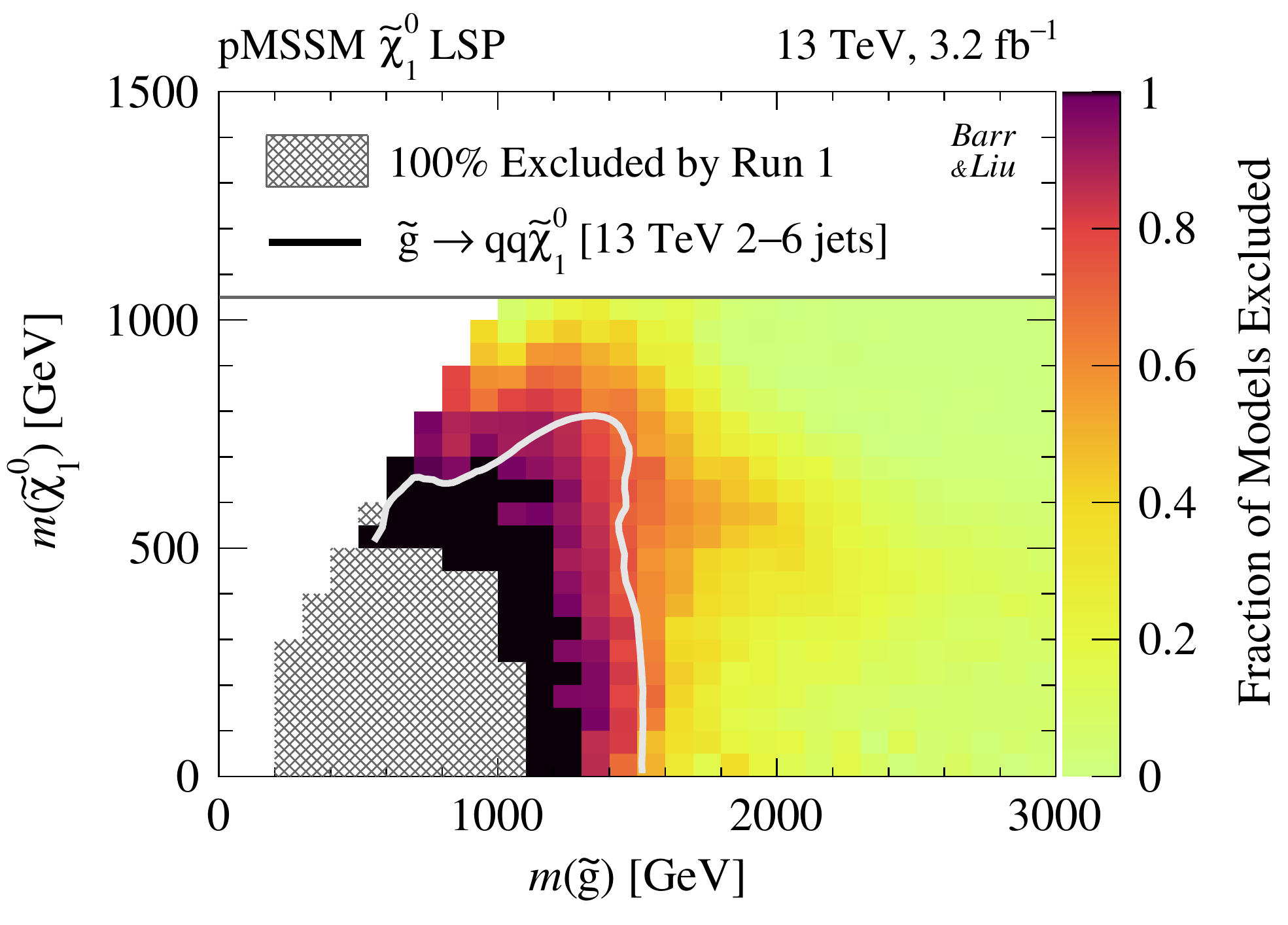}
          \caption{Gluino--LSP plane.}
        \end{subfigure}
        \begin{subfigure}[b]{0.49\textwidth}
          \includegraphics[width=\textwidth]{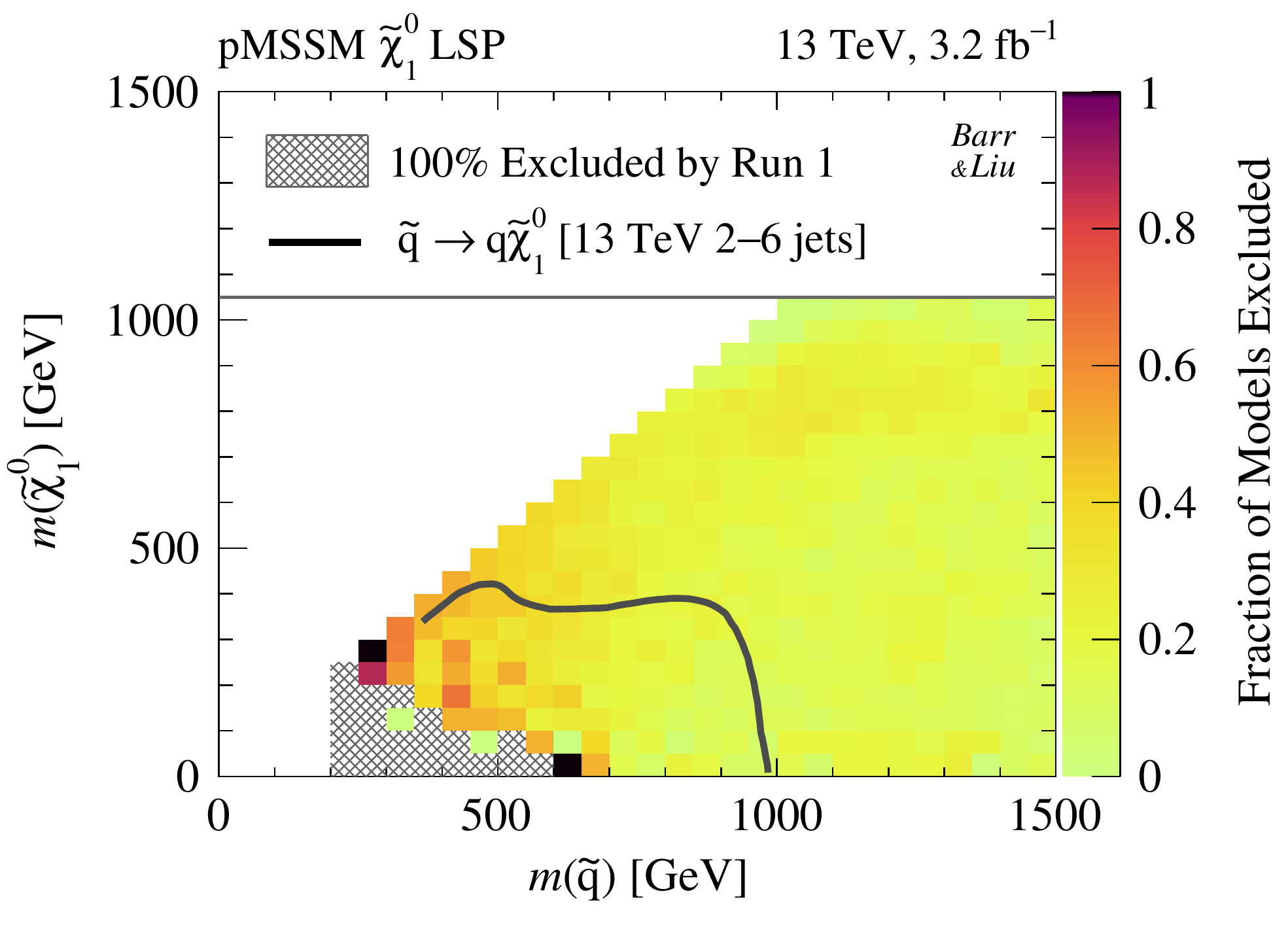}
          \caption{Squark--LSP plane.}
        \end{subfigure}
        \begin{subfigure}[b]{0.49\textwidth}
          \includegraphics[width=\textwidth]{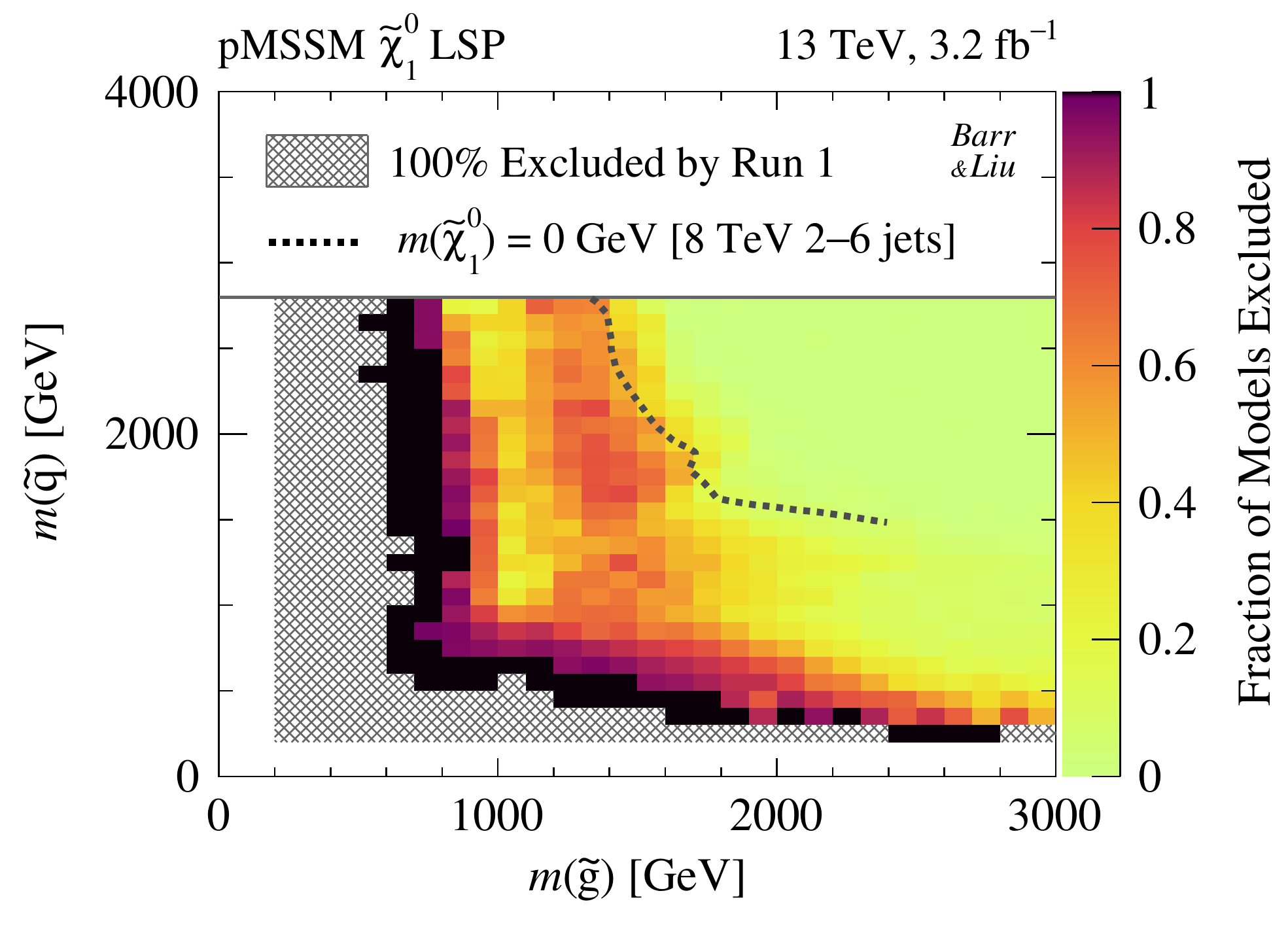}
          \caption{\label{fig:gsqk_fracExcl}Gluino--squark plane.}
        \end{subfigure}
    \caption{\label{fig:mass_plane_Gl_Sqk_chi10}Fraction of model points excluded by the combined constraints from the six early 13~TeV searches considered in Table~\ref{tab:listSearches}, out of the points that survived Run~1 constraints. In each mass bin, the colour scale denotes the fraction of points excluded at 95\% confidence level normalised to the number of points satisfying `Without long-lived' in Table~\ref{tab:generatedfrac}, where black indicates 100\% exclusion.  This is projected into the mass planes of the gluino--LSP $\tilde{g}$-$\tilde{\chi}_1^0$ (upper left) and squark--LSP $\tilde{q}$-$\tilde{\chi}_1^0$ (upper right). Here, $m(\tilde{q})$ is the mass of the lightest squark among the first two generations. White regions correspond to no models being produced by ATLAS due to prior non-LHC constraints. Hatched grey regions indicate bins where all points were excluded by Run~1 in Ref.~\cite{Aad:2015baa}. Overlayed grey solid lines are the simplified model limits from the 13~TeV 2--6 jets search~\cite{Aaboud:2016zdn} for gluinos $\tilde{g} \to qq\tilde{\chi}^0_1$ (upper left) and squarks $\tilde{q} \to q\tilde{\chi}^0_1$ (upper right). In the latter case, all eight squarks are of the first two generations and assumed to be mass-degenerate. For the gluino--squark plane (lower), the overlayed grey dashed line is taken from the `gluino--squark--LSP simplified pMSSM' scenario from the 8 TeV 2--6 jets search~\cite{Aad:2014wea}.
    }
    
\end{figure*}

\begin{figure*}[]
    \centering
         \includegraphics[width=0.9\textwidth]{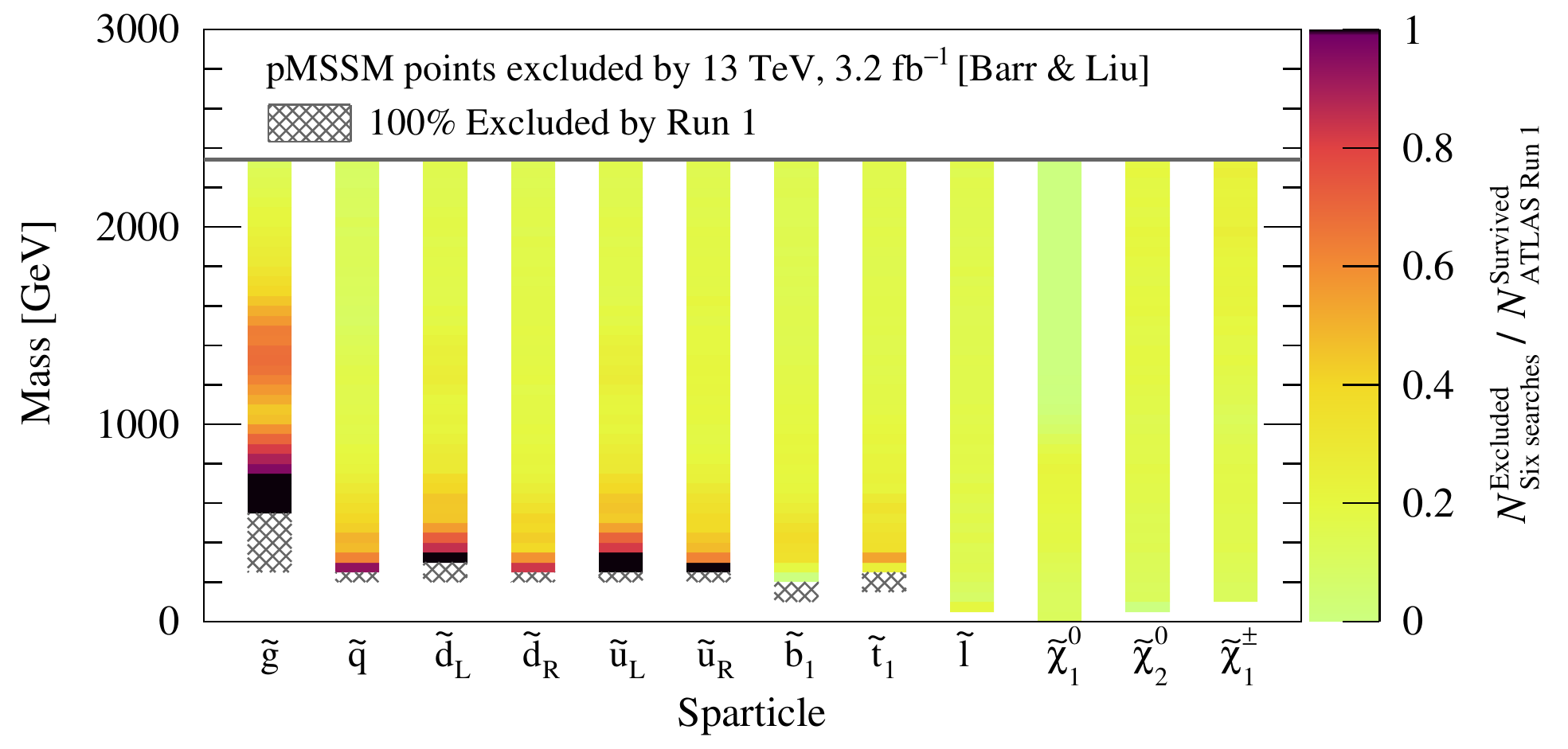}
          \caption{Fraction of models excluded out of those that survived Run~1 searches by the combined constraints from six 13~TeV analyses listed in Table~\ref{tab:listSearches} for different sparticle masses. The exclusion information is from Ref.\cite{Barr:2016inz}. The colour scale indicates the fraction of models excluded, normalised to the number satisfying `Without long-lived' in Table~\ref{tab:generatedfrac}. Black regions denote that all points were excluded in the mass bins of 50~GeV size. Here, $\tilde{q}$ ($\tilde{l}$) is the lightest squark (slepton) of either the first or second generations. The hatched grey regions indicate all models in the mass bins were excluded by Run~1 ATLAS searches from Ref.~\cite{Aad:2015baa}. Models with long-lived gluinos, squarks or sleptons are removed from these figures.
    }
    \label{fig:sparticle_summary}
\end{figure*} 

\begin{table*}[]
\centering
\resizebox{0.60\textwidth}{!}{%
\begin{tabular}{lcccccc}
\toprule
           & 2--6 jets & 7--10 jets & 1-lepton & Multi-b & SS/3L  & Monojet \\
\midrule
2--6 jets  & 100\%     & 3\%        & 5\%      & 13\%    & 0\%    & 10\% \\
7--10 jets & 76\%      & 100\%      & 59\%     & 91\%    & 4\%    & 6\% \\
1-lepton   & 65\%      & 34\%       & 100\%    & 55\%    & 8\%    & 7\% \\
Multi-b    & 39\%      & 12\%       & 13\%     & 100\%   & 1\%    & 1\% \\
SS/3L      & 10\%      & 5\%        & 17\%     & 6\%     & 100\%  & 3\% \\ 
Monojet    & 99\%      & 3\%        & 6\%      & 5\%     & 1\%    & 100\% \\ 
\bottomrule
\end{tabular}
}
\caption{\label{tab:overlap} Exclusion overlap: percentage of models excluded by a Run~2 analysis on each row that is also excluded by another in the columns. 100\% is reserved for complete overlap of models excluded. 
}
\end{table*}

Overall, out of 181.8k pMSSM points considered for 13~TeV (3.2~fb$^{-1}$) sensitivity, 15.7\% were excluded by the combined constraints. Figure~\ref{fig:mass_plane_Gl_Sqk_chi10} displays the fraction of the 181.8k points surviving Run 1 that are excluded by the six early 13~TeV searches. These are projected into the 2-dimensional planes in the masses of the gluino $\tilde{g}$, LSP $\tilde{\chi}^0_1$ and lightest squark of the first or second generation $\tilde{q}$. Moreover, figure~\ref{fig:sparticle_summary} summarises the fraction of models excluded by various the masses of several other sparticles. 

For gluinos $\tilde{g}$, the high fractions (above 60\%) of models excluded involve masses below 1~TeV, and to a lesser extent around 1.4~TeV, is unambiguous. Reduced sensitivity around 1.1~TeV is due primarily to mass splittings between the gluino and LSP being less than 200~GeV (Figure~\ref{fig:sparticle_summary}). This corresponds to a localised region of high exclusion around 1.4~TeV gluino and 1.8~TeV squark masses the gluino--squark plane (Figure~\ref{fig:mass_plane_Gl_Sqk_chi10}). Overall, good corroboration with the simplified models is observed.

The lightest squark $\tilde{q}$ of the first two generations shows a smaller but noticeable extension of sensitivity beyond Run~1 with 3.2~fb$^{-1}$ of integrated luminosity. 
This is due to the less advantageous scaling of cross-sections between 8 and 13~TeV compared with gluino production from LHC parton distribution functions. In the squark--LSP plane (Figure~\ref{fig:mass_plane_Gl_Sqk_chi10}), the simplified model limit assumes all eight squarks are of the first two generations are mass-degenerate and therefore over-constrains squark scenarios in the pMSSM.

Figure~\ref{fig:sparticle_summary} separates the masses of the squarks by the chirality of the partner quark. As the left-handed squarks $\tilde{d}_L, \tilde{u}_L$ form an SU(2) doublet, their mass splittings are typically negligible compared to those of the right-handed squarks $\tilde{d}_R, \tilde{u}_R$. In accord with Ref.~\cite{Aad:2015baa}, right-handed down-type squarks are least constrained of the four light flavour squarks. Analyses targeting third generation squarks were not considered and so there is little impact on the lightest sbottom $\tilde{b}_1$ and stop $\tilde{t}_1$ masses. 

Also shown in Figure~\ref{fig:sparticle_summary} are uncoloured sparticles, which can participate as intermediate states in cascade decays of the gluino or squark(s). However, it is seen that the sensitivity of the analyses considered in Table~\ref{tab:listSearches} have little to no correlation with these sleptons or electroweakinos. The lower mass bounds on charged electroweak sparticles are due to LEP searches. Here, slepton $\tilde{\ell}$ refers to the lightest superpartner of the left- or right-handed charged lepton of the first or second generation. 

Table~\ref{tab:overlap} shows the overlap matrix between the six searches we considered. This quantifies the fraction of points excluded by one analysis are also excluded by another. For example, of the points excluded by the 7--10 jets search, 76\% of them were also excluded by the 2--6 jets analysis. Overall, the sensitivity is complementary, with no analysis completely excluding the points probed by another search.

\subsection{\label{sec:prior_distro}Distributions prior to 13~TeV interpretation}

\begin{figure*}
    \centering
    \begin{subfigure}[b]{0.49\textwidth}
         \includegraphics[width=\textwidth]{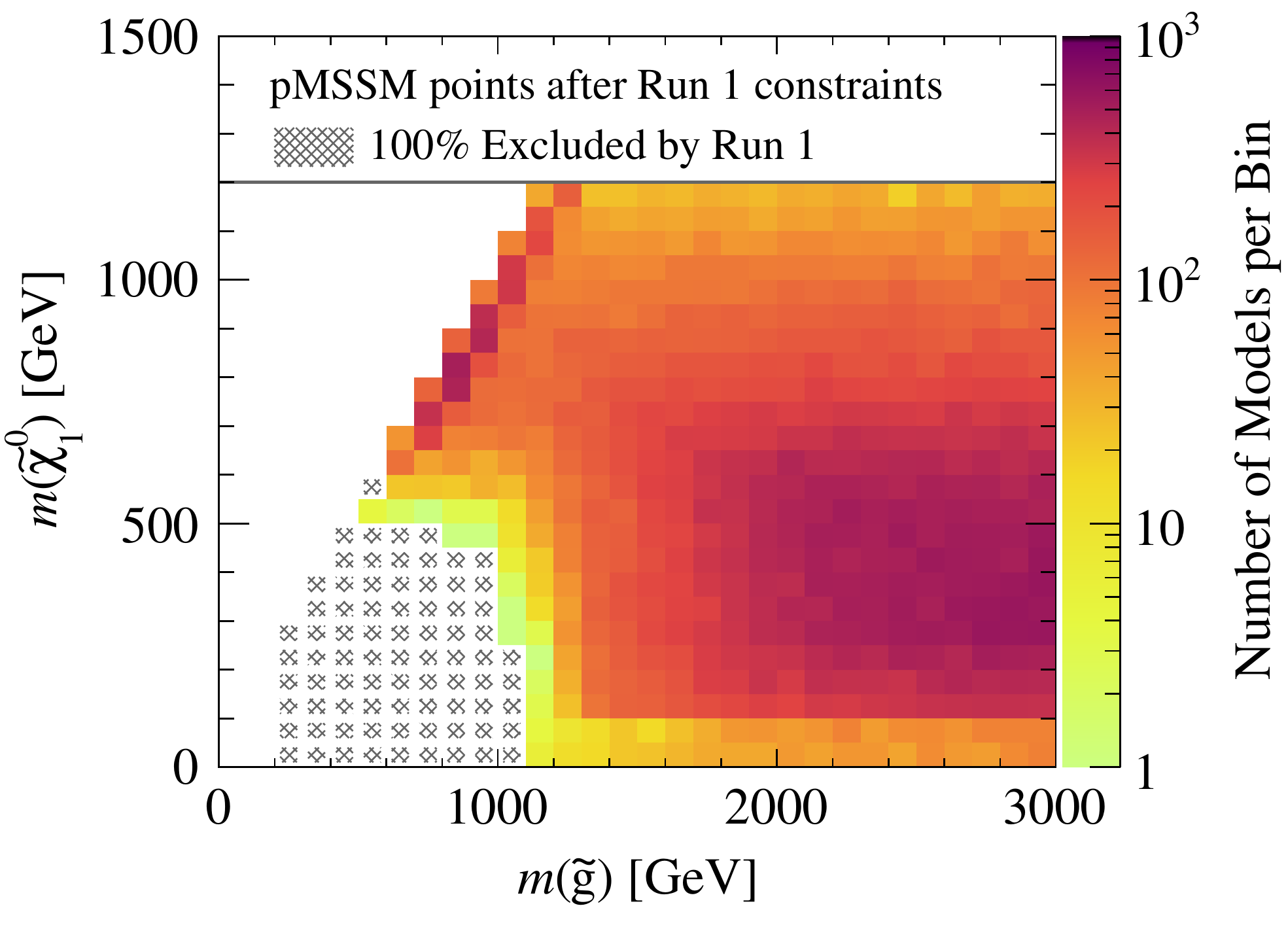}
          \caption{Gluino-LSP plane.}
        \end{subfigure}
        \begin{subfigure}[b]{0.49\textwidth}
         \includegraphics[width=\textwidth]{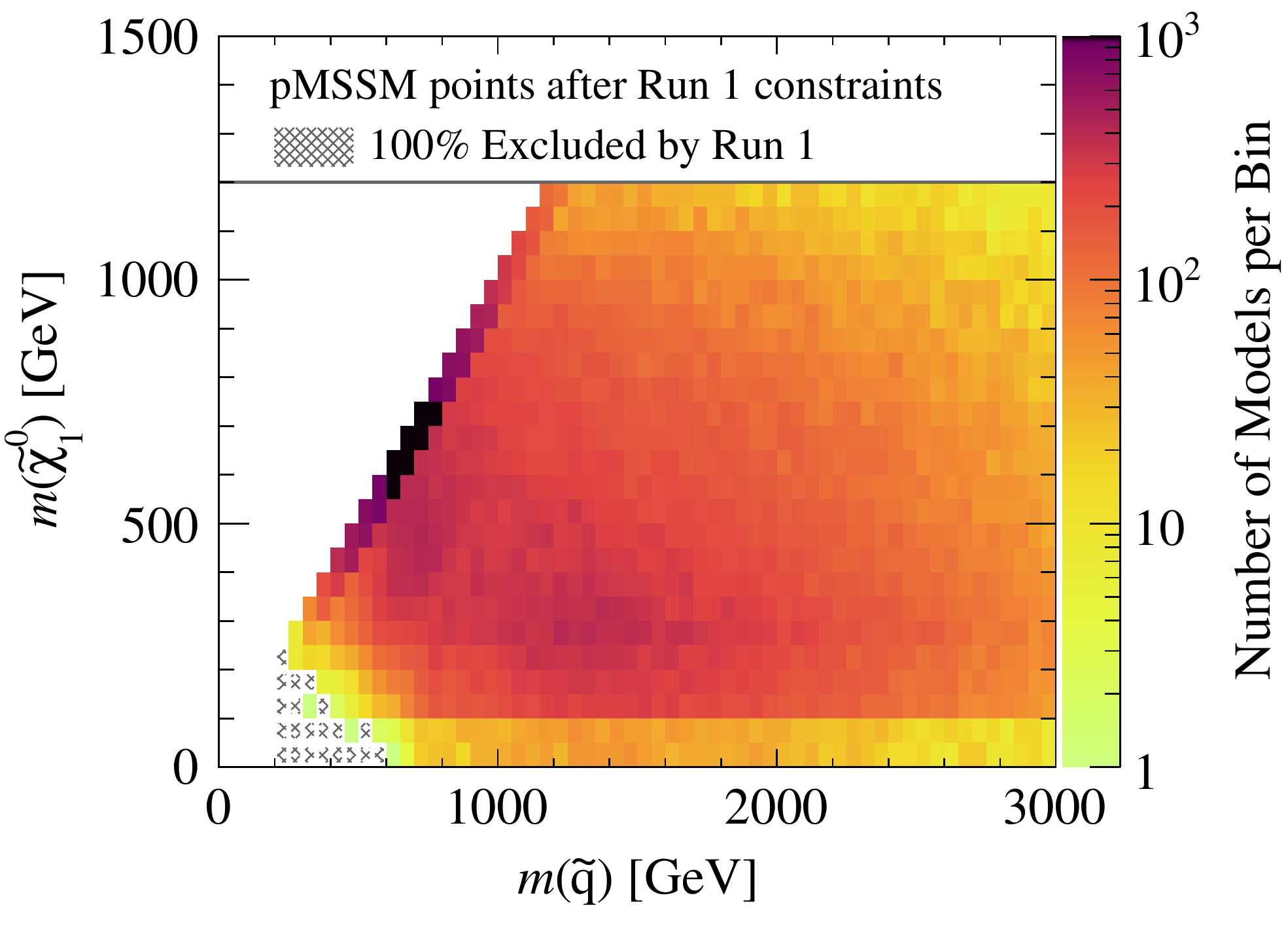}
          \caption{Squark-LSP plane.}
        \end{subfigure}
        \begin{subfigure}[b]{0.49\textwidth}
         \includegraphics[width=\textwidth]{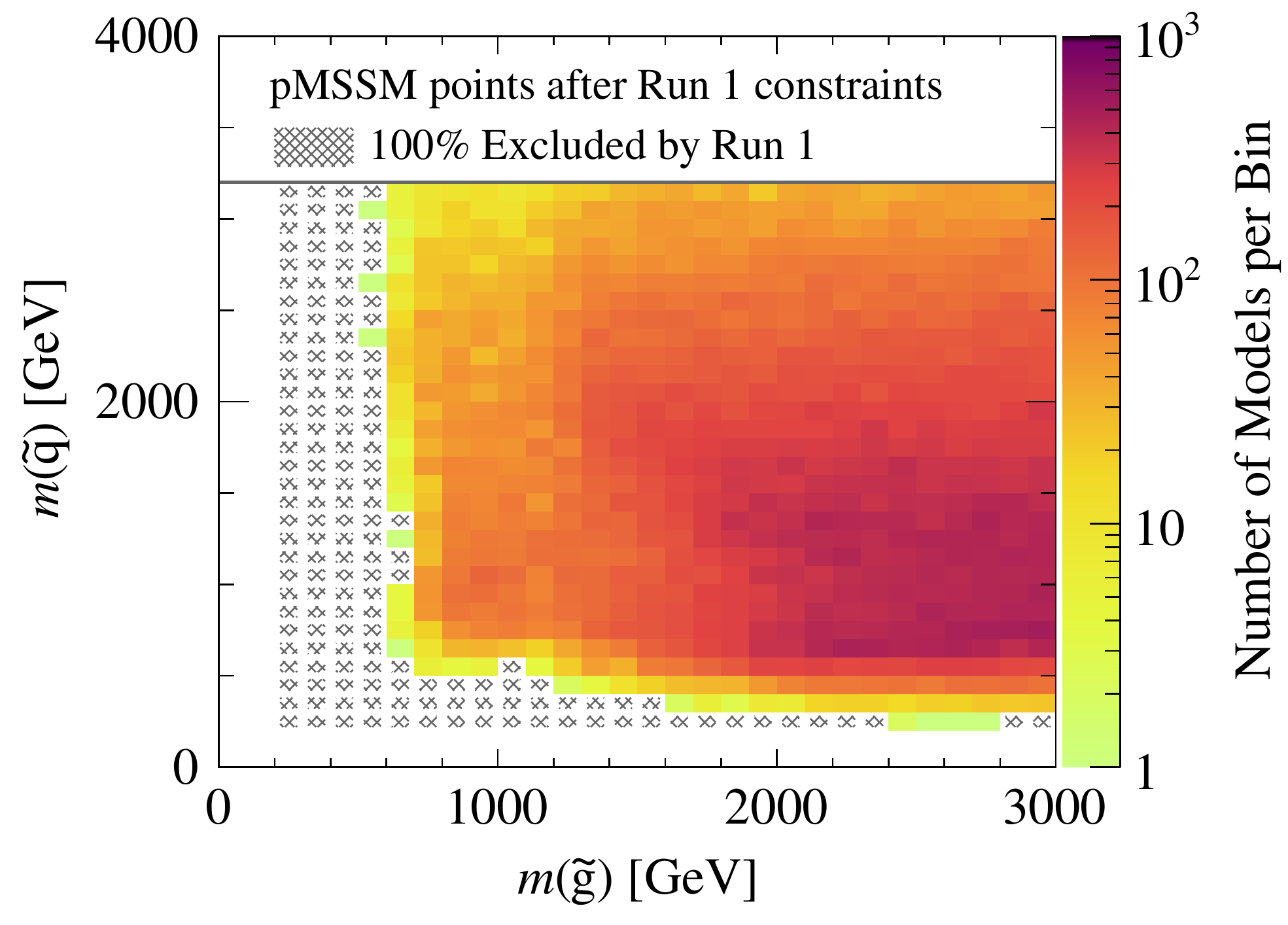}
          \caption{\label{fig:2D_prior_Gl_Sqk}Gluino-squark plane.}
        \end{subfigure}
    \caption{\label{fig:2Dprior_distro}Number of models that survived Run~1 ATLAS constraints, before applying our Run~2 analyses. White regions indicate no models. This is projected into the gluino-LSP (upper-left), squark-LSP (upper-right), and gluino-squark (lower) mass planes. Here $m(\tilde{q})$ is the mass of the lightest left- or right- handed squark of either the first or second generations. The colour scale has a maximum of 1000 models per bin, indicated by black. Hatched grey regions indicate mass bins where 100\% of the points are excluded by Run~1 searches in Ref.~\cite{Aad:2015baa}. Models with long-lived gluinos, squarks or sleptons are removed from these figures.
    }
    
\end{figure*}

When interpreting the results of this work, especially Figures~\ref{fig:mass_plane_Gl_Sqk_chi10}, it is instructive to examine the distribution of the 181.8k points that survived Run 1 constraints, before any 13~TeV constraints are applied. Figure \ref{fig:2Dprior_distro} displays these prior distributions projected into the three mass planes involving the gluino, squarks and LSP. Along the diagonal where the gluino or squark are similar in mass to the LSP, there is a larger abundance of models. These are primarily models with bino-like LSP, which require the LSP to coannihilate with a near-degenerate gluino to prevent oversaturation of the observed relic abundance. Their density were enhanced from the importance sampling used by ATLAS in the pre-selection of the pMSSM points~\cite{Aad:2015baa}.

Furthermore, the region below 100~GeV with substantially fewer models are also mainly bino-like LSP models. These correspond to where the LSP relic density is reduced by resonance annihilation through a $Z^0$ or $h^0$ boson through the so-called `funnel' mechanism. Models with Higgsino- or Wino-like LSPs typically have near-degenerate charginos which are constrained by LEP bounds, hence reducing the concentration of models in this region. These projections also show very few (less than 10) models in the mass bins coloured green close to the region where 100\% of models were excluded in Run~1.

\section{\label{sec:validation} Details of analysis validation}

This appendix details the validation of the analyses we implemented in this work. The purpose of the validation was the check our implementation of event selection in \textsc{MadAnalysis}. We display the cutflows (signal acceptance after each event selection)
for our implementation compared with those published in the supplementary material of each ATLAS analysis. We also present a simplified model limit comparison for each of the six analysis. 

To generate signal samples for validation, we used the same simulation setup as all the pMSSM points simulated in this study (outlined in Subection~\ref{sec:13TeV_method}). The main differences here are that we generate signals with up to 2 extra partons in the matrix element and the jet matching scale was set to one quarter of the produced sparticle mass, in accord with ATLAS. The event yields were normalised to the NLO squark or gluino cross sections from the LHC SUSY Cross Section Working Group~\cite{Borschensky:2014cia}.

In summary, the agreement is sufficient for the scope of this study. 

\subsection{\label{sec:vali_cutflows}Cutflow comparison}

For the cutflow comparison, we generated signal samples at a benchmark point presented by ATLAS in the public supplementary material of each analysis. The comparison ensures our implementations of each search in \textsc{MadAnalysis} are selecting the correct kinematic spaces in all signal regions. Tables~\ref{tab:cuflow_26J} to \ref{tab:cutflow_MonoJ} compare the cumulative percentage signal acceptances found by the collaboration `ATLAS' after each selection criterion `Cut' with ours `MA5 Recast'. In these tables, the model nomenclature and variables labelling each cut follows those of the ATLAS supplementary material for ease of comparison; see the references in Table~\ref{tab:listSearches} for full details. 

We achieved better than 30\% agreement in the final yields for the signal regions `SR' in each analysis. This is reasonable given the systematic uncertainties introduced by using a different version of \textsc{Pythia} for showering compared with ATLAS, together with application of the same object isolation, overlap removal, reconstruction and $b$-tagging efficiencies in the \textsc{Delphes} detector simulation for all six analyses (to reduce disk usage to a manageable level).

\begin{table}[]
\centering
\begin{tabular}{llll}
\toprule
                &                                    &\multicolumn{2}{l}{Acceptance [\%]} \\
SR                   & Cut                                         & ATLAS          & MA5 Recast                \\
\midrule
\multirow{7}{*}{2jl} & Initial                                     & 100            & 100                          \\
                     & $E_{\rm T}^{\rm miss}, p_{\rm T}^{j1} > 200$ GeV       & 90.2           & 91.4                       \\
                     & 2 jet multiplicity                          & 90.2           & 91.4                    \\
                     & $\Delta \phi(E_{\rm T}^{\rm miss}, {\rm jet})_\textrm{min}$ & 55.0           & 56.1                       \\
                     & $p_{\rm T}^{j2}>200$ GeV                            & 54.3           & 54.9                       \\
                     & $E_{\rm T}^{\rm miss}/\sqrt{H_{\rm T}}>15$ GeV$^{1/2}$     & 29.2           & 28.7                       \\
                     & $m_{\rm eff}>1200$ GeV                      & 29.1           & 28.7                       \\
\midrule
\multirow{7}{*}{2jm} & Initial                                     & 100          & 100                      \\
                     & $E_{\rm T}^{\rm miss} > 200$, $p_{\rm T}^{j1} > 300$ GeV       & 89.8           & 90.6                       \\
                     & 2 jet multiplicity                          & 89.8           & 90.6                       \\
                     & $\Delta \phi(E_{\rm T}^{\rm miss}, {\rm jet})_\textrm{min}$ & 72.2           & 73.0                       \\
                     & $p_{\rm T}^{j2}>50$ GeV                           & 72.2           & 73.0                       \\
                     & $E_{\rm T}^{\rm miss}/\sqrt{H_{\rm T}}>15$ GeV$^{1/2}$      & 34.8           & 34.6                       \\
                     & $m_{\rm eff}>1600$ GeV                               & 34.3           & 33.7                       \\
\midrule
\multirow{7}{*}{2jt} & Initial                                     & 100          & 100                      \\
                     & $E_{\rm T}^{\rm miss}, p_{\rm T}^{j1} > 200$ GeV       & 90.2           & 91.4                       \\
                     & 2 jet multiplicity                          & 90.2           & 91.4                       \\
                     & $\Delta \phi(E_{\rm T}^{\rm miss}, {\rm jet})_\textrm{min}$ & 55.0           & 56.1                       \\
                     & $p_{\rm T}^{j2} > 200$ GeV                            & 54.3           & 54.9                       \\
                     & $E_{\rm T}^{\rm miss}/\sqrt{H_{\rm T}}>20$ GeV$^{1/2}$ & 16.7           & 16.2                       \\
                     & $m_{\rm eff} > 2000$ GeV                               & 16.0           & 14.8                       \\
\midrule
\multirow{9}{*}{4jt} & Initial                                     & 100          & 100                      \\
                     & $E_{\rm T}^{\rm miss}, p_{\rm T}^{j1} > 200$ GeV       & 90.2           & 91.4                       \\
                     & 4 jet multiplicity                          & 85.7           & 86.6                  \\
                     & $\Delta \phi(E_{\rm T}^{\rm miss}, {\rm jet})_\textrm{min}$ & 59.7           & 60.1                       \\
                     & $p_{\rm T}^{j2}>100$ GeV                            & 59.7           & 60.1                       \\
                     & $p_{\rm T}^{j4}>100$ GeV                            & 52.3           & 51.6                       \\
                     & Aplanarity $> 0.04$                                  & 35.6           & 35.9                       \\
                     & $E_{\rm T}^{\rm miss}/m_{\rm eff}>0.2$          & 25.2           & 26.4                       \\
                     & $m_{\rm eff}>2200$ GeV                               & 21.4           & 20.4                       \\
\midrule
\multirow{9}{*}{5j}  & Initial                                     & 100          & 100                      \\
                     & $E_{\rm T}^{\rm miss}, p_{\rm T}^{j1} > 200$ GeV       & 90.2           & 91.4                       \\
                     & 5 jet multiplicity                          & 64.3           & 67.3                       \\
                     & $\Delta \phi(E_{\rm T}^{\rm miss}, {\rm jet})_\textrm{min}$ & 43.6           & 45.3                       \\
                     & $p_{\rm T}^{j2}>100$ GeV                            & 43.6           & 45.3                       \\
                     & $p_{\rm T}^{j4}>100$ GeV                            & 40.2           & 41.4                       \\
                     & Aplanarity $> 0.04$                                 & 28.6           & 30.1                       \\
                     & $E_{\rm T}^{\rm miss}/m_{\rm eff}>0.25$          & 14.0           & 15.1                       \\
                     & $m_{\rm eff}>1600$ GeV                               & 13.9           & 14.8                       \\
\midrule
\multirow{9}{*}{6jm} & Initial                                     & 100          & 100                      \\
                     & $E_{\rm T}^{\rm miss}, p_{\rm T}^{j1} > 200$ GeV       & 90.2           & 91.4                       \\
                     & 6 jet multiplicity                          & 36.8           & 42.3                       \\
                     & $\Delta \phi(E_{\rm T}^{\rm miss}, {\rm jet})_\textrm{min}$ & 23.7           & 27.3                       \\
                     & $p_{\rm T}^{j2}>100$ GeV                            & 23.7           & 27.3                       \\
                     & $p_{\rm T}^{j4}>100$ GeV                            & 22.6           & 26.0                       \\
                     & Aplanarity $>0.04$                                 & 16.9           & 20.0                       \\
                     & $E_{\rm T}^{\rm miss}/m_{\rm eff}>0.25$          & 7.6            & 8.9                        \\
                     & $m_{\rm eff}>1600$ GeV                               & 7.5            & 8.8                        \\
\midrule
\multirow{9}{*}{6jt} & Initial                                     & 100          & 100                      \\
                     & $E_{\rm T}^{\rm miss}, p_{\rm T}^{j1} > 200$ GeV       & 90.2           & 91.4                       \\
                     & 6 jet multiplicity                          & 36.8           & 42.3                       \\
                     & $\Delta \phi(E_{\rm T}^{\rm miss}, {\rm jet})_\textrm{min}$ & 23.7           & 27.3                       \\
                     & $p_{\rm T}^{j2}>100$ GeV                            & 23.7           & 27.3                       \\
                     & $p_{\rm T}^{j4}>100$ GeV                            & 22.6           & 26.0                       \\
                     & Aplanarity $>0.04$                                  & 16.9           & 20.0                       \\
                     & $E_{\rm T}^{\rm miss}/m_{\rm eff}>0.2$          & 10.6           & 13.0                       \\
                     & $m_{\rm eff}>2000$ GeV                               & 9.9            & 11.9                       \\
\bottomrule
\end{tabular}
\caption{Cutflow comparison for 2--6 jets analysis using gluino direct decay $\left(m(\tilde{g}), m(\tilde{\chi}^0_1)\right) = (1600, 0)$ GeV signal.}
\label{tab:cuflow_26J}
\end{table}

\begin{table}[]
\centering
\begin{tabular}{llll}
\toprule
               & \multicolumn{2}{l}{Acceptance [\%]} \\
Cut            & ATLAS          & MA5 Recast            \\
\midrule
Initial & 100          & 100                      \\
Preselection   & 99.9           & --                         \\
Event cleaning & 99.0           & --                         \\
MET cleaning   & 95.8           & --                         \\
Lepton veto    & 47.8           & 48.8                       \\
\midrule\midrule
6j50           & 42.4           & 44.1                       \\
Trigger        & 42.4           & --                         \\
\midrule
8j50           & 25.0           & 25.7                       \\
$E_{\rm T}^{\rm miss}/\sqrt{H_{\rm T}} > 4$ GeV$^{1/2}$  & 17.9           & 18.5                       \\
$N_\textrm{b-jet} \geq 0$        & 17.9           & 18.5                       \\
$N_\textrm{b-jet} \geq 1$        & 17.3           & 17.9                       \\
$N_\textrm{b-jet} \geq 2$        & 14.2           & 15.5                       \\
\midrule
9j50           & 16.5           & 16.6                       \\
$E_{\rm T}^{\rm miss}/\sqrt{H_{\rm T}} > 4$ GeV$^{1/2}$  & 11.3           & 11.6                       \\
$N_\textrm{b-jet} \geq 0$        & 11.3           & 11.6                       \\
$N_\textrm{b-jet} \geq 1$        & 11.0           & 11.3                       \\
$N_\textrm{b-jet} \geq 2$        & 9.3            & 10.1                       \\
\midrule
10j50          & 9.1            & 9.3                        \\
$E_{\rm T}^{\rm miss}/\sqrt{H_{\rm T}} > 4$ GeV$^{1/2}$  & 6.0            & 6.2                        \\
$N_\textrm{b-jet} \geq 0$        & 6.0            & 6.2                        \\
$N_\textrm{b-jet} \geq 1$        & 5.8            & 6.1                        \\
$N_\textrm{b-jet} \geq 2$        & 5.1            & 5.5                        \\
\midrule\midrule
5j80           & 40.5           & 43.5                       \\
Trigger        & 40.5           & --                         \\
\midrule
7j80           & 19.7           & 18.7                       \\
$E_{\rm T}^{\rm miss}/\sqrt{H_{\rm T}} > 4$ GeV$^{1/2}$  & 13.6           & 13.2                       \\
$N_\textrm{b-jet} \geq 0$        & 13.6           & 13.2                       \\
$N_\textrm{b-jet} \geq 1$        & 13.2           & 12.8                       \\
$N_\textrm{b-jet} \geq 2$        & 10.8           & 11.1                       \\
\midrule
8j80           & 10.3           & 9.7                        \\
$E_{\rm T}^{\rm miss}/\sqrt{H_{\rm T}} > 4$ GeV$^{1/2}$  & 6.8            & 6.6                        \\
$N_\textrm{b-jet} \geq 0$        & 6.8            & 6.6                        \\
$N_\textrm{b-jet} \geq 1$        & 6.7            & 6.4                        \\
$N_\textrm{b-jet} \geq 2$        & 5.5            & 5.8                        \\
\bottomrule
\end{tabular}
\caption{Cutflow comparison for 7--10 jets analysis using pMSSM slice signal with gluino and lightest chargino masses given by $\left(m(\tilde{g}), m(\tilde{\chi}^\pm_1)\right) = (1300, 200)$ GeV. Signal regions are defined by the jet multiplicities and b-tagging requirements.}
\label{tab:cutflow_710J}
\end{table}

\begin{table}[]
\centering
\begin{tabular}{llll}
\toprule 
                           &                              & \multicolumn{2}{l}{Acceptance [\%]} \\
SR                         & Cuts                         & ATLAS             & MA5 Recast                 \\
\midrule\midrule
   \multicolumn{4}{l}{Gluino 1-step $\left(m(\tilde{g}), m(\tilde{\chi}^\pm_1), m(\tilde{\chi}^0_1)\right) = (1385, 705, 25)$ GeV} \\
\midrule
\multirow{8}{*}{Preselect} & Initial                        & 100             & 100                             \\
                           & Cleaning cuts                  & 88.5              & --                                \\
                           & 1 base lepton                  & 30.8              & 29.8                              \\
                           & 1 signal lepton                & 25.1              & 26.5                              \\
                           & $N_\textrm{jet} \geq 3$        & 25.1              & 26.5                              \\
                           & $E_{\rm T}^{\rm miss} > 100$ GeV  & 24.6              & 25.5                              \\
                           & $E_{\rm T}^{\rm miss}$ trigger & 24.4              & --                                \\
                           & $p_{\rm T}^\ell >  35 $ GeV       & 23.3              & 25.5                              \\
\midrule
\multirow{8}{*}{5j}        & $E_{\rm T}^{\rm miss} > 250$ GeV & 19.2              & 20.6                              \\
                           & $N_\textrm{jet} \geq 5$       & 18.2              & 19.7                              \\
                           & $p_{\rm T}^{j1} > 225$ GeV       & 18.2              & 19.4                              \\
                           & $p_{\rm T}^{j5} > 50$ GeV        & 16.0              & 16.9                              \\
                           & Aplanarity $>0.04$          & 12.7              & 13.0                              \\
                           & $m_{\rm T} > 275$ GeV            & 8.5               & 8.5                               \\
                           & $E_{\rm T}^{\rm miss}/m_{\rm eff} > 0.1$  & 8.3   & 8.4                             \\
                           & $m_{\rm eff} > 1800$ GeV         & 7.6               & 7.2                               \\
\midrule
\multirow{8}{*}{6j}        & $E_{\rm T}^{\rm miss} > 250$ GeV  & 19.2              & 20.6                              \\
                           & $N_\textrm{jet} \geq 6$      & 15.1              & 17.5                              \\
                           & $p_{\rm T}^{j1} > 125$ GeV        & 15.1              & 17.5                              \\
                           & $p_{\rm T}^{j6} > 30$ GeV         & 15.1              & 16.5                              \\
                           & Aplanarity $>0.04$          & 12.0              & 12.8                              \\
                           & $m_{\rm T} > 225 $ GeV           & 9.0               & 9.2                               \\
                           & $E_{\rm T}^{\rm miss}/m_{\rm eff} > 0.2$  & 5.3       & 5.5                           \\
                           & $m_{\rm eff} > 1000$ GeV         & 5.3               & 5.5                               \\
\midrule
\multirow{7}{*}{4j low-x}  & $E_{\rm T}^{\rm miss} > 200$ GeV  & 21.6              & 22.5                              \\
                           & $N_\textrm{jet} \geq 4$        & 21.4              & 22.4                              \\
                           & $p_{\rm T}^{j1} > 325$ GeV        & 19.4              & 19.1                              \\
                           & $p_{\rm T}^{j4} > 150$ GeV        & 10.5              & 10.3                              \\
                           & Aplanarity $ >0.04$         & 8.9               & 8.4                               \\
                           & $m_{\rm T} > 125 $ GeV           & 7.7               & 6.9                               \\
                           & $m_{\rm eff} > 2000$ GeV          & 6.3               & 5.9                               \\
\midrule
\multirow{7}{*}{4j high-x} & $E_{\rm T}^{\rm miss} > 200$ GeV  & 21.6              & 22.5                              \\
                           & $N_\textrm{jet} \geq 4$         & 21.4              & 22.4                              \\
                           & $p_{\rm T}^{j1} > 325$ GeV         & 19.4              & 19.1                              \\
                           & $p_{\rm T}^{j4} > 30$ GeV        & 19.4              & 19.1                              \\
                           & $m_{\rm T} > 425$ GeV            & --                & 7.6                               \\
                           & $E_{\rm T}^{\rm miss}/m_{\rm eff} > 0.3 $   & 1.5       & 1.5                        \\
                           & $m_{\rm eff} > 1800 $ GeV         & 1.4               & 1.3                               \\
\midrule\midrule
   \multicolumn{4}{l}{Gluino 1-step $\left(m(\tilde{g}), m(\tilde{\chi}^\pm_1), m(\tilde{\chi}^0_1)\right) = (1000, 110, 60)$ GeV}                \\
\midrule
\multirow{7}{*}{Preselect} & Initial                          & 100             & 100                             \\
                           & Cleaning cuts                    & 85.9              & --                                \\
                           & 1 base lepton                    & 36.2              & --                                \\
                           & 1 signal lepton                  & 30.3              & 35.1                             \\
                           & $N_\textrm{jet} \geq 2$          & 30.3              & 26.5                             \\
                           & $p_{\rm T}^\ell < 35 $ GeV       & 9.9               & 9.4                              \\
                           & $E_{\rm T}^{\rm miss} > 200$ GeV & 6.8               & 6.5                              \\
\midrule
\multirow{5}{*}{2j}        & $E_{\rm T}^{\rm miss} > 530$ GeV  & 0.8               & 0.9                              \\
                           & $N_\textrm{jet} \geq 2$           & 0.8               & 0.9                              \\
                           & $p_{\rm T}^{j1} > 180$ GeV        & 0.8               & 0.9                              \\
                           & $m_{\rm T} > 100$ GeV            & 0.2               & 0.2                              \\
                           & $E_{\rm T}^{\rm miss}/m_{\rm eff} > 0.3$     & 0.0               & 0.03                              \\
\midrule
\multirow{7}{*}{5j}        & $E_{\rm T}^{\rm miss} > 375$ GeV & 2.5               & 2.6                              \\
                           & $N_\textrm{jet} \geq 5$          & 2.3               & 2.6                              \\
                           & $p_{\rm T}^{j1} > 200$ GeV       & 2.3               & 2.4                              \\
                           & $p_{\rm T}^{j2} > 200$ GeV       & 2.1               & 2.4                              \\
                           & $p_{\rm T}^{j3} > 200$ GeV       & 1.4               & 2.1                              \\
                           & Aplanarity $> 0.02$            & 1.3               & 1.4                              \\
                           & $H_{\rm T} > 1100$ GeV           & 1.2               & 1.3                               \\
\bottomrule
\end{tabular}

\caption{Cutflow comparison for 1-lepton analysis.}
\label{tab:cutflow_1L}
\end{table}

\begin{table}[]
\centering
\begin{tabular}{lllll}
\toprule
                                  &                                  & \multicolumn{2}{l}{Acceptance [\%]} \\
SR                                & Cut                             & ATLAS       & MA5 Recast            \\
\midrule\midrule
   \multicolumn{4}{c}{Gtt $\left(m(\tilde{g}), m(\tilde{\chi}^0_1)\right) = (1700, 200)$ GeV}                       \\
\midrule
\multirow{6}{*}{Preselect}        & Initial                       & 100       & 100                   \\
                                  & Trigger                         & 92.1        & --                    \\
                                  & Cleaning                        & 90.7        & --                    \\
                                  & $N_\textrm{jets} \geq 4$        & 90.7        & 94.5                  \\
                                  & $N_\textrm{b-jets} \geq 3$      & 68.2        & 70.3                  \\
                                  & $E_\textrm{T}^{\rm miss} > 200$ GeV    & 63.5        & 63.9                  \\
\midrule                              
\multirow{4}{*}{Gtt-0L}       & Lepton veto                     & 35.2        & 34.7                  \\
                                  & $\Delta \phi^{4j}_{\rm min}$    & 26.3        & 26.1                  \\
                                  & $N_\textrm{jets} \geq 8$  & 22.0    & 22.5                  \\
                                  & $m_\textrm{T, min}^\textrm{b-jets} > 80$ GeV & 19.8        & 20.2                  \\
\midrule
\multirow{4}{*}{Gtt-0L-A}         & $m_{\rm eff}^{\rm incl} > 1700$ GeV   & 18.5        & 18.7                  \\
                                  & $E_\textrm{T}^{\rm miss} > 400 $ GeV   & 15.4        & 15.4                  \\
                                  & $N_\textrm{b-jets} \geq 3$        & 15.4        & 15.4                  \\
                                  & $N_\textrm{top} \geq 1$             & 15.2        & 14.4                  \\
\midrule
\multirow{4}{*}{Gtt-0L-B}         & $m_{\rm eff}^{\rm incl} > 1250$ GeV   & 19.7        & 20.0                  \\
                                  & $E_\textrm{T}^{\rm miss} > 350 $ GeV   & 17.2        & 17.3                  \\
                                  & $N_\textrm{b-jets} \geq 4$        & 11.1        & 11.4                  \\
                                  & $N_\textrm{top} \geq 1$             & 10.9        & 10.9                  \\
\midrule
\multirow{3}{*}{Gtt-0L-C}         & $m_{\rm eff}^{\rm incl} > 1250$ GeV   & 19.7        & 20.0                  \\
                                  & $E_\textrm{T}^{\rm miss} > 350 $ GeV   & 17.2        & 17.3                  \\
                                  & $N_\textrm{b-jets} \geq 4$        & 11.1        & 11.4                  \\
\midrule
\multirow{4}{*}{Gtt-1L}         & 1 signal lepton                   & 28.3        & 29.2                  \\
                                  & $N_\textrm{b-jets} \geq 6$      & 26.5        & 27.7                  \\
                                  & $m_\textrm{T, min}^\textrm{b-jets} > 80$ GeV & 17.1        & 17.8                  \\
                                  & $m_{\rm T} > $ 150 GeV            & 14.2        & 14.7                  \\
\midrule
\multirow{3}{*}{Gtt-1L-A}         & $m_{\rm eff}^{\rm incl} > 1100$ GeV & 14.2        & 14.7                  \\
                                  & $E_\textrm{T}^{\rm miss} > 200$ GeV    & 14.2        & 14.7                  \\
                                  & $N_\textrm{top} \geq 1$             & 13.5        & 13.6                  \\
\midrule
\multirow{2}{*}{Gtt-1L-B}         & $m_{\rm eff}^{\rm incl} > 1100$ GeV & 14.2        & 14.7                  \\
                                  & $E_\textrm{T}^{\rm miss} > 300$ GeV    & 13.3        & 13.8                  \\
\midrule\midrule
   \multicolumn{4}{c}{Gbb $\left(m(\tilde{g}), m(\tilde{\chi}^0_1)\right) = (1700, 200)$ GeV} \\
\midrule
\multirow{6}{*}{Preselect}        & Initial                       & 100       & 100                \\
                                  & Trigger                         & 96.0        & --                 \\
                                  & Cleaning                        & 94.7        & --                   \\
                                  & $N_\textrm{jets} \geq 4$        & 92.1        & 92.2                 \\
                                  & $N_\textrm{b-jets} \geq 3$      & 58.9        & 60.4                 \\
                                  & $E_\textrm{T}^{\rm miss} > 200$ GeV    & 55.6        & 55.6                 \\
\midrule
\multirow{2}{*}{Gbb-0L}           & Lepton veto                     & 55.0        & 55.6                 \\
                                  & $\Delta \phi^{4j}_{\rm min}$    & 39.5        & 40.6                 \\
\midrule
\multirow{4}{*}{Gbb-0L-A}         & $m_{\rm eff}^{4j} > 1600$ GeV       & 36.6        & 36.3                 \\
                                  & $E_\textrm{T}^{\rm miss} > 350$ GeV    & 33.0        & 32.5                 \\
                                  & $N_\textrm{jets}^{p_{\rm T} > 90} \geq 4$   & 29.4        & 29.1                 \\
                                  & $N_\textrm{b-jets}^{p_{\rm T} > 90} \geq 3$ & 25.5        & 24.9                 \\
\midrule
\multirow{4}{*}{Gbb-0L-B}         & $m_{\rm eff}^{4j} > 1400$ GeV       & 38.5       & 38.8                 \\
                                  & $E_\textrm{T}^{\rm miss} > 450$ GeV        & 30.2        & 30.1                 \\
                                  & $N_\textrm{jets}^{p_{\rm T} > 90} \geq 4$   & 26.7        & 26.6                 \\
                                  & $N_\textrm{b-jets}^{p_{\rm T} > 90} \geq 3$ & 23.1        & 22.7                 \\
\midrule
\multirow{2}{*}{Gbb-0L-C}         & $m_{\rm eff}^{4j} > 1400$ GeV       & 38.5        & 38.8                 \\
                                  & $E_\textrm{T}^{\rm miss} > 300$ GeV    & 28.1        & 27.6                 \\
\bottomrule
\end{tabular}

\caption{Cutflow comparison for Multi-b analysis.}
\label{tab:cutflow_Multib}
\end{table}

\begin{table}[]
\centering
\begin{tabular}{llll}
\toprule
                        &                              & \multicolumn{2}{l}{Acceptance [\%]}           \\
SR                      & Cut                          & ATLAS          & MA5 Recast                \\
\midrule\midrule
   \multicolumn{4}{c}{Gluino slepton $\left(m(\tilde{g}), m(\tilde{\chi}^0_1)\right) = (1300, 500)$ GeV} \\
\midrule
\multirow{7}{*}{SR0b3j} & Initial                      & 100            & 100                    \\
                        & $N_{\ell}\geq3$              & 9.7            & 10.1                     \\
                        & Trigger                      & 9.7            & --                       \\
                        & $N_\textrm{b-jet}=0$ & 7.9       & 8.6                       \\
                        & $N_\textrm{jet}\geq3$& 7.4       & 8.1                       \\
                        & $E_{\rm T}^{\rm miss} > 200$ GeV & 5.1            & 5.9                      \\
                        & $m_{\rm eff} > 550$ GeV          & 5.1            & 5.9                      \\
\midrule\midrule
   \multicolumn{4}{c}{Gluino 2-step $\left(m(\tilde{g}), m(\tilde{\chi}^0_1)\right) = (1100, 400)$ GeV}   \\
\midrule
\multirow{7}{*}{SR0b5j} & Initial                      & 100            & 100                    \\
                        & $N_{\ell}^{\rm SS}\geq2$         & 3.6          & 3.9                      \\
                        & Trigger                      & 3.4            & --                       \\
                        & $N_\textrm{b-jet}=0$& 2.2        & 2.5                       \\
                        & $N_\textrm{jet}\geq5$& 1.5       & 2.0                       \\
                        & $E_{\rm T}^{\rm miss} > 125$ GeV & 1.3            & 1.6                      \\
                        & $m_{\rm eff} > 650$ GeV          & 1.3            & 1.6                      \\
\midrule\midrule 
   \multicolumn{4}{c}{Sbottom 1-step $\left(m(\tilde{b}_1), m(\tilde{\chi}^0_1)\right) = (600, 50)$ GeV}     \\
\midrule
\multirow{7}{*}{SR1b}   & Initial                      & 100            & 100                    \\
                        & $N_{\ell}^{\rm SS}\geq2$         & 4.6            & 4.68                     \\
                        & Trigger                      & 4.1            & --                       \\
                        & $N_\textrm{b-jet}\geq1$          & 3.6            & 3.92                     \\
                        & $N_\textrm{jet}\geq4$            & 2.0            & 2.31                     \\
                        & $E_{\rm T}^{\rm miss} > 150$ GeV & 1.3            & 1.44                     \\
                        & $m_{\rm eff} > 550$ GeV          & 1.3            & 1.44                     \\
\midrule\midrule
   \multicolumn{4}{c}{Gtt $\left(m(\tilde{g}), m(\tilde{\chi}^0_1)\right) = (1200, 700)$ GeV}        \\
\midrule
\multirow{6}{*}{SR3b}   & Initial                      & 100            & 100                    \\
                        & $N_{\ell}^{\rm SS}\geq2$       & 3.4          & 3.6                      \\
                        & Trigger                      & 3.1            & --                       \\
                        & $N_\textrm{b-jet}\geq3$        & 1.5            & 1.8                      \\
                        & $E_{\rm T}^{\rm miss} > 125$ GeV & 1.2            & 1.4                      \\
                        & $m_{\rm eff} > 650$ GeV          & 1.2            & 1.4                      \\
\bottomrule
\end{tabular}

\caption{Cutflow comparison for SS/3L analysis.}
\label{tab:cutflow_SS3L}
\end{table}

\begin{table}[]
\centering
\begin{tabular}{llll}
\toprule
                           &                                        & \multicolumn{2}{l}{Acceptance [\%]} \\
SR                         & Cut                                    & ATLAS            & MA5 Recast           \\
\midrule
\multirow{9}{*}{Preselect} & Initial events                         & 100            & 100                \\
                           & Trigger                                & 83.7           & --                   \\
                           & Cleaning cut                           & 83.0           & --                   \\
                           & Lepton veto                            & 83.0           & 100                \\
                           & $N_{\rm jets} \leq 4$                  & 77.3           & 93.8               \\
                           & $\Delta \phi(E_{\rm T}^{\rm miss}, {\rm jet}) > 0.4$         & 73.5           & 88.4               \\
                           & Jet quality                            & 70.1           & --                   \\
                           & $p_{\rm T}^{j1} > 250$ GeV         & 22.7           & 18.8               \\
                           & $E_{\rm T}^{\rm miss} > 250$           & 21.1           & 16.7               \\
\midrule
EM1                        & $250 < E_{\rm T}^{\rm miss} < 300$ GeV & 3.0            & 2.4                \\
EM2                        & $300 < E_{\rm T}^{\rm miss} < 350$ GeV & 3.4            & 2.5                \\
EM3                        & $350 < E_{\rm T}^{\rm miss} < 400$ GeV & 3.1            & 2.6                \\
EM4                        & $400 < E_{\rm T}^{\rm miss} < 500$ GeV & 4.4            & 4.0                \\
EM5                        & $500 < E_{\rm T}^{\rm miss} < 600$ GeV & 2.8            & 2.4                \\
EM6                        & $600 < E_{\rm T}^{\rm miss} < 700$ GeV & 1.8            & 1.2                \\
IM7                        & $E_{\rm T}^{\rm miss} > 700$       GeV & 2.6            & 1.5               \\
\bottomrule
\end{tabular}
\caption{Cutflow comparison for Monojet analysis using the squark direct decay model using the $\left(m(\tilde{q}), m(\tilde{\chi}^0_1)\right) = (650, 645)$ GeV signal.}
\label{tab:cutflow_MonoJ}
\end{table}

\subsection{\label{sec:vali_simpMods}Simplified model limits}

\begin{figure*}
    \centering
    \begin{subfigure}[b]{0.49\textwidth}
         \includegraphics[width=\textwidth]{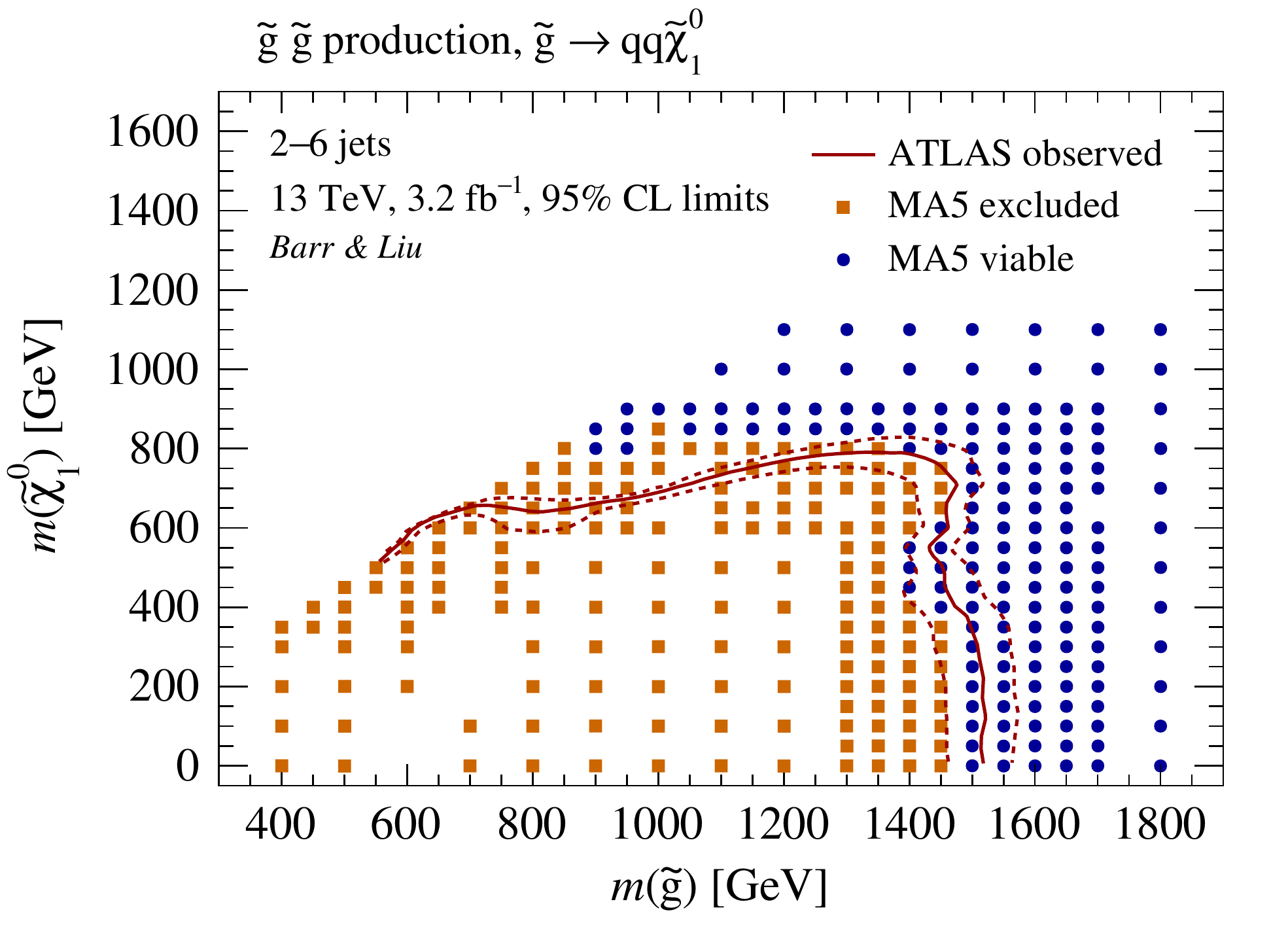}
          \caption{2--6 jets analysis, $\tilde{g}\rightarrow qq\tilde{\chi}^0_1$ direct decay model.}
        \end{subfigure}
        \begin{subfigure}[b]{0.49\textwidth}
         \includegraphics[width=\textwidth]{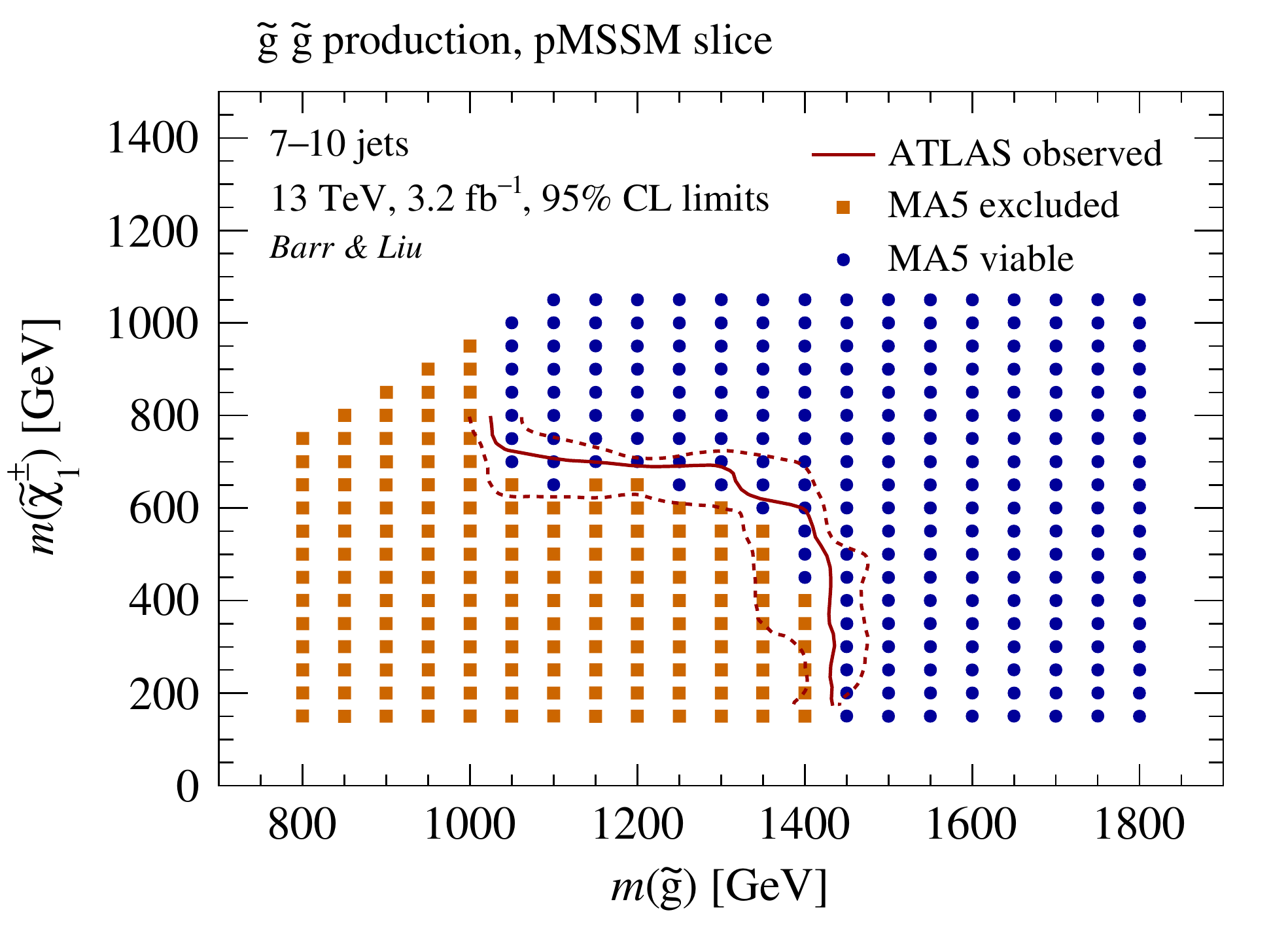}
          \caption{7--10 jets analysis, pMSSM slice model.}
        \end{subfigure}

        \begin{subfigure}[b]{0.49\textwidth}
         \includegraphics[width=\textwidth]{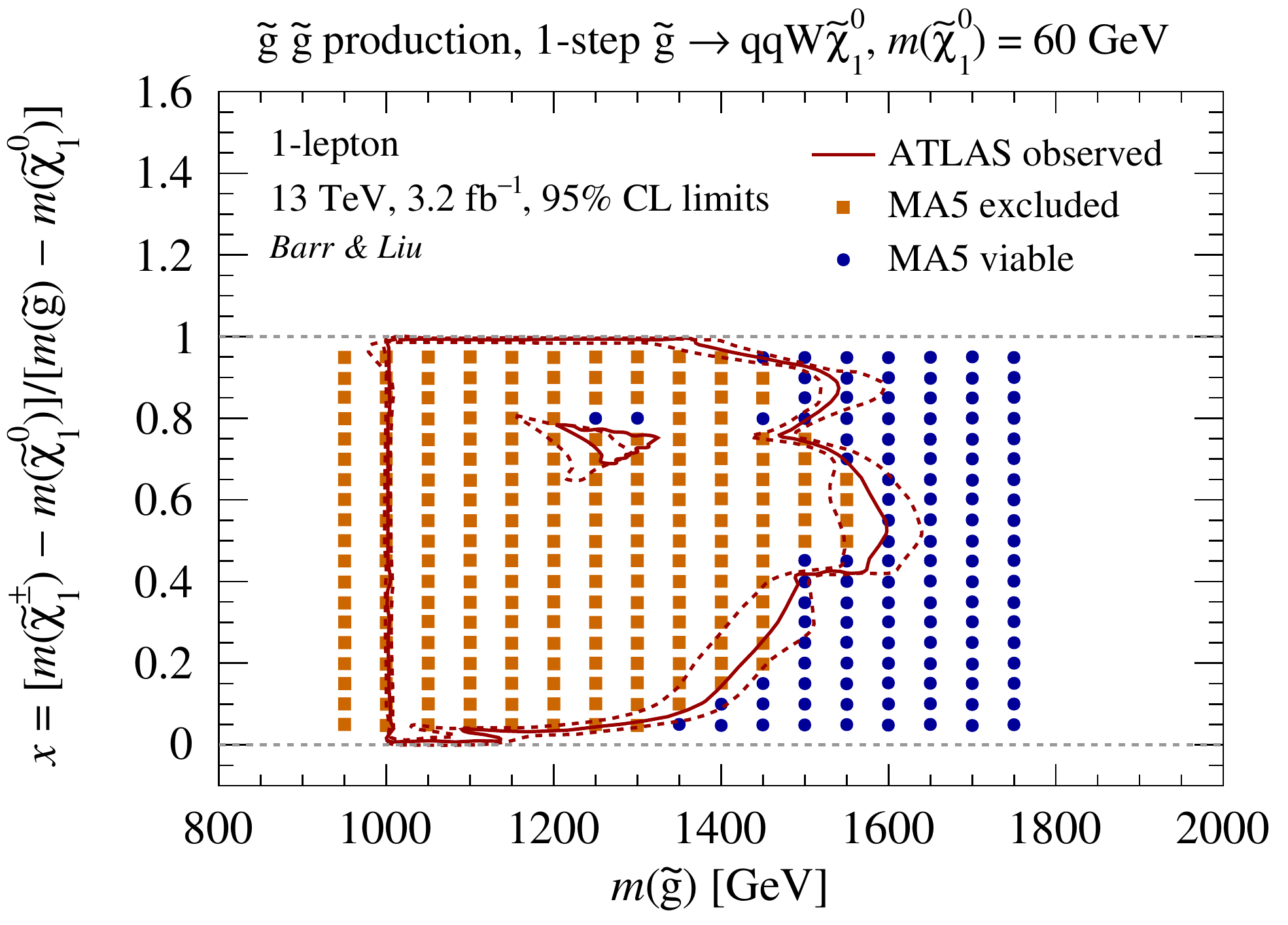}
          \caption{1 lepton analysis, $\tilde{g}\rightarrow qqW\tilde{\chi}^0_1$ model, fixed $m(\tilde{\chi}^0_1) = 60$ GeV.}
        \end{subfigure}
        \begin{subfigure}[b]{0.49\textwidth}
         \includegraphics[width=\textwidth]{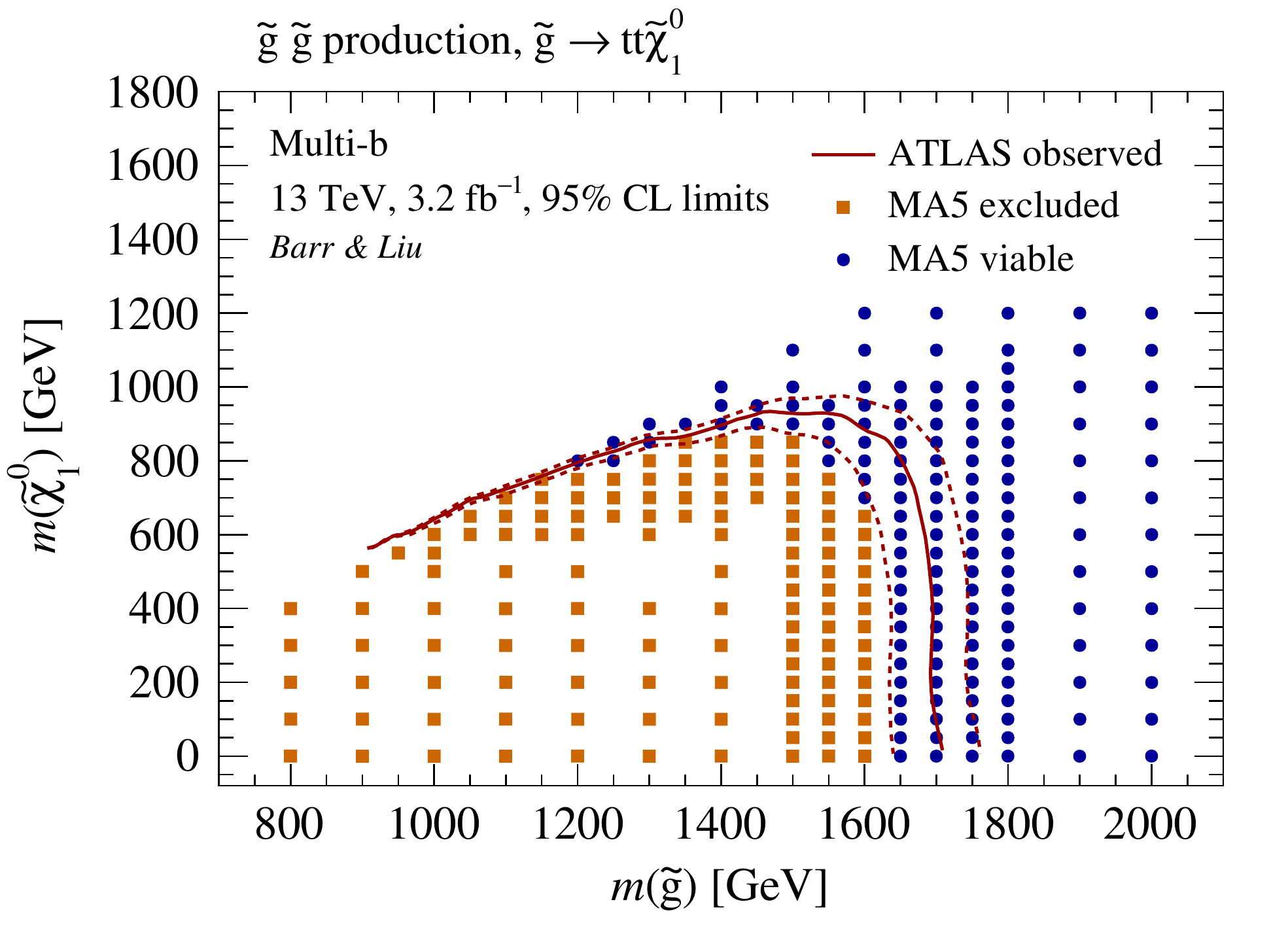}
          \caption{Multi-b analysis, $\tilde{g}\rightarrow tt\tilde{\chi}^0_1$ model.}
        \end{subfigure}

        \begin{subfigure}[b]{0.49\textwidth}
         \includegraphics[width=\textwidth]{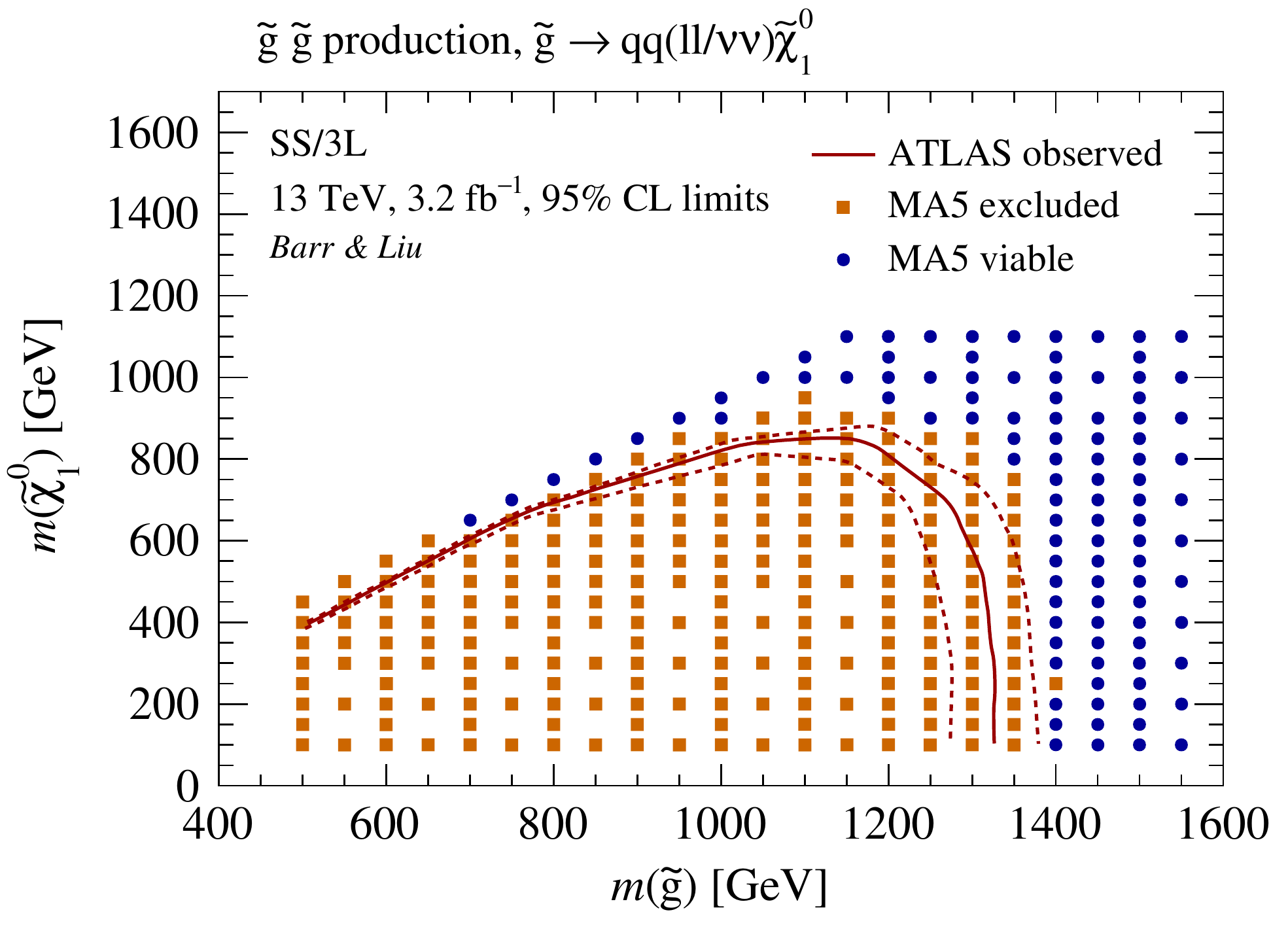}
          \caption{SS/3L analysis, $\tilde{g}\rightarrow qq(\ell\ell/\nu\nu)\tilde{\chi}^0_1$ model.}
        \end{subfigure}
        \begin{subfigure}[b]{0.49\textwidth}
         \includegraphics[width=\textwidth]{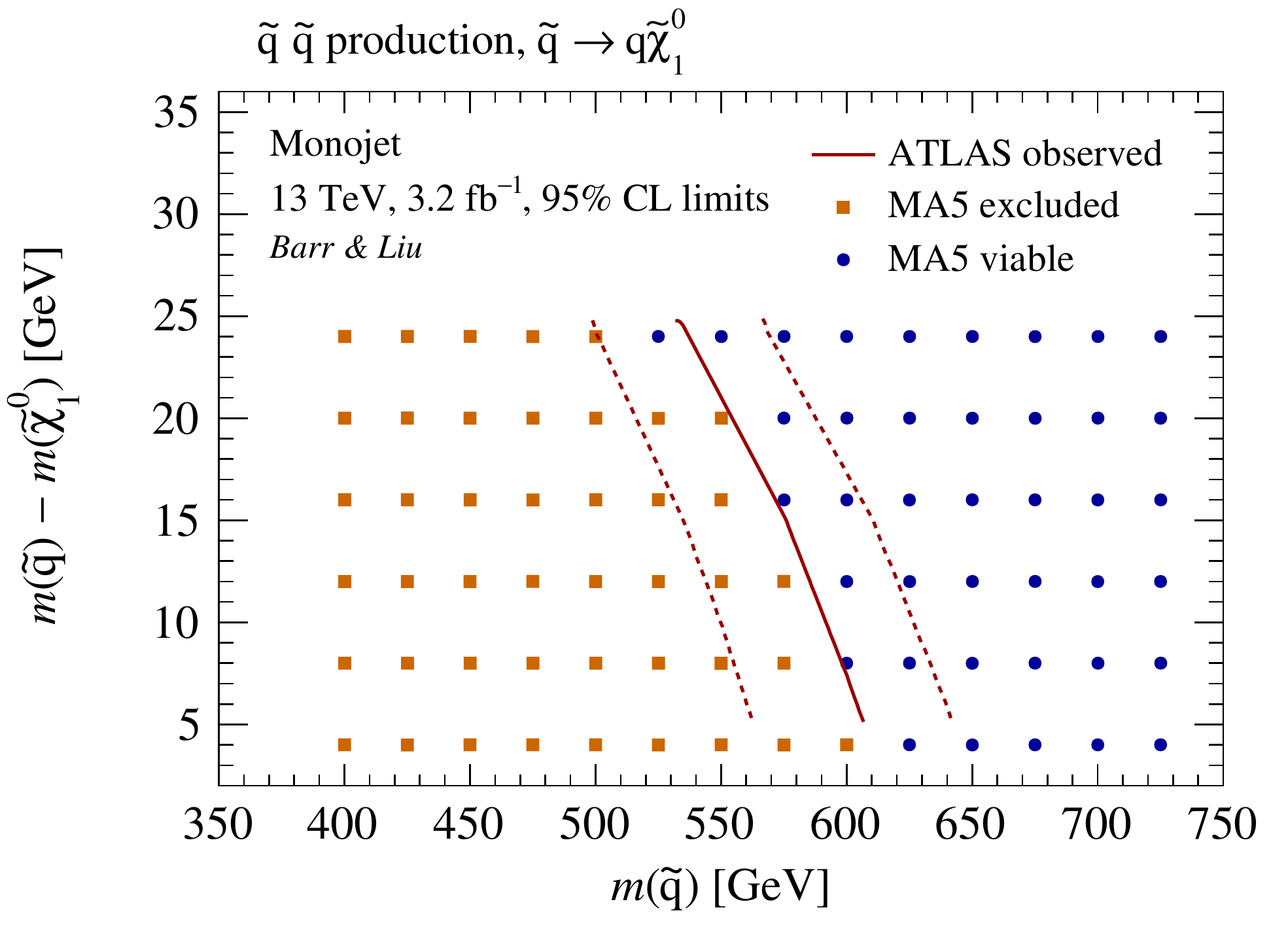}
          \caption{Monojet analysis, $\tilde{q}\rightarrow q\tilde{\chi}^0_1$ model.}
        \end{subfigure}
    \caption{\label{fig:vali_simpMods}
    Comparison of representative simplified model limits reproduced by our \textsc{MadAnalysis} implementation and those published by ATLAS for each analysis considered (Table~\ref{tab:listSearches}). The red solid line is the 95\% confidence level limit observed by ATLAS with the dashed lines either side indicating the reported one standard deviation theoretical uncertainties. Orange squares indicate points generated by us excluded at 95\% confidence level `MA5 excluded', while blue circles are those not excluded `MA5 viable'. The models follow the nomenclature used by ATLAS for ease of comparison; see the references in Table~\ref{tab:listSearches} for full details.
    }
    
\end{figure*}

Figure~\ref{fig:vali_simpMods} shows comparisons of exclusion limits published by ATLAS (solid lines) with our own excluded points (orange squares) using one of the simplified model scenarios considered in each analysis. We varied the appropriate masses in each model SLHA card and generated signal samples for each point. The \textsc{RecastingTools} routine in \textsc{MadAnalysis} was used to set 95\% confidence level limits. 

Small discrepancies arose between our \textsc{MadAnalysis} implementation and those of ATLAS due to various sources of systematic uncertainty in our simulation setup, such as different \textsc{Pythia} versions and fast detector simulation setup. Nonetheless, the general corroboration in the shape of our \textsc{MadAnalysis} exclusions across the simplified model parameter spaces, together with overall cutflow agreement, gives confidence that the event selections were implemented correctly.

\section{\label{sec:spec_plots}Full mass spectra}

\begin{figure*}
    \centering
        \begin{subfigure}[b]{0.49\textwidth}
         \includegraphics[width=0.95\textwidth]{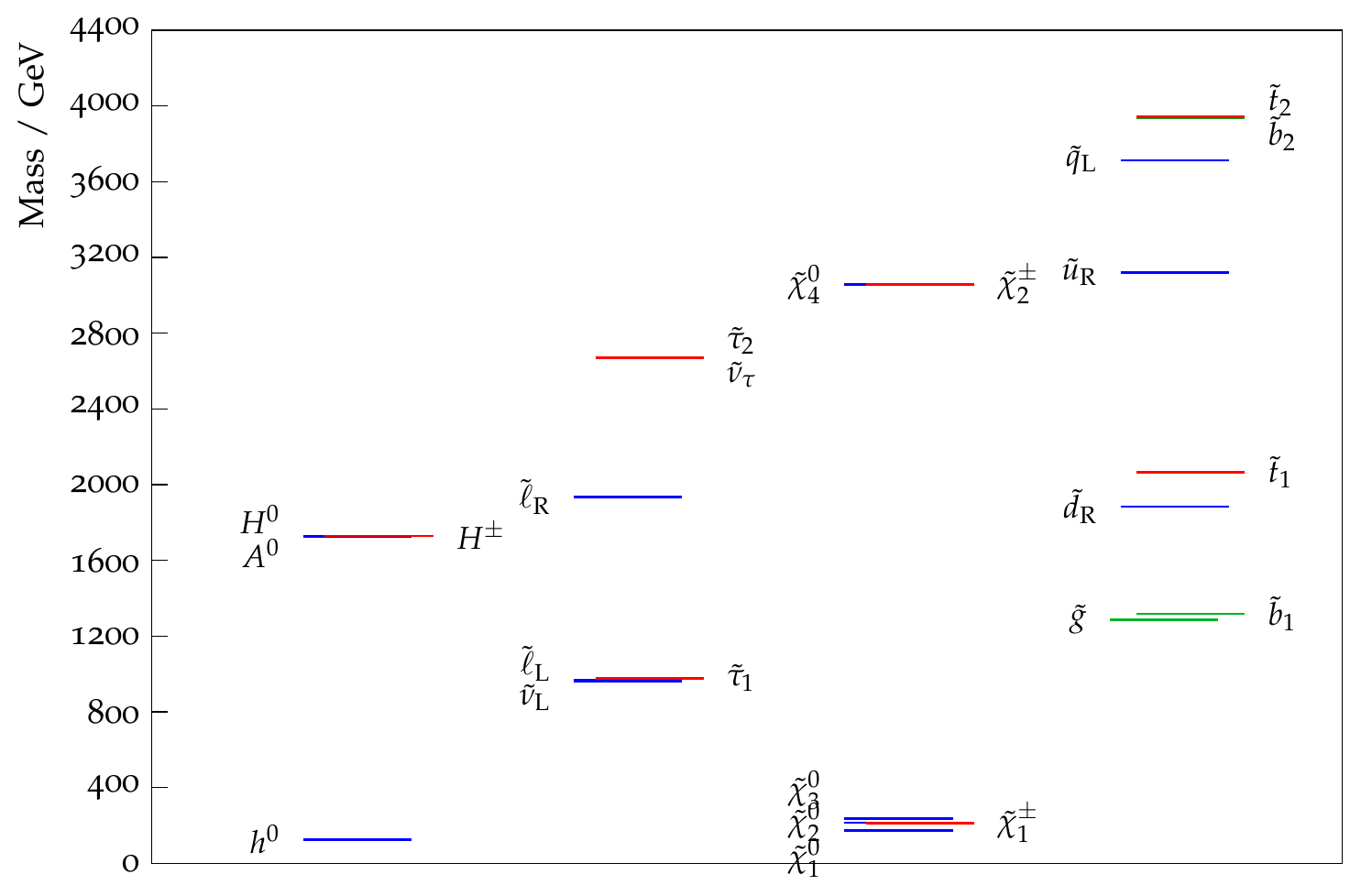}
         \caption{\label{fig:full_Multi_b_model}Model 148229034 (Multi-b).}
        \end{subfigure}
        \begin{subfigure}[b]{0.49\textwidth}
         \includegraphics[width=0.95\textwidth]{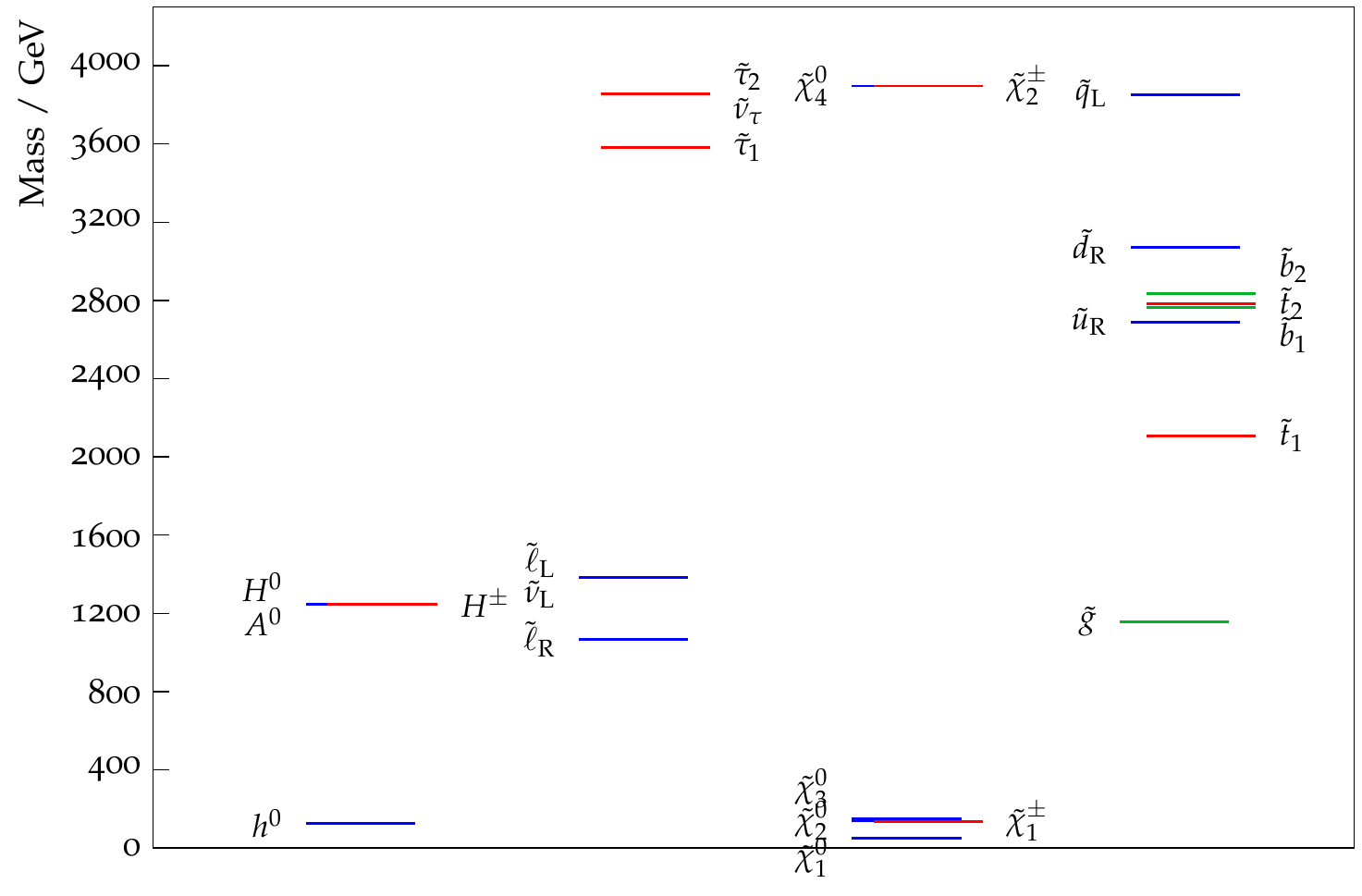}
         \caption{\label{fig:full_7_10_jets_model}Model 227558023 (7--10 jets).}
        \end{subfigure}
        \\[0.8cm]
        \begin{subfigure}[b]{0.49\textwidth}
         \includegraphics[width=0.95\textwidth]{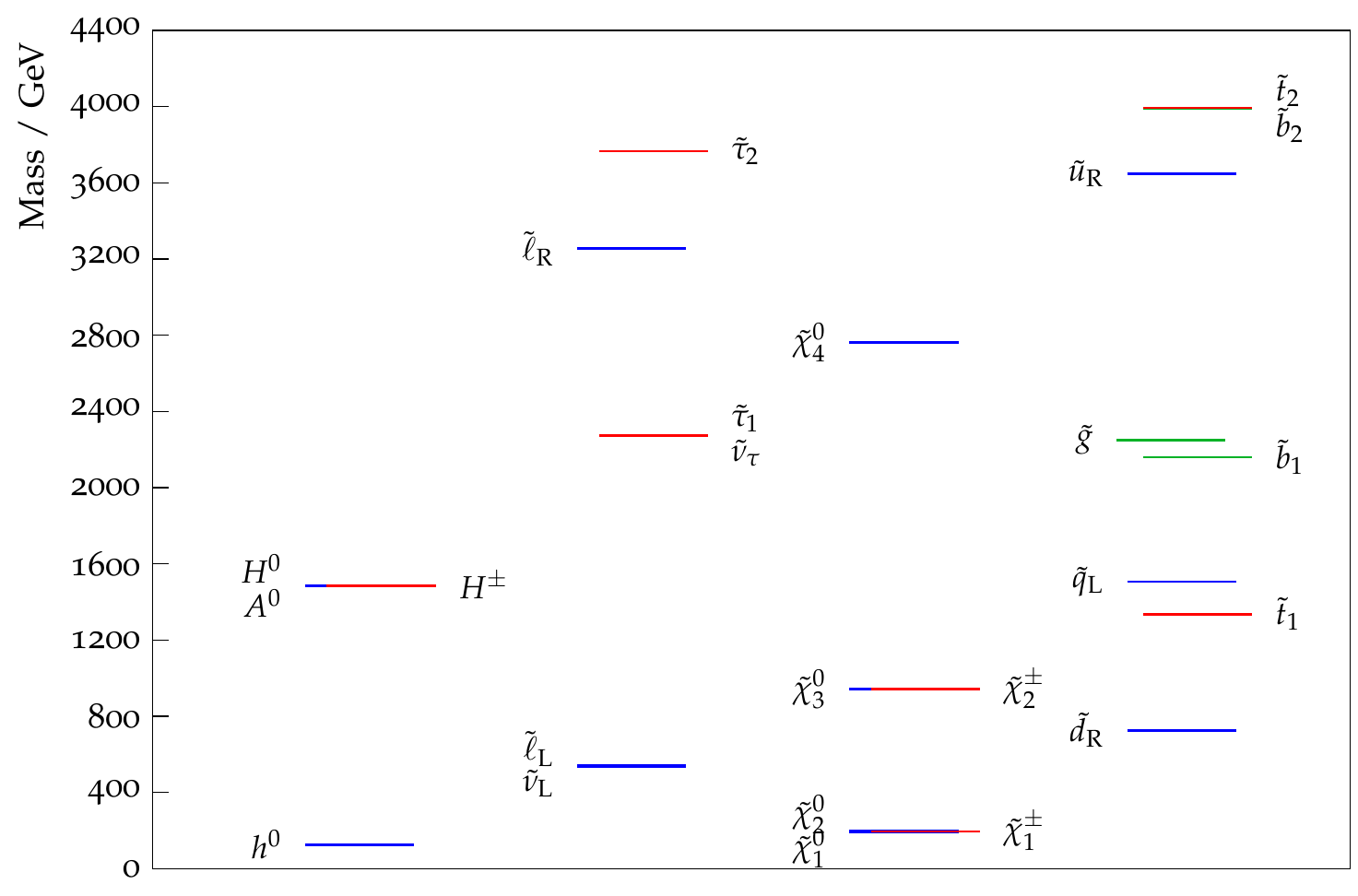}
         \caption{\label{fig:full_1_lepton_model}Model 13382371 (1-lepton).}
        \end{subfigure}
        \begin{subfigure}[b]{0.49\textwidth}
         \includegraphics[width=0.95\textwidth]{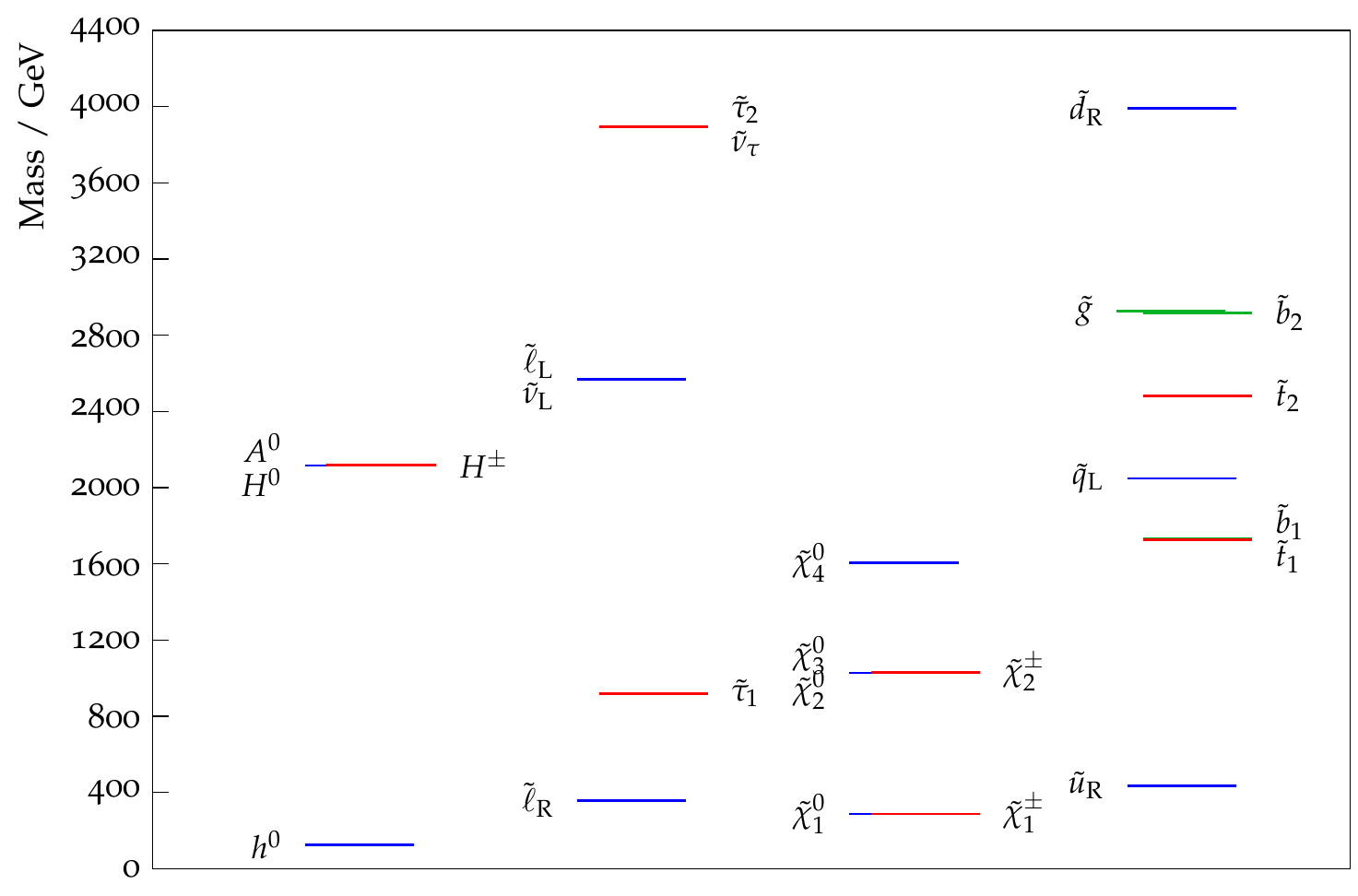}
         \caption{\label{fig:full_SS3L_light_Sqk}Model 11733067 (SS/3L).}
        \end{subfigure}
        
    \caption{The same model points as those in figure~\ref{fig:mass_spec} with full mass spectra displayed but without branching ratios for clarity.
    }
    \label{fig:full_mass_spec}
\end{figure*}

Figure~\ref{fig:full_mass_spec} displays the full mass spectra of the model points corresponding to those in Figure~\ref{fig:mass_spec} of Section~\ref{sec:compl_searches}.

\clearpage

\end{document}